\documentclass{jfm}
\usepackage{graphicx}
\usepackage{float}

\usepackage{xpatch}

\usepackage{amsmath} 
\usepackage{scalerel}
\usepackage{tikz}
\usetikzlibrary{svg.path}

\definecolor{orcidlogocol}{HTML}{A6CE39}
\tikzset{
  orcidlogo/.pic={
    \fill[orcidlogocol] svg{M256,128c0,70.7-57.3,128-128,128C57.3,256,0,198.7,0,128C0,57.3,57.3,0,128,0C198.7,0,256,57.3,256,128z};
    \fill[white] svg{M86.3,186.2H70.9V79.1h15.4v48.4V186.2z}
                 svg{M108.9,79.1h41.6c39.6,0,57,28.3,57,53.6c0,27.5-21.5,53.6-56.8,53.6h-41.8V79.1z M124.3,172.4h24.5c34.9,0,42.9-26.5,42.9-39.7c0-21.5-13.7-39.7-43.7-39.7h-23.7V172.4z}
                 svg{M88.7,56.8c0,5.5-4.5,10.1-10.1,10.1c-5.6,0-10.1-4.6-10.1-10.1c0-5.6,4.5-10.1,10.1-10.1C84.2,46.7,88.7,51.3,88.7,56.8z};
  }
}

\newcommand\orcidicon[1]{\href{https://orcid.org/#1}{\mbox{\scalerel*{
\begin{tikzpicture}[yscale=-1,transform shape]
\pic{orcidlogo};
\end{tikzpicture}
}{|}}}}
\usepackage[colorlinks=true, allcolors=blue]{hyperref}
\usepackage{caption}
\usepackage{subcaption}
\usepackage{tikz}
\usepackage[export]{adjustbox}
\usepackage[normalem]{ulem}

\newcommand{\flavio}[1]{\textcolor{black}{#1}}

\usepackage{epstopdf, epsfig}
\usepackage{xcolor}

\shorttitle{Optimal transitional mechanisms of separated shear layers}
\shortauthor{Savarino, Sipp \& Rigas}

\title{Optimal transitional mechanisms of incompressible separated shear layers subject to external disturbances}

\author{Flavio Savarino\aff{1}\corresp{\email{flavio.savarino17@imperial.ac.uk}}, Denis Sipp\aff{2} and Georgios Rigas\aff{1}}

\affiliation{\aff{1} Department of Aeronautics, Imperial College London, Exhibition Rd, London SW7 2AZ, UK
             \aff{2} DAAA, ONERA, Institut Polytechnique de Paris, 92190, Meudon, France}

\begin{document}

\maketitle

\begin{abstract}
Optimal transitional mechanisms are analysed for an incompressible shear layer developing over a short, pressure gradient-induced laminar separation bubble (LSB) with peak reversed flow of 2\%. Although the bubble remains globally stable, the shear layer destabilises due to the amplification of external time- and spanwise-periodic disturbances. Using linear resolvent analysis (RA), we demonstrate that the pressure gradient modifies boundary layer receptivity, shifting from Tollmien-Schlichting (T-S) waves and streaks in a zero pressure gradient (ZPG) environment to Kelvin-Helmholtz (K-H) and centrifugal instabilities in the presence of the LSB. To characterise the non-linear evolution of these disturbances, we employ the Harmonic-Balanced Navier-Stokes (HBNS) framework, solving the Navier-Stokes equations in spectral space with a finite number of Fourier harmonics. Additionally, adjoint optimisation is incorporated to identify \flavio{forcing disturbances that maximise the mean skin friction drag, conveniently chosen as the cost function for the optimisation problem since it is commonly observed to increase in the transitional stage}. Compared to attached boundary layers, this transition scenario exhibits both similarities and differences. While oblique T-S instability is replaced by oblique K-H instability, both induce streamwise rotational forcing through the quadratic non-linearity of the N-S equations. However, in separated boundary layers, centrifugal instability first generates strong streamwise vortices due to multiple centrifugal resolvent modes, which then develop into streaks via lift-up. Finally, we show that the progressive distortion and disintegration of K-H rollers, driven by streamwise vortices, lead to \flavio{the breakdown of large coherent structures}.
\end{abstract}


\section{Introduction\label{sec:introduction}}
Laminar separation bubbles (LSBs) appear in low-to-medium Reynolds number flows ($\mathrm{Re} \sim 10^3–10^5$ based on the chord of a typical wing section) at both low-speed incompressible and high-speed compressible flow regimes. Low-speed scenarios include unmanned aerial vehicles (UAVs), ground vehicles, gas turbines and turbofan engine compressors, while high-speed examples are supersonic engine intakes and super/hypersonic vehicles where the impingement of shock waves on aerodynamic surfaces may give rise to shock-induced flow separation \citep{babinsky_harvey_2011}. Their presence has been associated to important changes in the aerodynamic performance of engineering systems such as drag and acoustic noise increase, structural vibrations and degradation of stability characteristics. For such reasons of engineering interest, much research has focused on the description of the LSB physics, including the mean bubble topology, the study of stability of the shear layer, its breakdown to turbulence and ways to control the separation length (see review by \citet{Jahanmiri2011LaminarSB} and references therein). While the 2D mean bubble characteristics, namely the separation and reattachment points, the dead-air and core vortex regions in the fore and aft parts of separation, and the formation of a laminar separated shear layer, are well established thanks to the seminal work of \citet{gaster1963,tani1964,horton1968,o_meara1987}, the three-dimensional spatio-temporal dynamics of this complex flow is not fully understood, especially due to the non-linear, multi-modal and multi-scale behaviour of the shear layer  \citep{alam1999,alam_sandham_2000}.

\subsection{Linear Stability studies}\label{subsec:linear stability studies}
Several studies have focused on the receptivity of the shear layer to free-stream disturbances \citep{rist_maucher_2002,rist2002,marxen_henningson_2011,Simoni2013experimentalwakes,Marxen2013vortex,Simoni2014experimentalLSB,Michelis2017impulsiveforcing,Yarusevych2017steady,Michelis2018spanwise,Michelis2018origin,Zhang2019input-output,hosseinverdi_fasel_2019,Karp2020suppression}, since it constitutes the main site where fluctuations are amplified, providing the noise-amplifier behaviour of the flow. For this reason, the fundamental mechanisms underlying the generation of unsteadiness in LSBs have been investigated by means of local linear stability theory (LST) \citep{Yarusevych2017steady,Michelis2017impulsiveforcing,Ziade2018shear,Michelis2018spanwise,Michelis2018origin} and global approaches accounting for both true and pseudo-resonances of the linearised N-S operator, the former by means of the generalised eigenvalue problem \citep{theofilis2000,Gallaire2007three-dimensional,Rodriguez_gennaro_juniper_2013} and the latter through the resolvent operator \citep{Zhang2019input-output,yeh_benton_taira_garmann2020,Cura_Hanifi_Cavalieri_Weiss_2024}. This distinction stems from the continuous effort of the community to explain the origins of unsteadiness in LSBs both in the presence of external disturbances and at low levels of free-stream turbulence where unsteadiness appears spontaneously.

In the literature, the classification of LSBs is commonly based on the maximum reversed flow parameter \flavio{$u_{rev}=\left |u_{\text{min}}\right |/U_{\infty}$, where $u_{\text{min}}$ is the minimum (most negative) value of streamwise velocity in the flow field and $U_{\infty}$ the free-stream velocity}. \citet{theofilis2000} have found that an intrinsic instability appears in LSBs without the presence of external disturbances for magnitudes far less than 15-20\%, which was originally the threshold found for the onset of self-sustained unsteadiness in the flow. \citet{theofilis2000} have shown that a faster route to transition involves the emergence of a primary unstable mode at peak reversed flow as low as 7-8\%. Much effort was made to explain the physical origin of this mode. Using the generalised eigenvalue problem approach for the detection of global instability mechanisms \citet{Gallaire2007three-dimensional,Rodriguez_gennaro_juniper_2013,Rodriguez2015secondary,Rodriguez2021self-excited} have found that the unstable mode is stationary in time and three-dimensional with spanwise periodicity. More specifically, the global mode displays spanwise-periodic patches of streamwise velocity fluctuations with characteristic wavelength of approximately the mean bubble height. \citet{Zhang2016biglobal} also found a similar mode in the separation bubble developing on a low Reynolds number NACA0012 aerofoil and observed a dependency on the reversed flow magnitude. They concluded that this intrinsic mechanism arises from the presence of local flow curvature, especially pronounced near the reattachment point, and ascribed it to a modal centrifugal instability. \citet{Rodriguez_gennaro_juniper_2013} have found that this instability breaks the two-dimensionality of the separation bubble by producing spanwise inflections that give rise to a characteristic peak-valley undulation of the bubble. A similar mode featuring low spanwise wavenumbers (corresponding to wavelengths of several separation bubble lengths) and low frequencies in the Strouhal number range of 0.01 was discovered even in turbulent separation bubbles \citep{Cura_Hanifi_Cavalieri_Weiss_2024} and ascribed to the low-frequency breathing dynamics of the bubble. 

Conversely, globally stable LSBs ($u_{rev}<7\%$) do not display any self-excited instability mechanism, and, therefore, their route to turbulence is determined by the amplification of external disturbances through the receptivity process of the shear layer. In this convectively unstable regime, both local (LST) and global (RA) methods are viable tools to describe the linear stage of the disturbance field dynamics.

\flavio{Standard LST is based on the assumption of locally parallel flow, which is not representative of the topology of the LSB flow due to strong mean-flow gradients arising from the presence of separation. Nonetheless, a number of studies have successfully applied the ``local'' assumption (at each streamwise station the flow appears approximately parallel)} for the identification of selective frequencies at which external disturbances are optimally amplified. For example, \citet{Yarusevych2017steady} found that the LSB forming on a NACA0012 aerofoil at 2 degrees incidence excites high-frequency disturbances in the range of the natural shear layer shedding frequency. Similar results were obtained by \citet{Michelis2017impulsiveforcing}, who captured the linear evolution of the most unstable instability of the shear layer accurately with LST but observed its little success in the aft part of the bubble, where non-linear effects and 3D structures are dominant. \citet{Ziade2018shear,Michelis2018spanwise,Michelis2018origin} recovered the frequency of the K-H instability in the shear layer and found a clear conformity with the natural shedding frequency, concluding that the main vortex shedding mechanism is driven by the K-H mode. These results were also confirmed by experimental data from \citet{Simoni2013experimentalwakes,Yarusevych2017steady}. 

Additional insight into the most unstable convective mechanism of the shear layer is provided by the linear resolvent studies of \citet{Zhang2019input-output,yeh_benton_taira_garmann2020}. The advantage of the global resolvent approach is that both modal and non-modal mechanisms and non-parallel effects are accounted for. In the convectively unstable regime, it is insightful to treat the fluid flow system as a transfer function that maps a given input to its corresponding output according to the system dynamics and seek for optimal amplification mechanisms. In other words, an input/output (I/O) relationship can be formulated between the external disturbance (input) and response of flow (output) based on the linearised N-S equations around an equilibrium (fixed-point) solution of the steady N-S equations \citep{schmid_henningson2001stability}. \citet{Zhang2019input-output} has shown that the resolvent approach is able to identify the optimal forcing-response mechanisms of forced LSBs. Forcing must be optimally applied at the natural shedding frequency in order to trigger maximal response of the flow in terms of its energy. The forcing location being upstream of separation and the response peaking in the aft part of the LSB close to reattachment highlight the tendency of the shear layer to amplify external disturbances throughout the separation region.

Linear analysis has also been applied to control the LSB dynamics and ultimately the separation distance. \citet{Michelis2017impulsiveforcing} have studied the effects of impulsive forcing on convectively unstable, short bubbles and have observed changes in their stability properties, therefore presenting a viable option to control the flapping and bursting mechanisms. \citet{Michelis2018spanwise} have found that the natural shedding frequency recovered by LST from the unforced baseline constitutes the optimal forcing frequency at which the coherence of the spanwise rollers shed by the shear layer is enhanced. This phenomenon resulted in the suppression of separation. With a similar goal, \citet{Karp2020suppression} applied linear transient growth analysis to find an initial perturbation shape to minimise the bubble length. Although different to the optimal non-linear forcing, linear analysis provided a good estimate to reduce separation. 

\subsection{Non-linear dynamics and transition}\label{subsec:non-linear dynamics and transition}
While the evolution of the primary K-H mode over the fore part of the separation region is substantially linear \citep{diwan_ramesh_2009}, non-linear interactions and saturation occur near the reattachment point, where disturbances reach finite amplitude and 3D structures appear. From the experimental investigations of \citet{Simoni2013experimentalwakes,Simoni2014experimentalLSB,Yarusevych2017steady}, the natural transition scenario of LSBs involves the breakdown of the large-scale, spanwise-uniform vortices shed by the shear layer into small scales. The process according to which coherent vortices break up is dominated by non-linearities requiring methods such as large eddy simulation (LES) and direct numerical simulation (DNS) to capture the transition. 

Systematic DNS investigations were carried by \citet{alam1999,alam_sandham_2000,rist_maucher_2002}. The fastest route to turbulence was obtained with the interaction of a pair of oblique waves $(\pm1\beta,1\omega)$ generating longitudinal vortices in the reattachment zone, results later confirmed by \citet{rist2002}. The simulations of \citet{marxen_henningson_2011,Marxen2013vortex} elucidated the mechanisms underlying the spanwise deformation of coherent large-scale vortices. While the substantially 2D shedding of vortices was attributed to the K-H instability, a new stationary mode $(2\beta,0\omega)$ was found to induce a spanwise modulation in the separation region due to its strong action on the mean-flow at high amplitude. Transition was achieved by forcing the LSB with $(0\beta,1\omega)$ and $(\pm1\beta,1\omega)$ waves, similar to the forcing scenario employed by \citet{Michelis2018spanwise} to explain the origins of vortex deformation along the spanwise direction. In this case, they proposed a simple model based on the synergistic action of planar and oblique waves to describe the amount of spanwise deformation of the main vortex. More recently, \citet{hosseinverdi_fasel_2019,Hosseinverdi2020onset} studied the transitional scenarios of LSBs subject to various levels of free-stream turbulence using DNS. In the unforced case, they found that the natural transition of the shear layer involves the breakdown of largely 2D vortex filaments, in agreement with \citet{marxen_henningson_2011}. With increasing levels of perturbations they observed the appearance of Klebanoff modes \citep{klebanoff_tidstrom_sargent_1962} on top of the 2D/weakly oblique K-H waves. These modes deteriorate the spanwise uniformity of the K-H rollers and promote the development of more complex 3D vortices which become unstable and breakdown to turbulence. Further analysis via Proper Orthogonal Decomposition (POD) revealed the appearance of a largely steady (very low-frequency), streamwise-elongated mode in the POD spectrum with smaller energy content compared to the two dominant modes representing the 2D rollers. The authors found striking similarity between the steady POD mode and the Klebanoff mode appearing in the DNS calculations.

Overall, it is apparent that the transition process in LSBs essentially requires the breakup of large spanwise uniform structures shed by the shear layer, meaning that primary 2D disturbances must interact with secondary 3D/oblique instabilities. While there is consensus on the origins and physical description of the primary (K-H) shear layer instability, there is still debate on the secondary mechanisms that initiate the disintegration of the K-H rollers, which is ubiquitous in separated shear layers under low/medium disturbance levels \citep{yang2019}. Recently, \citet{mohamed_aniffa_caesar_dabaria_mandal_2023} argued that transition to turbulence in detached shear layers is not due to streak instability/breakdown as in attached boundary layers, but rather that the interaction of streaks with K-H rollers leads to spanwise vortex distortion and formation of $\Lambda$-structures, which precede breakdown into small-scale turbulence. 

With the present research, we aim at unravelling the dominant instability mechanisms that lead to turbulence in convectively unstable separated shear layers. Our analysis seeks optimal mechanisms over the entire transition process, with the primary focus being the explanation of the non-linear interactions across the multitude of perturbation scales and mechanisms at play. In this way, we address:
\begin{itemize}
    \item The optimal spatial structure and selective frequency range of external disturbances that lead to maximal amplification of instabilities within the shear layer, about which we are not aware of existing studies in the literature.
    \item Secondary instabilities leading to the distortion of large-scale K-H rollers \citep{Marxen2013vortex,yang2019,kumar2023}.
    \item The final stages of transition involving the breakdown of large structures shed by the shear layer. Several numerical (LES/DNS) \citep{alam_sandham_2000,rist_maucher_2002,rist2002,Jones2010stabilityaerofoil,Marxen2013vortex,Ziade2018shear,hosseinverdi_fasel_2019,Hosseinverdi2020onset,kumar2023} and experimental (particle image velocimetry, PIV) \citep{Simoni2013experimentalwakes,Simoni2014experimentalLSB,mohamed_aniffa_caesar_dabaria_mandal_2023,dellacasagrande_lengani_simoni_ubaldi2024} studies have explored the transitional/turbulent dynamical stages of the shear layer but with predetermined forcing, which is not the case of our work since we optimise the forcing to discover efficient pathways to turbulence.
\end{itemize}

Since linear stability methods can elucidate only the initial stages of amplification of perturbations, we seek optimal routes of laminar-turbulent transition by means of a gradient-based non-linear optimisation framework. To achieve this, we follow the framework introduced by \citet{rigas2021HBM}, applied to the non-linear N-S equations projected on a finite and tractable number of spanwise and frequency harmonics, known as HBNS. 
By progressively increasing the amplitude of perturbations, we track the evolution of the shear layer state progressively from linear to weakly non-linear to highly non-linear dynamics. This approach enables the identification of the optimal underlying instability mechanisms and the harmonics at play, using a self-contained computational tool which makes optimisation on such a large-scale problem tractable, while the range of spatio-temporal scales is progressively increased to facilitate interpretation of the underlying mechanisms.

\subsection{Structure of the paper}\label{subsec:structure of the paper}
The manuscript is organised as follows.
In \S\ref{sec:methodology}, we outline the methodology and provide the theory for the study of the linear and non-linear evolution of disturbances. 
The computational set-up follows in \S\ref{sec:config_numerics_algo}. 
In \S\ref{sec:base-flow}, we present the laminar base-flow calculation for variable adverse pressure gradient (APG) magnitude and briefly outline the global stability of the LSB. 
In \S\ref{sec:Effect of adverse pressure gradient and separation on linear convective instabilities}, we present the optimal linear receptivity mechanisms of APG boundary layers with and without separation.
In \S\ref{sec:optimal non-linear disturbances}, we show results from the non-linear analysis and explain the various instability mechanisms responsible for the transition of the shear layer.
Finally, conclusions are drawn in \S\ref{sec:conclusions}.

\section{Methodology\label{sec:methodology}}
In this section we provide an overview of the theory underlying the study of the evolution of both linear and non-linear perturbations in the LSB flow. We first introduce the governing equations followed by formulations for linear input/output (resolvent) and non-linear input/output analyses.

\subsection{Governing equations\label{subsec:governing equations}}
For the study of stability and transition of an incompressible APG boundary layer forced by external disturbances, we consider the N-S equations,
\begin{equation}
    \partial_t \boldsymbol{u} + \boldsymbol{u} \cdot \nabla \boldsymbol{u} + \nabla p - \nu \nabla^2 \boldsymbol{u} = A \boldsymbol{f}'(\boldsymbol{x},t), \;\;\;\;
    \nabla \cdot \boldsymbol{u} = 0,
    \label{eq:NS}
\end{equation}
\noindent where $\boldsymbol{u}=[u,v,w]^{\top}$ is the velocity vector with components in the streamwise ($x$), wall-normal ($y$) and spanwise ($z$) directions, $p$ is the pressure, $\nu$ is the kinematic viscosity of the fluid and $\boldsymbol{f}'=[f_{x}',f_{y}',f_{z}']^{\top}$ is the external volumetric forcing vector of amplitude $A$ (the $'$ denotes the fluctuation around the zero mean).

\flavio{Equations \eqref{eq:NS} are solved on a rectangular domain where we impose a laminar Blasius velocity profile at inlet, zero-stress boundary condition at the outlet, no-slip at the plate wall and a zero-net mass-flux (ZNMF) suction-blowing velocity profile at the top boundary (more details provided in \S\ref{sec:config_numerics_algo}).}


\flavio{Introducing the velocity-pressure state vector $\boldsymbol{w}=[\boldsymbol{u},p]^{\top}$, treating $x$ and $y$ as discrete spatial directions, while the $z$ spatial direction and time $t$ as continuous, we arrive at the semi-discrete N-S equations,
\begin{equation}
    \mathsfbi{M} \partial_t \boldsymbol{w} + \mathsfbi{L}\boldsymbol{w} + \mathsfbi{N}(\boldsymbol{w},\boldsymbol{w})/2= A \mathsfbi{M}\mathsfbi{P} \boldsymbol{f}'(z,t),
    \label{eq:semi-discrete NS}
\end{equation}}
\noindent where $\mathsfbi{M}=\begin{pmatrix} \mathsfbi{M}' & \textbf{0} \\ \textbf{0} & 0 \end{pmatrix}$ is the mass matrix accounting for the spatial discretisation, $\mathsfbi{L}=\begin{pmatrix} -\nu \nabla^2 () & \nabla() \\ \nabla \cdot () & 0 \end{pmatrix}$ is the linear Stokes operator, $\mathsfbi{N}=\begin{pmatrix} \boldsymbol{u}_{1} \cdot \nabla \boldsymbol{u}_{2} + \boldsymbol{u}_{2} \cdot \nabla \boldsymbol{u}_{1} \\ 0 \end{pmatrix}$ is the symmetrised non-linear convection operator and $\mathsfbi{P}$ is the prolongation operator mapping the velocity state $[u,v,w]^{\top}$ to the full velocity-pressure state $[u,v,w,p]^{\top}$ by adding a zero component on the pressure. \flavio{This stems from the fact that forcing acts only on the three momentum equations but not on the continuity equation. Since the flow response is constituted by 4 states (velocities $u$, $v$ and $w$ and pressure $p$), the prolongation operator becomes necessary to match the size of the left and right hand sides of \eqref{eq:semi-discrete NS}.}

The amplitude of the external volumetric forcing $A$ is used as a discriminant between linear and non-linear analyses described in the next sections:
\begin{itemize}
    \item If $A\ll1$, which implies that the forced system behaves linearly, and provided the operator of the linearised system dynamics is stable, we can derive an input/output relationship between the perturbations and the external forcing to find selective frequencies over which the energy of the perturbations is maximised -- RA \S\S\ref{subsec:linear resolvent analysis};
    \item If $A\sim \mathcal{O}(1)$, the system behaves non-linearly and we need to solve the N-S equations to account for non-linear interactions at perturbation level and also between the perturbations and the mean-flow -- HBNS \S\S\ref{subsec:nonlinear input-output analysis with HBNS}.
\end{itemize}

\makeatletter
\let\@float@original\@float
\xpatchcmd{\@float}{\csname fps@#1\endcsname}{h!}{}{}
\makeatother


\subsection{Linear Resolvent Analysis\label{subsec:linear resolvent analysis}}
The linear dynamics in presence of a small amplitude forcing $A\ll1$ 
can be studied by linearising the N-S equations \eqref{eq:semi-discrete NS} around a fixed (or equilibrium) point of the system, which, in the context of flow stability, we call base-flow. We therefore decompose the state as
\begin{equation}    
    \boldsymbol{w}(z,t) = \boldsymbol{w}_{0,0} + A \boldsymbol{w}'(z,t) + \mathcal{O}(A^2),
    \label{eq:decomposition}
\end{equation}
where the subscripts of $\boldsymbol{w}_{m,n}$ are referred to the spanwise wavenumber $m\beta$ and to the angular frequency $n\omega$ to denote a specific scale (or harmonic) of the flow and will be used throughout the manuscript. Because the base-flow is steady and two-dimensional $m=n=0$. If we now inject \eqref{eq:decomposition} in \eqref{eq:semi-discrete NS}, we derive the steady N-S equations for the laminar base-flow at $A^0$,
\begin{equation}
     \mathsfbi{L}_0\boldsymbol{w}_{0,0} +  \mathsfbi{N}_0^0(\boldsymbol{w}_{0,0},\boldsymbol{w}_{0,0})/2= 0,
     \label{eq:steady NS}
\end{equation}
\noindent where $\mathsfbi{L}_{m}$ is extracted for the harmonic $m=0$ and $\mathsfbi{N}_{m_{1}}^{m_{2}}(\boldsymbol{w}_{m_{1},n_{1}},\boldsymbol{w}_{m_{2},n_{2}})$ is calculated here on the base-flow such that $m_{1}=m_{2}=0$.

We can analyse the system's linear response to the applied forcing according to
\begin{equation}
    \mathsfbi{M} \partial_t \boldsymbol{w}' + \mathsfbi{L}\boldsymbol{w}' + \mathsfbi{N}_{0}(\boldsymbol{w}_{0,0},\boldsymbol{w}') = A \mathsfbi{M}\mathsfbi{P} \boldsymbol{f}'(z,t),
    \label{eq:linearised forced NS}
\end{equation}
\noindent which are the linearised N-S equations with the zero-mean forcing term appearing in the right-hand side. Again, assuming modal behaviour for both the forcing and the response,
\begin{equation} 
    \boldsymbol{f}'(z,t) = \hat{\boldsymbol{f}}(\beta,\omega) \exp\left [ \mathrm{i}\left (\beta z + \omega t \right ) \right ], \;\;\; 
    \boldsymbol{w}'(z,t) = \hat{\boldsymbol{w}}(\beta,\omega) \exp\left [ \mathrm{i}\left (\beta z + \omega t \right ) \right ], 
    \label{eq:modal expansion resolvent}
    \end{equation}
\noindent where in this case $\omega$ is a real scalar denoting the angular frequency, and assuming the amplitude of the response $A$ is of the same order of $A$, we can inject \eqref{eq:modal expansion resolvent} into \eqref{eq:linearised forced NS} and obtain
\begin{equation}
    \hat{\boldsymbol{w}}(\beta,\omega) = \mathsfbi{H}(\beta,\omega) \mathsfbi{M}\mathsfbi{P} \hat{\boldsymbol{f}}(\beta,\omega),
    \label{eq:linear resolvent}
\end{equation}
\noindent where the transfer function between the forcing disturbance (input) and the response (output), $\mathsfbi{H}(\beta,\omega) = \left [ \mathrm{i} \omega \mathsfbi{M} - \mathsfbi{A}_{\mathrm{i}\beta}(\boldsymbol{w}_{0,0}) \right ]^{-1}$, is the resolvent operator.

Through the input/output formulation of the resolvent, we can study the receptivity of the boundary layer to infinitesimal external disturbances. That means, we can identify selective frequencies and spanwise wavenumbers where the external forcing produces maximal energy of the response. This is particularly insightful in globally stable noise-amplifier flows, which can transition and sustain turbulence only if external disturbances are supplied to the flow.

The optimal forcing shape $\hat{\boldsymbol{f}}$ that leads to maximal energy of the response at a given frequency $\omega$ and spanwise wavenumber $\beta$ is found through an optimisation procedure that employs the (squared) energy gain
\flavio{\begin{equation}
    \underset{\hat{\boldsymbol{f}}}{\text{max}} \left \{ \mathcal{G}^2 (\hat{\boldsymbol{f}}; \beta, \omega) = \frac{\left \langle \hat{\boldsymbol{w}}(\hat{\boldsymbol{f}}),\hat{\boldsymbol{w}}(\hat{\boldsymbol{f}}) \right \rangle_{\Omega_{\hat{\boldsymbol{w}}}}}{\left \langle \hat{\boldsymbol{f}},\hat{\boldsymbol{f}} \right \rangle_{\Omega_{\hat{\boldsymbol{f}}}}}\right \},
    \label{eq:energy gain}
\end{equation}}
\noindent as the cost function to be maximised. The energy norms of the forcing and response modes are defined as
\begin{equation}
    \left \langle \hat{\boldsymbol{f}},\hat{\boldsymbol{f}} \right \rangle_{\Omega_{\hat{\boldsymbol{f}}}} = \hat{\boldsymbol{f}}^{\mathrm{H}} \mathsfbi{Q}_{\hat{\boldsymbol{f}}} \hat{\boldsymbol{f}}, \;\;\; \quad
    \left \langle \hat{\boldsymbol{w}},\hat{\boldsymbol{w}} \right \rangle_{\Omega_{\hat{\boldsymbol{w}}}} = \hat{\boldsymbol{w}}^{\mathrm{H}} \mathsfbi{Q}_{\hat{\boldsymbol{w}}} \hat{\boldsymbol{w}}.
    \label{eq:energy norms}
\end{equation}
\noindent where $\mathsfbi{Q}_{(\cdot)}$ are the mass matrices associated to the inner products and are chosen such that the forcing modes are unit norm, i.e. $\left \langle \hat{\boldsymbol{f}},\hat{\boldsymbol{f}} \right \rangle_{\Omega_{\hat{\boldsymbol{f}}}}=1$, and the response modes have norm $\left \langle \hat{\boldsymbol{w}},\hat{\boldsymbol{w}} \right \rangle_{\Omega_{\hat{\boldsymbol{w}}}}=\mathcal{G}^2$. The regions of the flow where the response and forcing modes live are indicated by $\Omega_{(\cdot)}$. In the case of the response, this corresponds to the entire flow domain, while forcing modes can be restricted spatially to mimic any particular scenario if required. The symbol $^{\mathrm{H}}$ denotes the Hermitian operation.

It can be shown that \eqref{eq:energy gain} leads to a generalised EVP,
\begin{equation}
    \mathsfbi{P}^{\mathrm{H}}\mathsfbi{M}\mathsfbi{H}^{\mathrm{H}}\mathsfbi{Q}_{\hat{\boldsymbol{w}}}\mathsfbi{H}\mathsfbi{M}\mathsfbi{P} \hat{\boldsymbol{f}} = \mathcal{G}^2 \mathsfbi{Q}_{\hat{\boldsymbol{f}}} \hat{\boldsymbol{f}},
    \label{eq:generalised EVP resolvent}
\end{equation}
\noindent where $\mathcal{G}_{i}^{2}=\lambda_{i}$ are the set of eigenvalues and $\hat{\boldsymbol{f}}_{i}$ are the set of orthonormal eigenvectors ranked in descending order of $\mathcal{G}_{i}^{2}$ ($i=1,2,3,\dots$). The leading eigenvector is therefore the optimal forcing mode at a given frequency and spanwise wavenumber pair, followed by eigenvectors associated with smaller eigenvalues which are referred to as sub-optimal forcing modes.

The framework described is global, in the sense that it is valid for non-parallel flows such as APG boundary layers and separated boundary layers. It is also capable of analysing non-modal/transient instability mechanisms, which the generalised EVP cannot. Finally, the optimal forcing can be quite general since no initial assumption is made neither on its shape nor on its spatial region of action. The limitation of linear RA is that the response of the system can only be calculated at the same frequency and spanwise wavenumber as the external forcing. This is because a linear framework cannot capture the interaction between different frequencies in the response, which is a feature of non-linearity. In \S\S\ref{subsec:nonlinear input-output analysis with HBNS} we retain the non-linear terms of the governing N-S equations since we relax the assumption of infinitesimal disturbances.

\subsection{Non-linear Input/Output Analysis with HBNS}\label{subsec:nonlinear input-output analysis with HBNS}
\flavio{For finite amplitude $A \sim \mathcal{O}(1)$ forcing disturbances, we have to consider the non-linear N-S equations \eqref{eq:semi-discrete NS}. In this case we consider the forcing and the velocity-pressure state to be the sum of several Fourier modes,
\begin{equation} 
    \boldsymbol{f}(z,t) = \hat{\boldsymbol{f}}_{0,0} + \sum_{\substack{m=-M,n=-N, \\ (m,n) \neq (0,0)}}^{M,N} \hat{\boldsymbol{f}}_{m,n}(x,y;m\beta,n\omega) \exp\left [ \mathrm{i}\left ( m \beta z + n \omega t \right ) \right ], 
    \label{eq:Fourier expansion}
\end{equation}
\noindent with a similar expansion for $\boldsymbol{w}(z,t)$, and where the symbol ${\hat{(\cdot)}}$ denotes the complex Fourier modes (or coefficients), $M$ and $N$ indicate the maximum order of spanwise and frequency harmonics, respectively, and $\beta$ and $\omega$ are the fundamental spanwise wavenumber and frequency appearing in the wave term $\exp\left [ \mathrm{i}\left ( m \beta z + n \omega t \right ) \right ]$. By construction of this expansion, an arbitrary number of multiples of the fundamental harmonic $\hat{\boldsymbol{w}}_{m=1,n=1}$ can be considered. The higher the number, the higher is the order of the non-linear system. The zeroth harmonic $\hat{\boldsymbol{w}}_{0,0}$ is the time- and spanwise-averaged velocity-pressure state, which we will be sometimes referred to as the mean-flow for brevity. While this is non-zero in the response, it is zero in the forcing expansion because we only apply forcing at the perturbation level, as initially expressed in the governing equations \eqref{eq:NS}. Furthermore, because the system's state, $\boldsymbol{w}$, is real, the symmetry condition $\hat{\boldsymbol{w}}_{-m,-n}=\hat{\boldsymbol{w}}^{*}_{m,n} $,
holds for all $(m,n)$, which therefore implies that $\hat{\boldsymbol{w}}_{0,0}$ is real. The symbol $^{*}$ denotes the complex conjugate. The same symmetry applies to the forcing.}

We now inject the expansion \eqref{eq:Fourier expansion} into \eqref{eq:semi-discrete NS} to derive the non-linear HBNS system,
\begin{equation} 
    \mathsfbi{R}(\hat{\boldsymbol{w}}) = A \mathsfbi{M}\mathsfbi{P}\hat{\boldsymbol{f}},
    \label{eq:HBNS system compact}
\end{equation}
\noindent where $\mathsfbi{R}(\hat{\boldsymbol{w}})$ contains all linear and non-linear terms. This system is obtained by balancing the harmonics in each equation, meaning the first equation describes the evolution of the mean-flow as it departs from the base-flow due to the modification imparted by the non-linear interactions between the perturbations and the mean-flow, the second equation satisfies the evolution of the $(0,1)$ harmonic, the third equation the $(0,2)$ harmonic, and so on until the $(M,N)$ harmonic. Because the system must have finite size to be solved computationally, higher order harmonics not included in the expansion constitute a truncation error that affects the accuracy of the discrete system.

The present framework, introduced by \citet{rigas2021HBM} for incompressible ZPG boundary layers, is a non-linear extension of the linear input/output (resolvent) analysis, since it allows the calculation of the multi-harmonic response of the system to any given finite amplitude forcing disturbance. For this reason, it is suitable for optimisation similarly to RA, except that the multi-harmonic response solution makes it possible to formulate more complex cost functions than the perturbation energy.

The optimal forcing shape $\hat{\boldsymbol{f}}$ is calculated by augmenting the HBNS framework with a constrained optimisation procedure. In this regard, we define a positive, real-valued, scalar cost functional $J(\hat{\boldsymbol{w}})$ to be maximised subject to the constraint that the response, $\hat{\boldsymbol{w}}$, resulting from the applied forcing, must be a solution of the HBNS system \eqref{eq:HBNS system compact}. 

In the context of transition to turbulence, we choose the cost functional to be the squared shear strain of the mean-flow modification integrated over the flat plate length,
\begin{equation}
    J(\hat{\boldsymbol{w}}) = J(\hat{\boldsymbol{w}}_{0,0}) = (\hat{\boldsymbol{w}}_{0,0} - \boldsymbol{w}_{0,0})^{\top}\mathsfbi{C}^{\top}\mathsfbi{C}(\hat{\boldsymbol{w}}_{0,0} - \boldsymbol{w}_{0,0}),
    \label{eq:cost functional}
\end{equation}
\noindent where $\mathsfbi{C}(\hat{\boldsymbol{w}}_{0,0} - \boldsymbol{w}_{0,0}) = \int (\partial (\hat{\boldsymbol{w}}_{0,0} - \boldsymbol{w}_{0,0})/\partial y)_{y=0}\:\mathrm{d}x$. It follows that the cost functional is related to the change in the overall skin friction drag on the plate as
\begin{equation}
    \Delta C_D = \frac{\nu J^{0.5}}{1/2U_{\infty}^{2}L_p},
    \label{eq:drag change}
\end{equation}
\noindent where $\nu$, $U_{\infty}$ and $L_{p}$ are the kinematic viscosity, free-stream velocity and plate length, respectively. This quantity is known to increase during transition due to the increased wall shear of the turbulent boundary layer, hence its physical relevance in our optimisation strategy. Refer to \citet{rigas2021HBM} for more details.


\section{Configuration, numerical discretisation and algorithms\label{sec:config_numerics_algo}}
\begin{figure}
    \centering
    \includegraphics[width=0.7\textwidth]{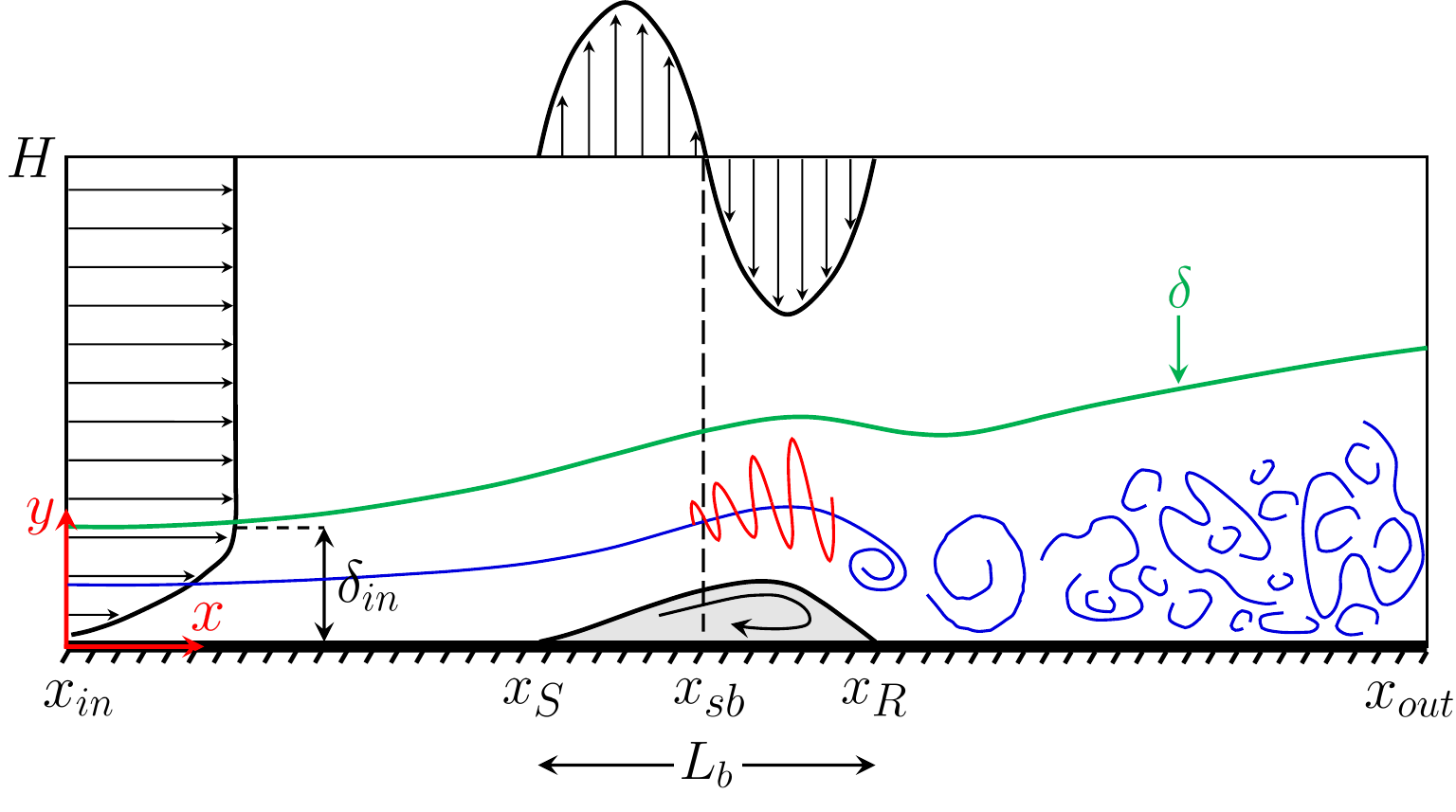}
    \caption{\flavio{Configuration for the calculation of the LSB. $x_S=$ separation point. $x_R=$ reattachment point. $L_b=x_R-x_S$ is the bubble length. Boundary layer thickness, $\delta$, and shear layer are qualitatively drawn in green and blue, respectively. Figure not to scale.}}
    \label{fig:domain}
\end{figure}

The set-up for the 2D base-flow calculation is shown in figure \ref{fig:domain}. Velocities are non-dimensionalised by the free-stream velocity $U_{\infty}$ and lengths by $L_{\mathrm{ref}} = \nu/U_{\infty}$, yielding the non-dimensional spatial coordinates $(x,y,z)\mapsto(\mathrm{Re}_x,\mathrm{Re}_y,\mathrm{Re}_z)$. The computational domain is rectangular where inlet and outlet boundaries are positioned at $x_{in} = 0.3\times10^5$ and $x_{out} = 1.8\times10^5$ with respect to the flat plate leading-edge. The top boundary is at $H = 0.24\times10^5$ away from the plate wall. Overall, the domain is $177 \delta_{in} \times 28 \delta_{in}$, where $\delta_{in} = 848$ is the $0.99U_{\infty}$ boundary layer thickness at inlet. The Reynolds number based on the displacement thickness, $\mathrm{Re}_{\delta^{*}}$, is 298 at inlet and 730 at the outlet. In order to calculate APG boundary layers, we impose a ZNMF suction-blowing velocity profile at the top boundary, qualitatively shown in figure \ref{fig:domain}. \flavio{This is a well-established boundary condition to generate model LSBs numerically \citep{Na_Moin_1998,Zhang2019input-output,hosseinverdi_fasel_2019} which essentially mimics experimental wind tunnel set-ups using displacement bodies or ZNMF actuators that impose an APG on the plate underneath \citep{diwan_ramesh_2009,griffin_oyarzun_cattafesta_tu_rowley_2013,Michelis2017impulsiveforcing,Yarusevych2017steady,Borgmann_Hosseinverdi_Little_Fasel_2025}.} The suction-blowing function adopted in our configuration reads
\begin{equation}
    V_{SB}\left ( x,y=H \right ) = -v_0 \: \overline{x} \: \exp \left ( \frac{1-\overline{x}^2}{2}  \right ),
    \label{eq:suction-blowing}
\end{equation}
\noindent whose formulation is identical to \citet{Karp2020suppression}, where $v_0$ is the amplitude and $\overline{x} = \frac{x - x_{sb}}{\Delta x_{sb} \exp (-\frac{1}{2})}$ is a non-dimensional parameter containing the coordinate location, $x_{sb} = \frac{x_2 + x_1}{2}$, and the width, $\Delta x_{sb}$, of the suction-blowing profile. Here, $x_2$ and $x_1$ are chosen to match the configuration of \citet{Karp2020suppression} for base-flow validation purposes (see \S\ref{app:validation of the LSB numerical set-up and comparison with literature}), such that $x_{sb} = 1.06\times10^5$. This coordinate effectively fixes the streamwise location of the bubble along the plate.

A 2D finite element discretisation of equation \eqref{eq:steady NS} is performed in \texttt{FreeFem++} \citep{freefemHecht}, where P1-bubble and P1 triangular elements are chosen for velocity and pressure states, respectively. A wall-normal stretching profile with 0.03\% stretching factor and $1/20 \delta_{in}$ first cell height is employed in the near-wall region ($0 \leq y \leq 4,000$), yielding 66 grid points of which approximately 20 inside the boundary layer to resolve the gradients. The remaining part of the domain ($4,000 < y \leq H$) is coarsely discretised with an unstructured mesh, since disturbances are unlikely in this region. Due to the strongly non-parallel nature of the LSB flow, large streamwise gradients are expected in the APG region, hence, a non-uniform streamwise discretisation is employed here. Elements of aspect ratio 10:1 (length $1/2 \delta_{in}$) are used upstream and downstream of the APG and a sine function is centred at $x=1.2\times10^{5}$ to progressively refine the streamwise grid spacing towards this coordinate, previously identified as the approximate location of the LSB vortex core. The finer element in this region of the mesh has aspect ratio 3:1 (length $3/20 \delta_{in}$). Overall, 415 points are used in the streamwise direction. A summary of the mesh parameters is provided in table \ref{tab:mesh}.

\begin{table}
  \begin{center}
    \def~{\hphantom{0}}
    \begin{tabular}{cccccccc}
      First element & Stretching & Max & Min & $N_x$ & $N_y$ & \# of elements & \\
      thickness & factor & aspect ratio & aspect ratio & & & & \\ 
       &  &  &  &  &  & Near-wall & Overall \\ [3pt]
      $1/20 \delta_{in}$ & 0.03\% & 10:1 & 3:1 & 415 & 66 & 53,820 & 68,694 \\
    \end{tabular}
    \caption{Mesh parameters for the base-flow and stability calculations.}
    \label{tab:mesh}
  \end{center}
\end{table}

We employ a root-finding Newton algorithm to compute the base-flow. The base velocity-pressure state is initialised with the Blasius boundary layer. By progressively increasing the pressure gradient through the boundary condition \eqref{eq:suction-blowing}, we obtain several solutions with and without separation.


For linear RA, we first construct the Jacobian and mass matrices for a given frequency-spanwise wavenumber pair $(\beta,\omega)$ and then invert the large sparse matrix $[\mathrm{i} \omega \mathsfbi{M} - \mathsfbi{A}_{\mathrm{i}\beta}(\boldsymbol{w}_{0,0})]$ to obtain the resolvent by calling the MUMPS. We perform a parametric study based on $\omega$ and $\beta$ to identify optimal linear 3D forcing-response mechanisms over a range of wavenumbers. We allow 3D disturbances to manifest everywhere in the domain but also impose the no-slip and inlet zero-perturbation boundary conditions and zero wall-normal perturbation at the far field boundary.

For non-linear I/O analysis we implement the iterative Newton algorithm on \eqref{eq:HBNS system compact} to solve for the system's non-linear response. This requires the calculation of the Jacobian $\mathsfbi{A} = \partial \mathsfbi{R}/ \partial \hat{\boldsymbol{w}}$ obtained by linearising the operator $\mathsfbi{R}(\hat{\boldsymbol{w}})$ around the time- and spanwise-averaged flow solution $\hat{\boldsymbol{w}}_{0,0,i}$ at the i-th iteration. The Jacobian is diagonally-dominant if the forcing amplitude $A$ is small since the leading diagonal blocks contain the linear terms, which are dominant at low amplitude, while the off-diagonal blocks contain the non-linear convective term. 
The diagonal dominance of $\mathsfbi{A}$ decreases as the non-linear term becomes progressively stronger at higher forcing amplitudes. Inversion of the Jacobian is performed using the block-Jacobi preconditioned Generalised Minimal RESidual (GMRES) algorithm from the PETSc library \citep{petsc}. For computational efficiency, the large HBNS system is handled by multiple cores using an MPI architecture, each core dealing with one harmonic. 

A Lagrange functional problem is formulated for the maximisation of the cost function in \eqref{eq:cost functional}, which naturally yields the linear system $\mathsfbi{A}^{\dagger} \Tilde{\boldsymbol{w}} = \mathrm{d}J/\mathrm{d}\hat{\boldsymbol{w}}$ to be solved for the adjoint state, $\Tilde{\boldsymbol{w}}$. This is used to evaluate the alignment of the adjoint state with the computed forcing vector $\hat{\boldsymbol{f}}$ which represents the convergence criterion (to a local maximum) of the optimisation problem. Convergence is met when the updated forcing is parallel to the adjoint, i.e. the angle between the two vectors satisfies $\cos \theta \rightarrow 1$. A modified Steepest Ascent Method is implemented to update the forcing at each iteration. More details provided in \citet{rigas2021HBM}.

\section{Base-flow\label{sec:base-flow}}

\subsection{Numerical base-flow solutions for different adverse pressure gradient magnitudes\label{subsec:numerical base-flow solutions for different adverse pressure gradient magnitudes}}
\begin{figure}
\centering
    \includegraphics[width=0.95\textwidth]{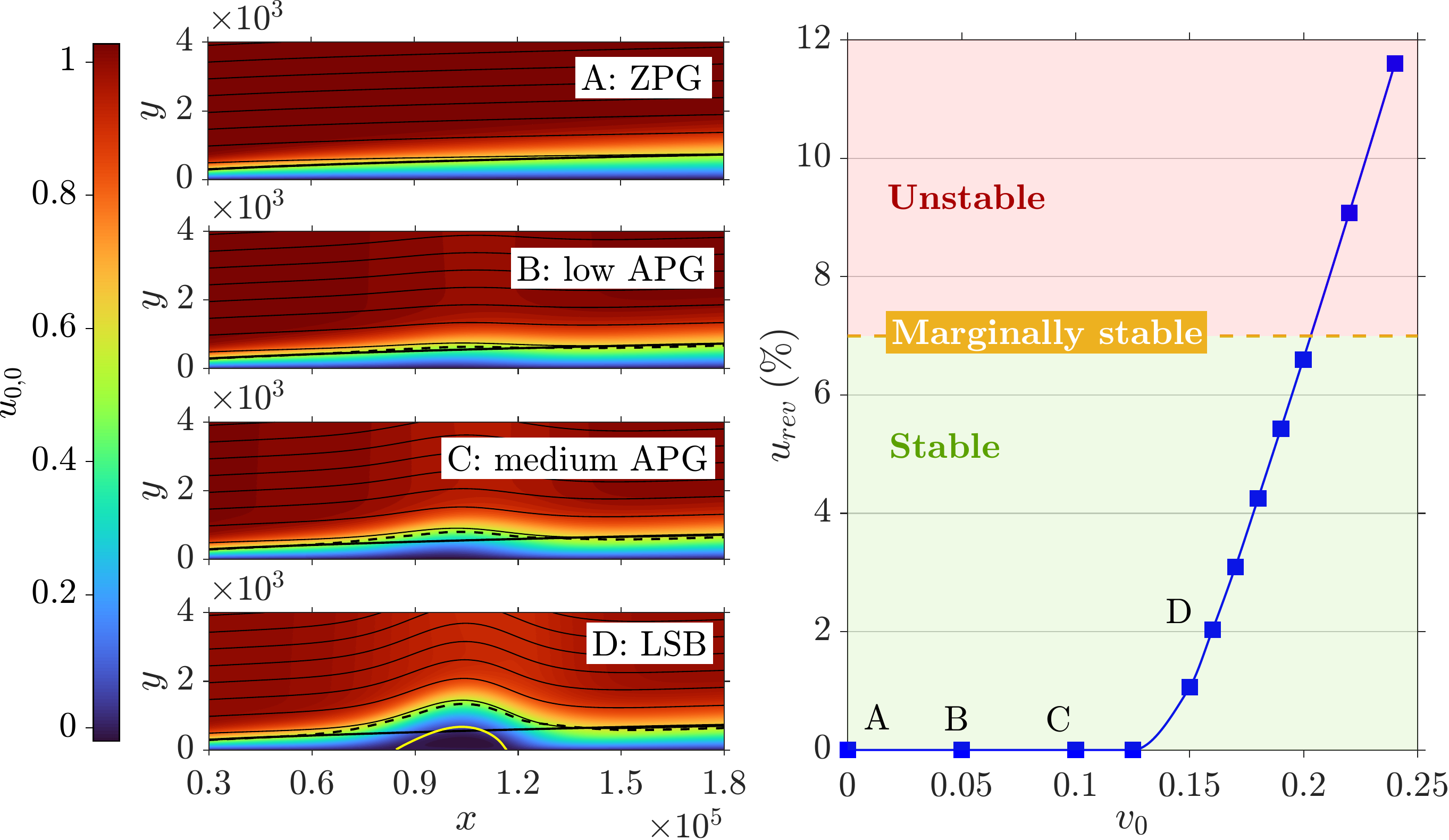}
    \caption{\flavio{(Left) Contours of streamwise velocity and streamlines from four base-flows computed at $v_0=0$ (A: ZPG boundary layer), $v_0=0.05$ (B: low APG boundary layer), $v_0=0.1$ (C: medium APG boundary layer) and $v_0=0.16$ (D: LSB with peak reversed flow $u_{rev}=2\%$). Displacement thickness from base-flow solution (black-dashed) and from Blasius solution (black-solid) superimposed. LSB dividing streamline (yellow-solid) is also shown in D. (Right) Change of $u_{rev}$ with varying $v_0$.}}
    \label{fig:base-flow}
\end{figure}

\flavio{Figure \ref{fig:base-flow} shows the computed laminar base-flow solutions for a range of suction-blowing amplitudes. As the amplitude $v_0$ is gradually increased, the boundary layer becomes more prone to separation. At $v_0=0.125$ we reach incipient separation and at $v_0=0.15$ we obtain the first LSB solution with $u_{rev}=1.06\%$ (see right panel of figure \ref{fig:base-flow}). The left panels display four boundary layers whose linear disturbances will be studied in \S\ref{sec:Effect of adverse pressure gradient and separation on linear convective instabilities}. Case A is the ZPG boundary layer reference, cases B and C are boundary layers under APG and case D shows a separated shear layer hosting a LSB with peak reversed flow $u_{rev}=2\%$. The boundary layer displacement thickness, $\delta^{*}$, and the streamlines show the influence of the APG starting at $x\approx 0.6\times10^5$, where the boundary layer deviates from the ZPG reference. Additional analysis, including validation and comparisons with the broader LSB literature, in \S\ref{app:validation of the LSB numerical set-up and comparison with literature} indicates that our model LSB is representative of a ``short'' bubble whose transition is typically driven by convective disturbances.}

\subsection{Global Stability\label{subsec:global stability}}
From our GSA calculations (see \S\ref{app:global stability}), we observe that the LSB becomes globally unstable due to the emergence of an unstable 3D, stationary (zero frequency) eigenmode for peak recirculation values beyond 6.60\%, in agreement with the existing literature reporting values in the range 7-8\% \citep{theofilis2000,Rodriguez_gennaro_juniper_2013,Rodriguez2021self-excited}. We add an indicative neutral stability margin in figure \ref{fig:base-flow} at 7\% to mark the threshold between the stable and unstable regimes. Further investigation on this mode (see \S\ref{app:global stability}) reveals the features of a modal centrifugal mechanism, occurring at much lower $u_{rev}$ than absolutely unstable 2D K-H waves, which were detected only above 15\% in flat plate \citep{Rodriguez_gennaro_juniper_2013} and bump \citep{ehrenstein_gallaire_2008} geometries. For any LSB base-flow computed within the $v_{0}$ range considered, we have not observed any globally unstable 2D K-H waves. This is in accordance with literature since we have generated bubbles up to $u_{rev} \approx 12\%$.

In \S\ref{sec:Effect of adverse pressure gradient and separation on linear convective instabilities} we study the optimal linear instability mechanisms of the four base-flows (A to D) using the resolvent approach in \S\S\ref{subsec:linear resolvent analysis}. Note that all of the four cases investigated further are globally stable, noise-amplifier-type flows.

\section{Effect of adverse pressure gradient and separation on linear convective instabilities\label{sec:Effect of adverse pressure gradient and separation on linear convective instabilities}}
In order to understand the origin of the different instability mechanisms of separated boundary layers hosting separation bubbles (case D in our study), we also analyse attached boundary layers subject to increasing APG magnitude (cases B and C). We reference the results to the well-known ZPG boundary layer benchmark (case A).

\subsection{Parametric study in the $(\beta,\omega)$ domain\label{subsec:Parametric study in the (beta,omega) domain}}
\begin{figure}
    \centering
    \includegraphics[width=0.9\textwidth]{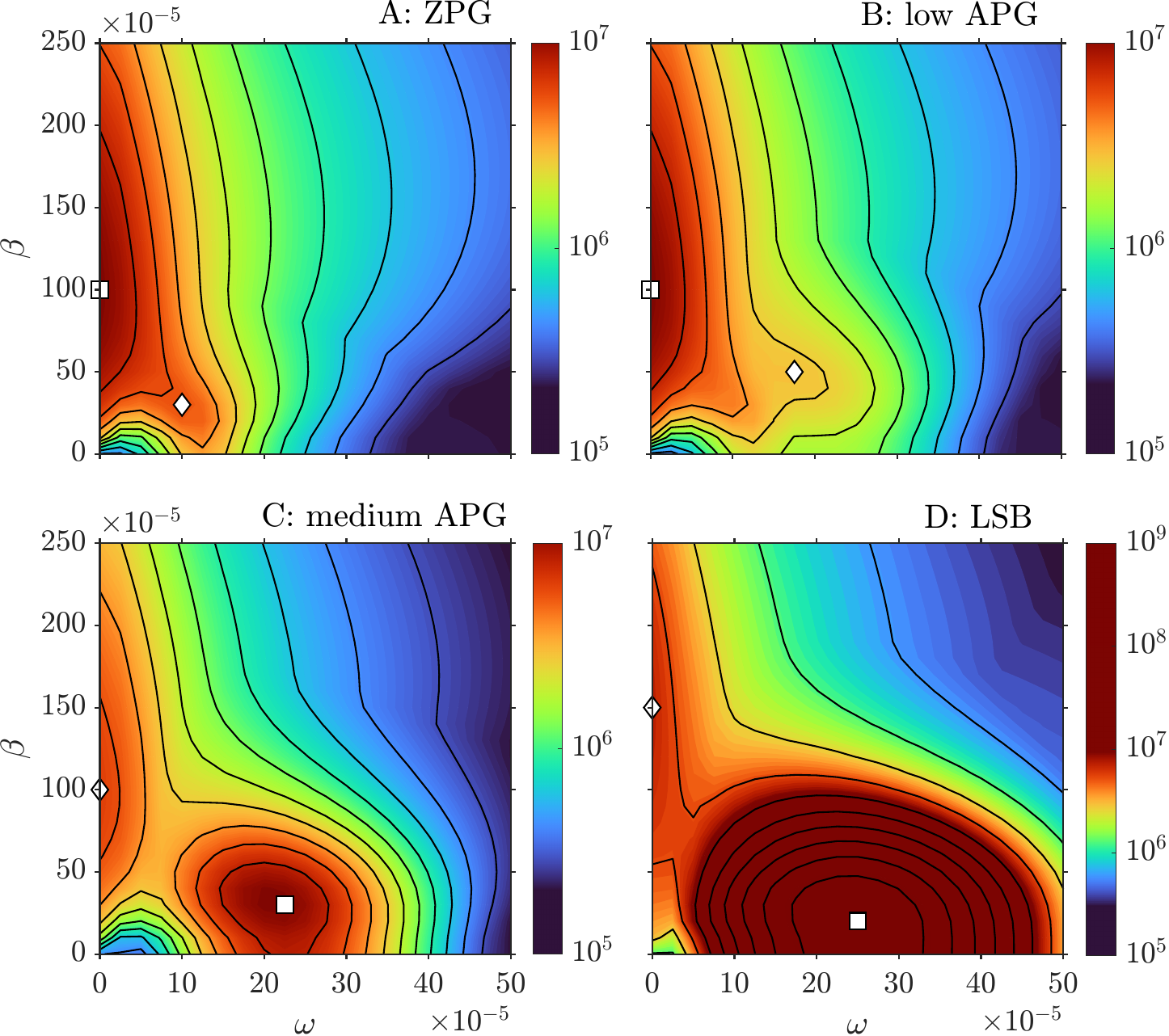}
    \vspace{1pt} \\
    \caption{\flavio{Contours of $\mathcal{G}^2(\beta,\omega)$. Square ($\square$) and diamond ($\lozenge$) markers refer to the most and second most amplified $(\beta,\omega )$ pair, respectively.}}
    \label{fig:gain maps}
\end{figure}

We perform a parametric study with RA by scanning a combination of frequencies $\omega=\left[0,50\right]\times10^{-5}$ and spanwise wavenumbers $\beta=\left[0,250\right]\times10^{-5}$ on cases A to D. In other words, we solve the optimisation problem \eqref{eq:energy gain} to extract the optimal linear forcing/response mode at each $(\beta,\omega)$ pair to unravel frequencies of high receptivity of the boundary layer to external disturbances and the physics of the mechanisms involved. Maps of the linear gain are shown in figure \ref{fig:gain maps} for cases A to D in the parametric space $(\beta,\omega)$. Two main classes of disturbances are identified at selective frequencies and spanwise wavenumbers: 
\begin{enumerate}
    \item 3D ($\beta \neq 0$) travelling-wave modes at frequency $\omega$ (diamond markers in cases A-B and square markers in cases C-D) and
    \item 3D ($\beta \neq 0$) steady ($\omega=0$) modes (squares in cases A-B and diamonds in cases C-D),
\end{enumerate}
which are listed in table \ref{tab:linear resolvent}. We examine the physics of these disturbances in the remainder of this section.

\begin{table}
  \begin{center}
    \def~{\hphantom{0}}
    \begin{tabular}{|c|c|c|ccc|ccc|}
      \multicolumn{3}{c}{} & \multicolumn{3}{c}{$\square\:(\times10^{-5})$} & \multicolumn{3}{c}{$\lozenge\:(\times10^{-5})$} \\
      CASE & Separation & $u_{rev}\:(\%)$ & $\beta$ & $\omega$ & Mode & $\beta$ & $\omega$ & Mode \\ [6pt]
      A: ZPG & no & 0 & 100 & 0 & Streaks & 30 & 10 & T-S \\ [3pt]
      B: low APG & no & 0 & 100 & 0 & Streaks & 50 & 17.5 & T-S \\ [3pt]
      C: medium APG & no & 0 & 30 & 22.5 & K-H & 100 & 0 & Streaks \\ [3pt]
      D: LSB & yes & 2 & 20 & 25 & K-H & 150 & 0 & Streaks \\ [3pt]
    \end{tabular}
    \caption{Summary of linear amplification mechanisms for cases A to D. The symbols have the same meaning as in figure \ref{fig:gain maps}.}
    \label{tab:linear resolvent}
  \end{center}
\end{table}

\subsection{From T-S to K-H dominated boundary layer instability\label{subsec:From T-S to K-H dominated boundary layer instability}}
\begin{figure}
    \centering
    \includegraphics[width=\textwidth]{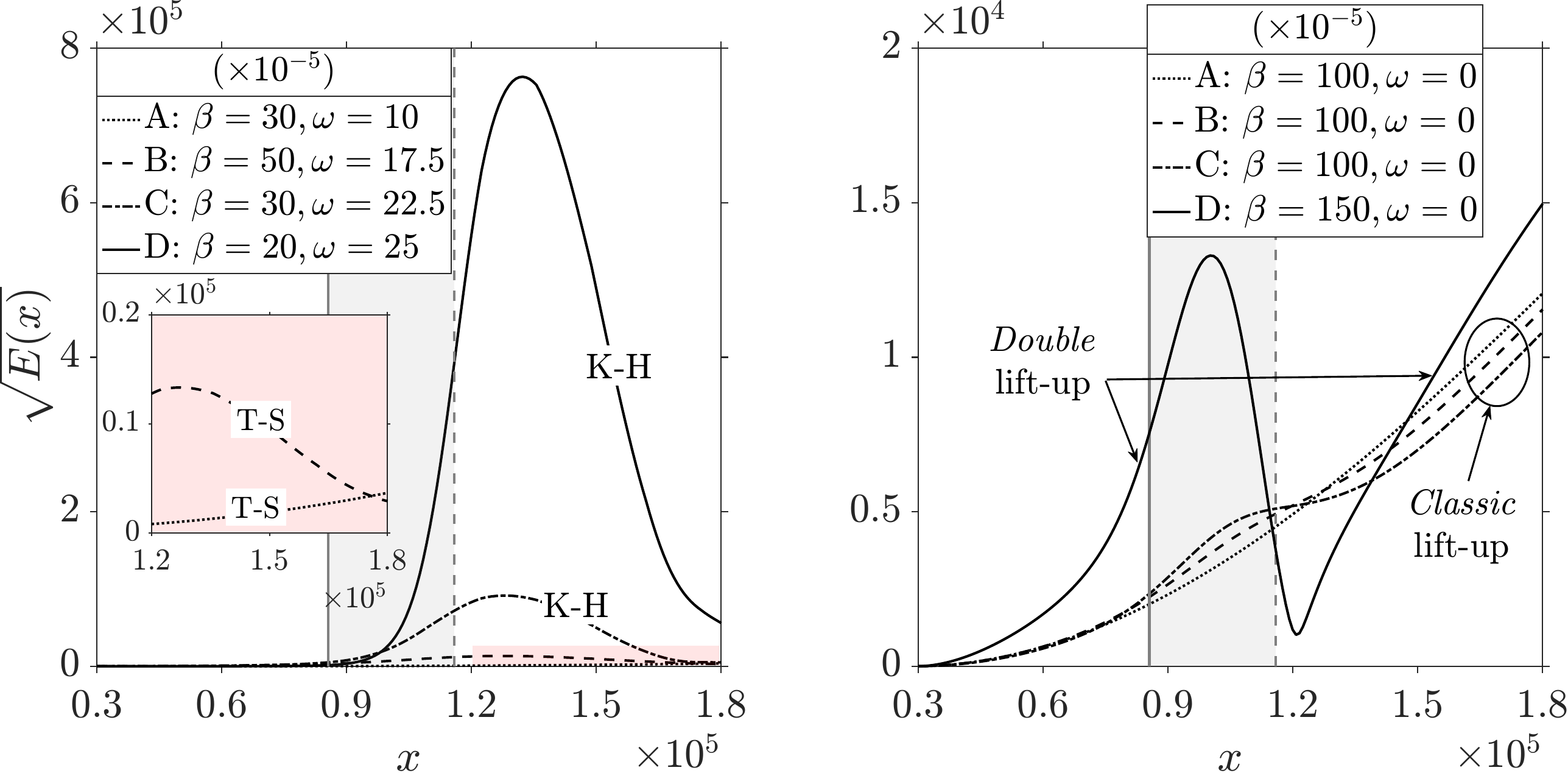}
    \caption{\flavio{Square root of the kinetic energy associated with the linear response modes marked in figure \ref{fig:gain maps}. (Left) Unsteady ($\omega \neq 0$) and (right) steady ($\omega=0$) modes. For case D, grey-solid line: separation; grey-dashed line: reattachment; shaded grey area: separation region. Case D (left) is down-scaled by one order of magnitude for clearer visualisation. The inset shows a close-up of cases A and B.}}
    \label{fig:linear amplitudes}
\end{figure}

\begin{figure}
    \centering
    \includegraphics[width=\textwidth]{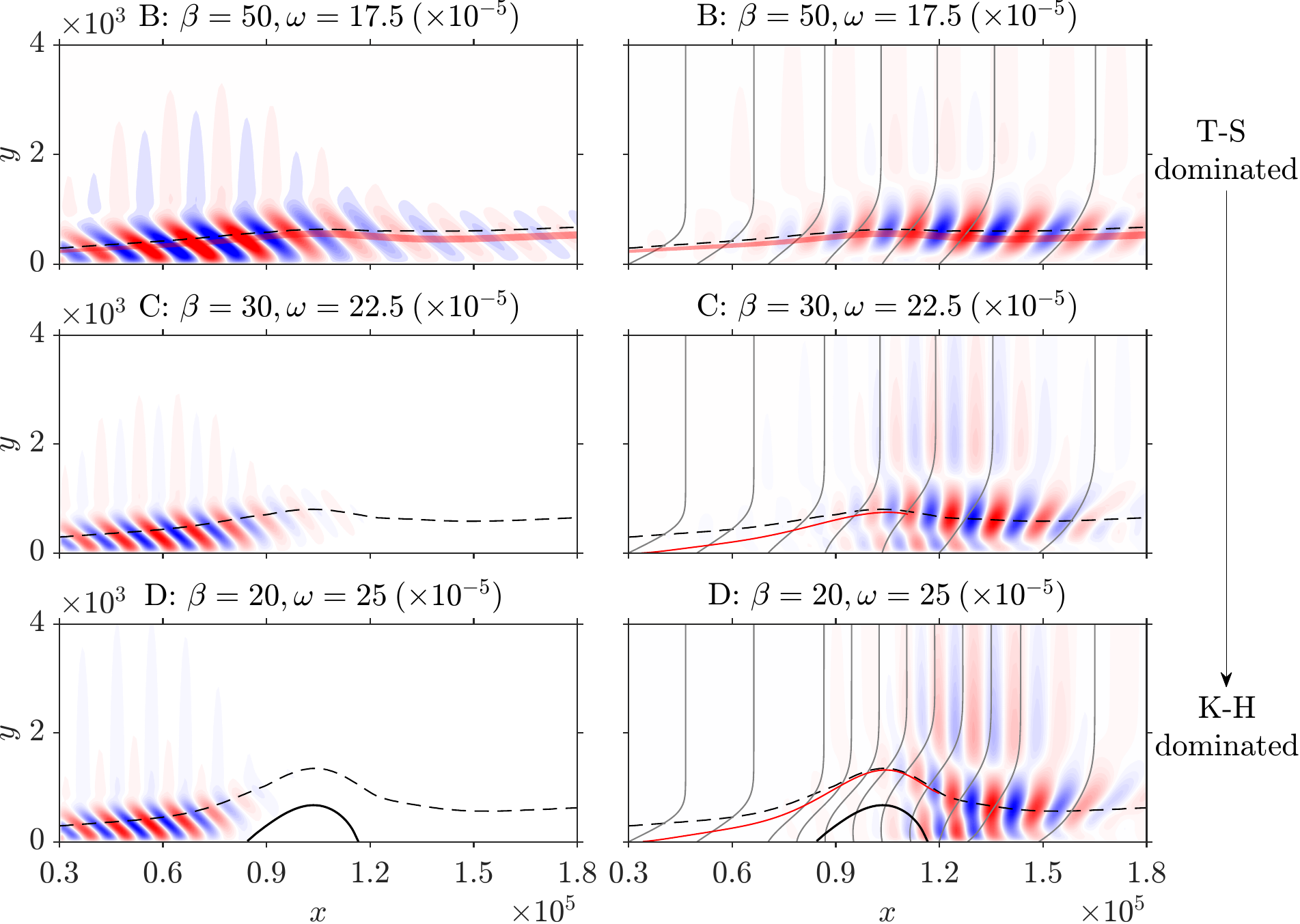}
    \caption{\flavio{Optimal unsteady resolvent modes showing the change of instability from T-S dominated to K-H dominated with increasing APG. Contours of (left) $\mathrm{Re}\left \{ \hat{f}_{x} \right \}$ and (right) $\mathrm{Re}\left \{ \hat{u} \right \}$. Black-dashed line: displacement thickness. Grey-solid lines: base-flow streamwise velocity profiles. Red-solid line: inflection line. Red shaded area: critical layer.}}
    \label{fig:T-S and K-H linear resolvent}
\end{figure}

Depending on the magnitude of the APG, the receptivity of the boundary layer to external disturbances changes both quantitatively and qualitatively, meaning that the pressure gradient affects both the energy/amplification and the characteristic physics of the underlying mechanisms. For the former, we refer to the gain in figure \ref{fig:gain maps} and for the latter we plot either the streamwise evolution of the square root of the perturbation kinetic energy, $\sqrt{E(x)}$, in figure \ref{fig:linear amplitudes}, where $E$ is
\begin{equation}
    E(x) = \int_{0}^{\infty} \left( |\hat{u}|^2 + |\hat{v}|^2 + |\hat{w}|^2\right) \: \mathrm{d}y
    \label{eq:perturbation energy}
\end{equation}
\noindent and a planar view of the spatial forcing/response modes in figure \ref{fig:T-S and K-H linear resolvent}.

From the linear gain in figure \ref{fig:gain maps}, we observe that ZPG (A) and low APG (B) attached boundary layers amplify T-S waves according to the Orr mechanism \citep{schmid1992new,schmid_henningson2001stability,rigas2021HBM}, which is more energetic for oblique rather than planar waves. We find $(\beta,\omega)=(30,10)\times10^{-5}$ for ZPG and $(\beta,\omega)=(50,17.5)\times10^{-5}$ for low APG (marked with diamonds in figure \ref{fig:gain maps} A-B) to be the frequency-spanwise wavenumber pairs where this mechanism achieves maximal amplification. While the gain of the oblique T-S wave mechanism is not affected dramatically by the APG, the shape of the spatial growth rate is, as shown in figure \ref{fig:linear amplitudes} (left). The inset indicates that in the ZPG boundary layer the energy of T-S waves increases monotonically with increasing $x$ ($=\mathrm{Re}_x$ in our non-dimensionalisation), while in the presence of low APG there is a peak of energy around $\mathrm{Re}_x \approx 1.2-1.3\times10^5$. In figure \ref{fig:T-S and K-H linear resolvent} (top) we show that the optimal forcing waves located in the first part of the pressure gradient region and within the critical layer excite T-S waves downstream. The spatial arrangement of the response, in agreement with the energy distribution in figure \ref{fig:linear amplitudes} (left), indicates that the amplification of T-S waves correlates with the pressure gradient distribution. This is because the pressure gradient enhances the viscous shear in the boundary layer, which is ultimately responsible for the Orr mechanism \citep{schmid_henningson2001stability}. Despite all the effects of the pressure gradient on T-S waves discussed thus far, the low APG boundary layer is still reminiscent of the same convective instability mechanisms of the ZPG boundary layer.

Stronger APG levels (case C) lead to (i) boundary layer thickening and (ii) inflectional velocity profiles. The latter is determinant for important changes in the dominant instability mechanism of the boundary layer. The energy of the most amplified mode found at $(\beta,\omega)=(30,22.5)\times10^{-5}$ (square marker in figure \ref{fig:gain maps}, case C) increases by 1 order of magnitude with respect to the T-S mode of case B but its distribution in $x$ is qualitatively similar -- see figure \ref{fig:linear amplitudes} (left). To detect any differences between these modes we compare their spatial structures in figure \ref{fig:T-S and K-H linear resolvent}. The middle panels show that the optimal forcing waves cluster in a narrower region of the domain more upstream compared to the classic T-S forcing mode and decay upstream of the boundary layer thickening, moreover, the response mode exhibits roller-type structures downstream of boundary layer thickening closely aligned with the inflection line. We ascribe these features to the inviscid K-H instability mechanism, which emerges when the inflection point becomes unstable for sufficiently large APG \citep{sandham_2008,diwan_ramesh_2009,marxen_henningson_2011,Zhang2019input-output}.

When the APG leads to flow separation (case D), the gain of the most amplified K-H mode at $(\beta,\omega)=(20,25)\times10^{-5}$ increases by 2 additional orders of magnitude due to the formation of a separated shear layer which is known to be more unstable than the attached counterpart \citep{yang2019}. The optimal forcing waves live upstream of the separation bubble and feed the response amplification occurring in the aft part of the separated shear layer (figure \ref{fig:T-S and K-H linear resolvent}, bottom), which is a critical site of shear layer instability \citep{sandham_2008,Jones2010stabilityaerofoil,Simoni2014experimentalLSB}. The peak response energy is reached shortly downstream of the reattachment point (figure \ref{fig:linear amplitudes}, left), where both velocity and pressure gradients are still significant. In case D it is even more evident than case C that the K-H rollers form two streets, one aligned with the inflection line and the other being very close to the wall. This is not observed in the shape of the T-S waves (case B). Because the most amplified K-H mode is 3D, even though 2D K-H waves exist in the gain spectrum \flavio{of figure} \ref{fig:gain maps} (case D), the angle of the waves relative to the spanwise direction can be computed as $\tan^{-1}(\beta/\alpha_{x})$, where $\alpha_{x}$ is the streamwise wavenumber. We find that it is approximately $8^{\circ}$, meaning that these waves are nearly 2D. These topological features agree with the broad literature on K-H instability of separated shear layers \citep{rist_maucher_2002,diwan_ramesh_2009,Simoni2013experimentalwakes,Simoni2014experimentalLSB,Michelis2018spanwise,Michelis2018origin,Zhang2019input-output,hosseinverdi_fasel_2019,Hosseinverdi2020onset}.

In summary, the APG, upon modification of the boundary layer velocity profile, enhances the spatial growth of convective disturbances and, if sufficiently strong, activates the K-H mechanism to the point the shear layer evolution is driven by this instability instead of classic T-S waves typical of ZPG and low APG cases. \flavio{We refer the reader to \S\ref{app:effect of adverse pressure gradient on Tollmien-Schlichting waves} for further discussion.}

\subsection{Effect of separation on the classic lift-up mechanism\label{subsec:Effect of separation on the classic lift-up mechanism}}
Another class of convective-type instabilities of boundary layers is steady $(\omega=0)$ and 3D $(\beta \neq 0)$. In ZPG (case A) and low APG (case B) boundary layers, there is a region in the gain spectrum extending over a large $\beta$ range where the classic lift-up mechanism stands out as the most amplified mode (figure \ref{fig:gain maps}, A and B). This non-modal mechanism is well-documented in the literature \citep{andersson_brandt_bottaro_henningson_2001,schmid_henningson2001stability,jacobs_durbin_2001,Symon2018nonnormality_resolvent,rigas2021HBM} and involves the transient growth of streamwise-oriented $u'$ streaks optimally forced by near-wall streamwise vortices (see figures \ref{fig:liftup APG forcing}-\ref{fig:liftup APG response} in \S\ref{app:lift-up mechanism affected by the adverse pressure gradient}). The physical process is characterised by the exchange of mean streamwise momentum across the boundary layer thickness due to the action of streamwise vorticity which pumps high-momentum fluid particles living in the outer shear layer deep into the boundary layer and, conversely, lifts low-momentum particles from the near-wall up to the outer layer. Notably, all the perturbation kinetic energy of the streaky mode plotted in figure \ref{fig:linear amplitudes} (right) comes from the $u'$ component, since $v'$ and $w'$ are typically nil -- refer to figure \ref{fig:liftup APG response}. The maximum gain of the lift-up mode occurs at $\beta=100\times10^{-5}$ and drops more steeply for lower $\beta$ than for higher $\beta$, indicating this instability is strongly 3D.

\begin{figure}
    \centering
    \includegraphics[width=\textwidth]{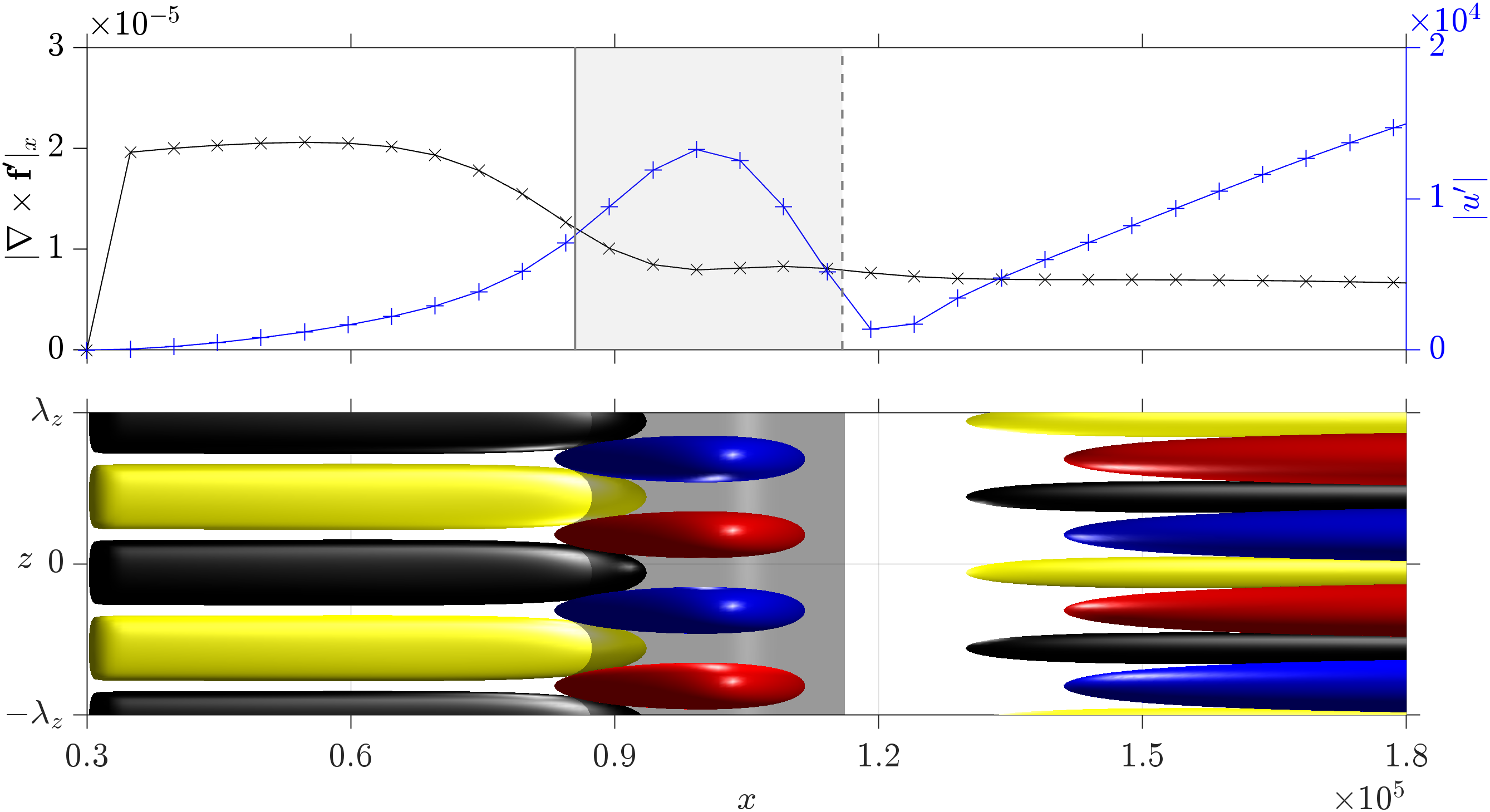}
    \caption{\flavio{``Double'' lift-up mechanism at $\beta=150\times10^{-5}$ (diamond marker in figure \ref{fig:gain maps}, D). (Top) Amplitudes of the streamwise component of the curl of forcing $(\nabla \times \boldsymbol{f}')_{x}$ (black-solid line with $\times$) and of the streamwise velocity of the response $u'$ (blue-solid line with $+$). (Bottom) Isosurfaces of $(\nabla \times \boldsymbol{f}')_{x}$ (coloured with yellow/black for positive/negative) and $u’$ (coloured with red/blue for positive/negative). The LSB is shown by the $u=0$ grey-coloured isosurface.}}
    \label{fig:doubleliftup_resolvent}
\end{figure}

The impact of the pressure gradient on the classic lift-up mechanism is now discussed. As long as the boundary layer remains attached, the process is continuous, meaning streamwise vorticity continually forces the $u'$ response whose spatial growth is almost monotonic -- and is perfectly monotonic in ZPG boundary layer (figures \ref{fig:liftup APG forcing}-\ref{fig:liftup APG response}). The local changes in spatial growth of cases B and C are due to the influence of the APG (figure \ref{fig:linear amplitudes}, right). Interestingly, when the separation bubble forms (case D), the spatially continuous action of the streamwise vortices is ``broken'' into two parts, one upstream of separation that creates local $u'$ disturbances over the bubble, and the other in the reattached shear layer downstream of the bubble leading to elongated streaks (see figure \ref{fig:doubleliftup_resolvent}). The $u'$ structures in the separation zone imply a local production of perturbation kinetic energy, as seen in figure \ref{fig:linear amplitudes} (right), where we introduce the terminology ``double'' lift-up to indicate the two-stage action of this process due to the presence of the bubble. We notice qualitative similarity to \citet{casacuberta_hickel_westerbeek_kotsonis_2022,casacuberta2024} as concerns the phase shift of the streaks over and after the separation, which the authors observed before and after the forward facing step. The mechanism described therein and termed ``reverse lift-up'' is different, though, since it involves the stabilisation of the streaks not present in our LSB set-up.

In brief, separation affects the spatial organisation of the structures typical of lift-up but the physical mechanism remains the same as the ZPG boundary layer's.

\subsection{Modal centrifugal instability mechanism in LSB\label{subsec:Modal centrifugal instability mechanism in LSB}}
\begin{figure}
    \centering
    \includegraphics[width=\linewidth]{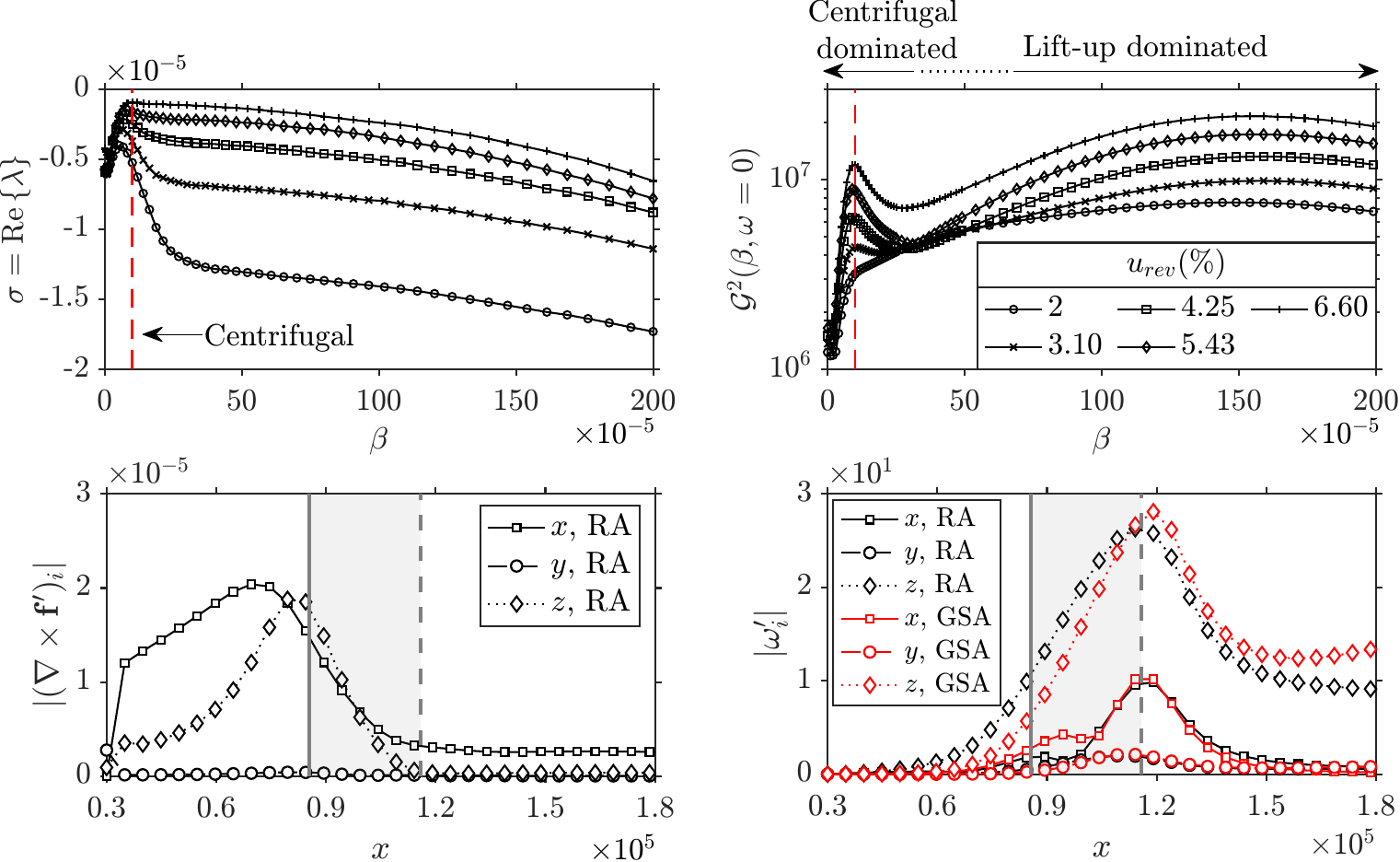}
    \caption{\flavio{(Top left) Temporal amplification of the least stable eigenmode from GSA and (top right) squared gain from RA at $\omega=0$ for several $\beta$ values and for five bubbles with increasing $u_{rev}$. (Bottom) Component-wise amplitudes of (left) curl of forcing computed from RA and (right) response vorticity computed from both RA and GSA at $\beta=10\times10^{-5}$ (marked on the top panels with red-dashed line) and $u_{rev}=2\%$ (case D). The eigenmode is scaled such that its total energy $\int E(x) \: \mathrm{d}x$ matches the resolvent response mode's.}}
    \label{fig:centrifugal GSA vs RA}
\end{figure}

A mechanism that is absent in attached boundary layers is the modal centrifugal instability related to the separation bubble. As reported in \S\S\ref{subsec:global stability} from the calculations presented in \S\ref{app:global stability}, this mechanism becomes globally unstable for sufficiently strong recirculating flow inside the bubble ($u_{rev}\approx7-8\%$), which is ultimately dictated by the magnitude of the APG. As shown in figure \ref{fig:centrifugal GSA vs RA} (top left), the temporal amplification factor of the least stable eigenvalue (which is stationary, $\mathrm{Im}  \left\{ \lambda \right\}=0$) computed over a range of $\beta$ for five LSB base-flows with $2\% \leq u_{rev} \leq 6.60\%$ is always negative, confirming that there is indeed no globally unstable mode in bubbles with $u_{rev}<7-8\%$. However, for $\beta$ values 1 to 2 orders of magnitude lower than those typical of streaks, the least stable eigenvalue is less damped. RA corroborates the GSA data. The gain in figure \ref{fig:centrifugal GSA vs RA} (top right) shows two regions of amplification, one with a peak at $\beta=10\times 10^{-5}$ that is very sensitive to the size of the bubble, and another at $\beta=150\times10^{-5}$ related to lift-up which has been already discussed in \S\S\ref{subsec:Effect of separation on the classic lift-up mechanism}. Remarkably, GSA fails to capture the lift-up mode because it is non-modal, while the agreement between RA and GSA on the presence of a modal mechanism for low $\beta$ is examined further in figure \ref{fig:centrifugal GSA vs RA} (bottom) for the $u_{rev}=2\%$ bubble (case D). The similarity between the optimal response from RA and the eigenmode from GSA in figure \ref{fig:centrifugal GSA vs RA} (bottom right) is striking. Either modes display the typical shape of the streamwise vorticity component $(\omega_{x}')$ of the centrifugal instability over the separation zone, which is especially intense at the separation and reattachment points of the shear layer. 

We highlight the important differences between the structure of the centrifugal and lift-up modes, since these may not be obvious. As far the forcing mode shape is concerned, we refer to figure \ref{fig:centrifugal GSA vs RA} (bottom left) and \ref{fig:liftup APG forcing} (fourth row), and note that:
\begin{itemize}
    \item In the centrifugal mechanism both $(\nabla \times \boldsymbol{f}')_{x}$ and $(\nabla \times \boldsymbol{f}')_{z}$ are present, while for lift-up it is only $(\nabla \times \boldsymbol{f}')_{x}$, and
    \item The optimal forcing of the centrifugal mode is particularly strong near the separation point, which is not observed in lift-up.
\end{itemize}
\noindent In terms of the response mode shape (figures \ref{fig:centrifugal GSA vs RA}, bottom right and \ref{fig:liftup APG response}, fourth row):
\begin{itemize}
    \item The wall-normal vorticity disturbance $(\omega_{y}')$ is weak in the centrifugal instability, which is not the case of the lift-up, where $\omega_{y}'\sim\omega_{z}'$, and
    \item Most importantly, $\omega_{x}'$ is virtually negligible in the lift-up mode, but is an essential feature of the centrifugal instability.
\end{itemize}

Overall, the centrifugal instability originates from the bubble, while lift-up does not. Centrifugal forcing optimally acts on the separation point and the corresponding response is maximal at reattachment. Therefore, this mechanism essentially occurs over one bubble length, as opposed to lift-up which manifests throughout the boundary layer development.

\flavio{In summary, linear analysis via both GSA and RA revealed that under APG conditions the K-H instability dominates over T-S waves and, when separation occurs, a new instability of centrifugal nature appears at low $\beta$. The lift-up mechanism leading to streaks is present for any pressure gradient magnitude. We now examine in \S\ref{sec:optimal non-linear disturbances} the non-linear, multi-modal interactions of the above disturbances to understand the sequence of mechanisms leading to turbulence.}

\section{Optimal non-linear disturbances\label{sec:optimal non-linear disturbances}}
In this section we present the non-linear mechanisms responsible for \flavio{laminar-turbulent transition} of the separated shear layer developing over the separation bubble with $u_{rev}=2\%$ (case D). 

Given the oblique K-H instability is the most amplified linear mechanism in the shear layer (refer to figure \ref{fig:gain maps}), we force the non-linear HBNS system at the same frequency and spanwise wavenumber found from RA ($\omega=25\times10^{-5}$ and $\beta=20\times10^{-5}$). We expect this mechanism to emerge strongly in the region of flow where the dynamics of the perturbations is still substantially linear, i.e. at incipient separation of the shear layer. Clearly, the non-linearities of the flow will also lead to new disturbance mechanisms which we are able to analyse with HBNS. The capability of the method to account for multi-harmonic forcing environments in the optimisation allows us to consider also planar waves $(0\beta,\pm1\omega)$ and steady disturbances $(\pm1\beta,0\omega)$ at fundamental frequency and spanwise wavenumber, similarly to \citet{rigas2021HBM}.

\subsection{Quasi-linear-in-$\omega$ and HBNS-in-$\beta$ systems}\label{subsec:quasi-linear-in-omega and HBNS-in-beta systems}
We explore HBNS systems of variable architecture by varying the number of harmonics in the Fourier expansion \eqref{eq:Fourier expansion} in order to identify and converge the leading modes that carry most of the energy of the disturbance field. To assess convergence of the solution, we track the behaviour of the total drag increase on the plate, $\Delta C_D$ in \eqref{eq:drag change}, resulting from the mean-flow modification.

\begin{figure}
    \centering
    \includegraphics[width=\linewidth]{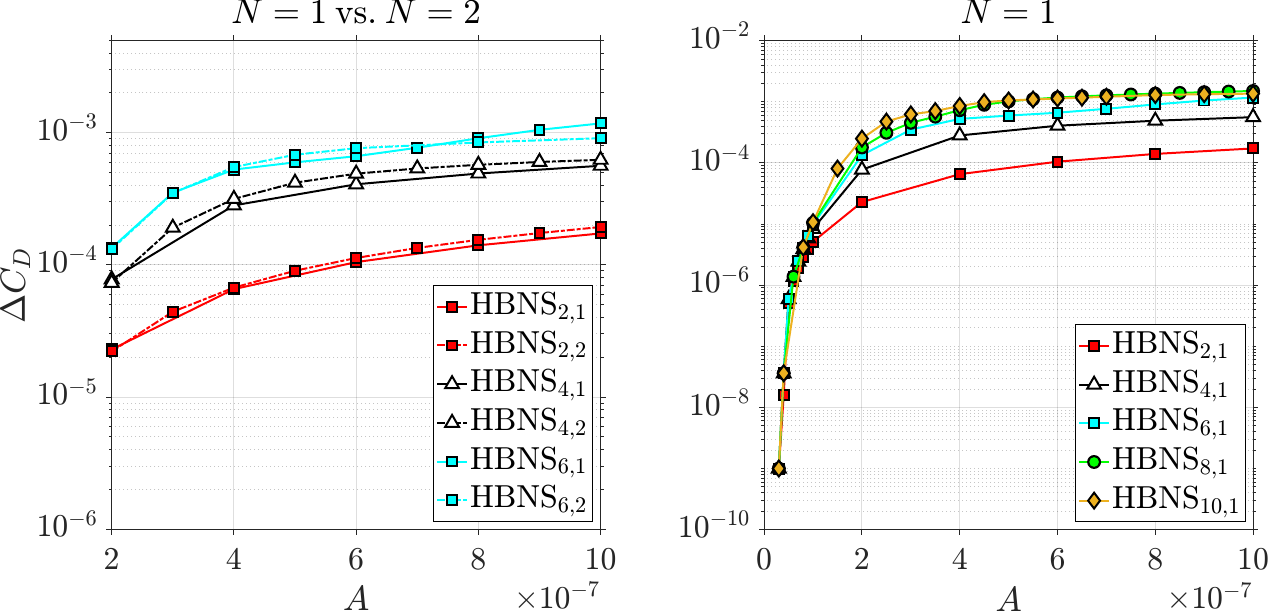}
    \caption{\flavio{(Left) Comparison of $\Delta C_D$ for systems with $N=1$ and $N=2$. (Right) $\Delta C_D$ of QL-in-$\omega$ and HBNS-in-$\beta$ systems.}}
    \label{fig:dCD}
\end{figure}

In figure \ref{fig:dCD} we show the $\Delta C_D$ parameter obtained in the range of forcing amplitudes $A=[0.01,10]\times10^{-7}$. In the left panel we compare systems with one ($N=1$) against two ($N=2$) time harmonics. We also vary the number of spanwise harmonics from a minimum of two ($M=2$) to a maximum of six ($M=6$). While the number of spanwise harmonics visibly affects the drag, systems with $N=2$ harmonics do not show significant difference to $N=1$. This implies that the contribution of harmonics with frequency $2\omega$ (and, consequently, also higher order harmonics $3\omega,4\omega,5\omega,\dots$) to the mean-flow modification is negligible.

The overall effect of varying $M$ (at constant $N=1$) on $\Delta C_D$ is seen in the right panel of figure \ref{fig:dCD}, where $M$ ranges from 2 to 10. We define these systems as quasi-linear (QL)-in-$\omega$ and HBNS-in-$\beta$ since they contain harmonics with at most $1\omega$ and a finite number of $m\beta$. 
Notably, convergence of $\Delta C_D$ is observed if $M=8$ at least, meaning we resolve the most energetic disturbances and underlying non-linear interactions within this range. We therefore show that the cascade of energy from low-order (large-scale) to high-order (small-scale) disturbances is dominant in space ($\beta$-modes) rather than in time ($\omega$-modes). For this reason we perform our analysis on QL-in-$\omega$ ($N=1$) systems for the remainder of the manuscript.

\subsection{Non-linear mechanisms at the early stages of transition\label{subsec:Non-linear mechanisms at the early stages of transition}}
We analyse the onset of non-linearity in the shear layer dynamics at weakly non-linear forcing amplitude $A=1\times10^{-7}$ where the base-flow modification is small but non-zero. This specific study is relevant to the early stages of transition, where the dynamics of the system can no longer be approximated as linear but, at the same time, only few harmonics (or scales) of the flow are active. The HBNS$_{2,1}$ system is suitable for this purpose since, by construction, it can calculate the first non-linear triadic interaction where $(\pm1\beta,\pm1\omega)$ waves produce the 3D steady $(\pm2\beta,0\omega)$ mode. Actually, the fundamental oblique mode may also produce planar $(0\beta,\pm2\omega)$ and oblique $(\pm2\beta,\pm2\omega)$ super-harmonics but these are shown to have little impact on $\Delta C_D$ in figure \ref{fig:dCD} (left). Furthermore, a low-order system such as HBNS$_{2,1}$ is sufficiently accurate at this forcing amplitude since all systems converge to the same $\Delta C_D$. Imposing symmetry about $z=0$ (i.e. $\hat{\boldsymbol{w}}_{-m,n}=[\hat{u}, \hat{v}, -\hat{w}]_{m,n}$ and $\hat{\boldsymbol{f}}_{-m,n}=[\hat{f}_x, \hat{f}_y, -\hat{f}_z]_{m,n}$), the number of coupled equations in this system amounts to $(M+1)\times(N+1)=6$, including the equation for the modification of the mean-flow by the disturbances.

\begin{figure}
    \centering
    \includegraphics[width=\linewidth]{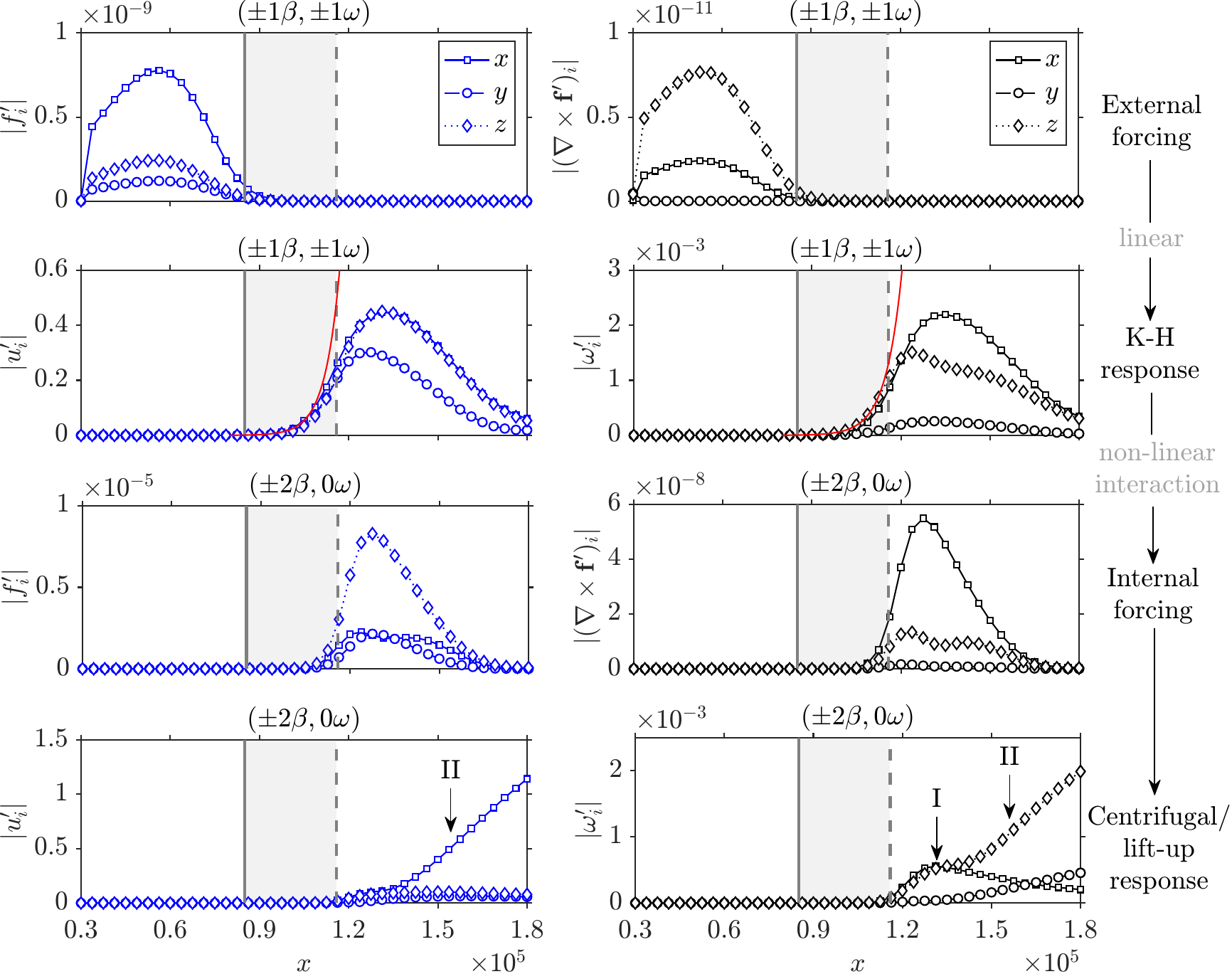}
    \caption{\flavio{Optimal non-linear mechanisms computed from HBNS$_{2,1}$ system by forcing at $(\beta,\omega)=(20,25)\times10^{-5}$ and $A=1\times10^{-7}$. Component-wise amplitudes of (first row, left) external forcing $(\pm1\beta,\pm1\omega)$ and (first row, right) curl, (second row, left) velocity and (second row, right) vorticity of the K-H response $(\pm1\beta,\pm1\omega)$, (third row, left) forcing and (third row, right) curl of forcing from non-linear triadic interaction $(\pm1\beta,-1\omega) + (\pm1\beta,+1\omega) = (\pm2\beta,0\omega)$, (fourth row, left) velocity and (fourth row, right) vorticity of the centrifugal/lift-up response $(\pm2\beta,0\omega)$. Time- and spanwise-averaged separation (grey-solid) and reattachment (grey-dashed) locations superimposed. Lines of best fit, $a\exp{(bx)}$, for $|u'|$ ($a=3.58\times10^{-11}, b=2.01\times10^{-4}, R^2 = 0.9999$) and $|\omega_{z}'|$ ($a=8.85\times10^{-13}, b=1.82\times10^{-4}, R^2 = 0.9977$) are plotted in red-solid.}}
    \label{fig:HBNS21_amplitudes}
\end{figure}

The optimal non-linear forcing/response solution in figure \ref{fig:HBNS21_amplitudes} shows that the shear layer initially amplifies the K-H mechanism over the separation region which then leads to the $(\pm2\beta,0\omega)$ mode through non-linearity. We now explain this scenario in more detail.

\begin{figure}
    \centering
    \includegraphics[width=\linewidth]{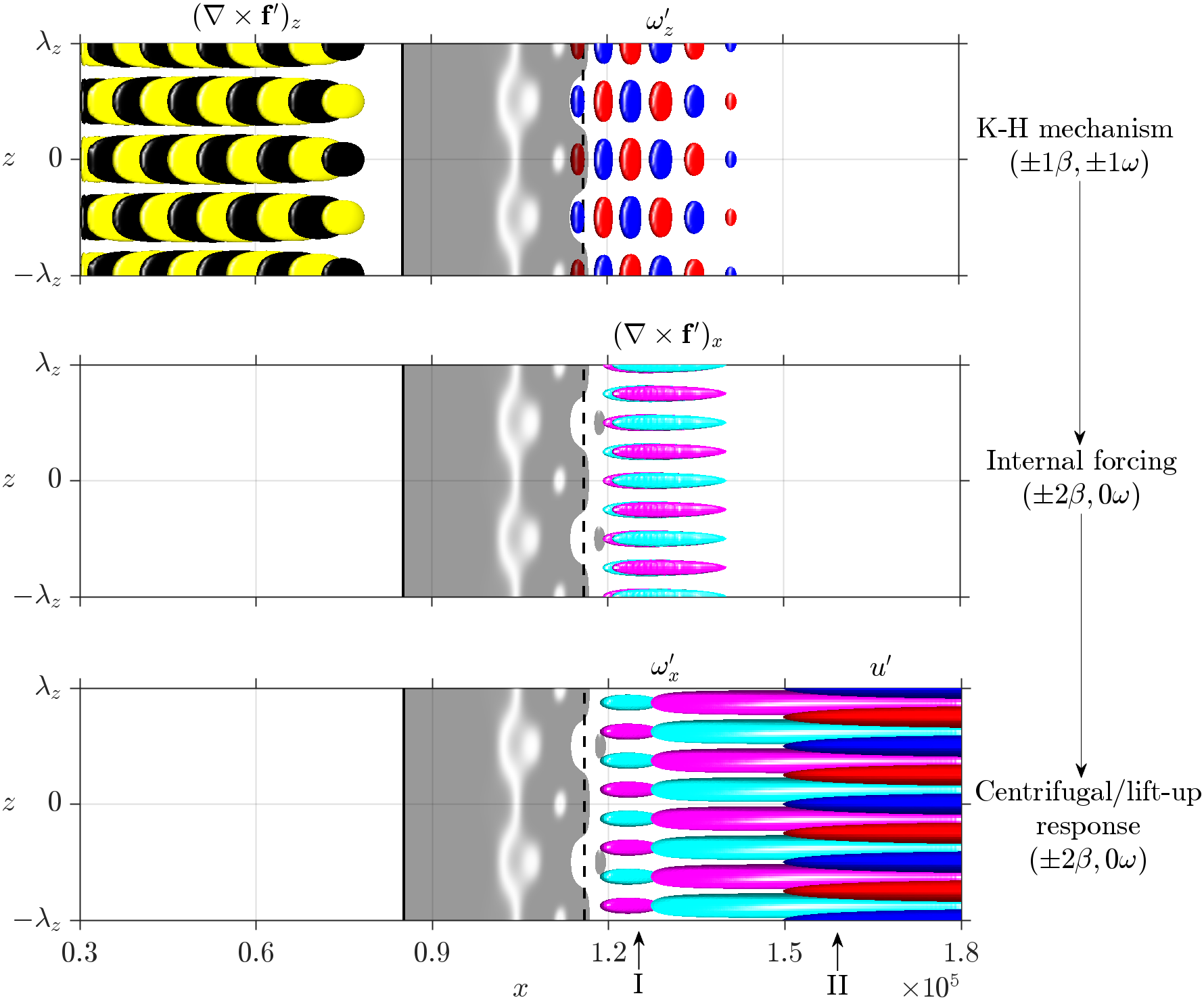}
    \caption{\flavio{3D structure of the modes in figure \ref{fig:HBNS21_amplitudes}. Isosurfaces of (top) $(\nabla \times \boldsymbol{f}')_{z}$ (yellow/black for positive/negative) and $\omega_{z}'$ (red/blue for positive/negative) from the K-H mechanism, (middle) $(\nabla \times \boldsymbol{f}')_{x}$ (magenta/cyan for positive/negative) from the non-linear internal forcing and (bottom) $\omega_{x}'$ (magenta/cyan for positive/negative) and $u'$ (red/blue for positive/negative) from the centrifugal/lift-up mechanisms. The instantaneous separation bubble is plotted with the $u=0$ isosurface.}}
    \label{fig:HBNS21_isosurf}
\end{figure}

Forcing optimally originates only from oblique waves (with planar wave and steady disturbance forcing harmonics being virtually zero, therefore not displayed) with a strong $(\nabla \times \boldsymbol{f}')_{z}$ component located upstream of the separation point -- see figure \ref{fig:HBNS21_amplitudes} (first row). Their spatial structure, plotted in the top panel of figure \ref{fig:HBNS21_isosurf}, is characterised by rollers of finite span tilted against the mean shear. The structure of the optimised forcing is visually similar to the linear resolvent forcing mode of figure \ref{fig:T-S and K-H linear resolvent} (bottom left). This result indicates that the 3D K-H instability is the key to trigger mean-flow modification, which is in the cost functional \eqref{eq:cost functional}, therefore justifying the observed similarity. This result also confirms the effectiveness of oblique waves in generating efficient mechanisms for transition to turbulence of both attached and separated shear flows \citep{schmid1992new,Marxen2013vortex,rigas2021HBM,dwivedi_sidharth_jovanovic_2022}. The K-H response grows in the separated shear layer with a spatial amplification rate that is well approximated by an exponential curve -- see figure \ref{fig:HBNS21_amplitudes} (second row), indicating the dynamics of the flow is essentially governed by a linear mechanism over the entire front portion of the separation bubble up to the location of the bubble's apex, in agreement with \citet{diwan_ramesh_2009}. Because these waves are spanwise-periodic, they form a chequerboard pattern of rollers with finite spanwise length, as shown in figure \ref{fig:HBNS21_isosurf} (top).

Near the mean reattachment location the K-H waves reach finite amplitude and saturation effects occur. Concurrently, the non-linear interaction of these waves governed by the $\mathsfbi{N}(\boldsymbol{w},\boldsymbol{w})$ operator in \eqref{eq:semi-discrete NS} produces an internal forcing term (equivalent to internal stresses) with a strong $(\nabla \times \boldsymbol{f}')_{x}$ component reaching its peak energy shortly downstream of reattachment -- figure \ref{fig:HBNS21_amplitudes} (third row). This term is computed in the equation for the $(2\beta,0\omega)$ harmonic ($m=2$, $n=0$) from the HBNS$_{2,1}$ system, 
\begin{equation}
    \left [ \mathsfbi{L}_{2} + \mathsfbi{N}_{0}^{2} \left ( \hat{\boldsymbol{w}}_{0,0}, \cdot \right ) \right ] \hat{\boldsymbol{w}}_{2,0} + \mathsfbi{N}_{1}^{1} \left ( \hat{\boldsymbol{w}}_{1,-1}, \hat{\boldsymbol{w}}_{1,1} \right ) = 0,
    \label{eq:(2,0) harmonic equation}
\end{equation} 
\noindent where $\mathsfbi{N}_{1}^{1} \left ( \hat{\boldsymbol{w}}_{1,-1}, \hat{\boldsymbol{w}}_{1,1} \right )$ governs the non-linear interaction of $(1\beta,-1\omega)$ with $(1\beta,1\omega)$ waves through the convective term of the N-S. From now onwards, $\hat{\boldsymbol{f}}_{\text{K-H}} = \left [ \hat{f}_{\text{K-H},x}, \hat{f}_{\text{K-H},y}, \hat{f}_{\text{K-H},z} \right ]^{\top} = -\mathsfbi{P}^{\top} \mathsfbi{N}_{1}^{1} \left ( \hat{\boldsymbol{w}}_{1,-1}, \hat{\boldsymbol{w}}_{1,1} \right )$ will be referred to as ``K-H forcing'' because it originates from the non-linear interaction of two K-H waves (for the explicit derivation see \S\ref{app:forcing produced by the non-linear interaction of a pair of K-H waves}). $\mathsfbi{P}^{\top}$ is the restriction operator mapping the full velocity-pressure state $[u,v,w,p]^{\top}$ to the velocity state $[u,v,w]^{\top}$.

The $(\pm2\beta,0\omega)$ response in figure \ref{fig:HBNS21_amplitudes} (fourth row) displays two characteristic regions. By comparison with the linear centrifugal and lift-up modes in figures \ref{fig:centrifugal GSA vs RA} (bottom right)-\ref{fig:liftup APG response} (fourth row), we notice that region I displays the features of centrifugal instability, which are that $\omega_{x}'$ peaks near the reattachment point and that it is of the same order of magnitude of $\omega_{z}'$, and region II is the site of lift-up where both $\omega_{y}'$ and $\omega_{z}'$ are amplified but the $\omega_{x}'$ response is weak. The structures of $\omega_{x}'$ in figure \ref{fig:HBNS21_isosurf} (bottom) further highlight the presence of two mechanisms in the $(\pm2\beta,0\omega)$ mode. The first set of streamwise vortices localised around the mean reattachment point is related to the centrifugal mode pertaining to the separation zone \citep{Sansica_Sandham_Hu_2016,Hildebrand2018,mauriello2022}. At reattachment, the flow is strongly curved due to the displacement effect of the bubble, which makes it a susceptive site for the development of centrifugal instabilities of Görtler type \citep{Gortler1941,Saric1994,li_malik_1995}. \flavio{A detailed study evaluating sufficient conditions for centrifugal instability is provided in \S\ref{app:rayleigh discriminant criterion for centrifugal instability}, where we essentially demonstrate that streamline curvature at separation and reattachment supports a locally unstable centrifugal mode.} On the other hand, the second set of streamwise-elongated vortices and $u'$ streaks are ascribed to lift-up \citep{jacobs_durbin_2001,Balamurugan_Mandal_2017,rigas2021HBM} taking place in the redeveloping boundary layer. 

In summary:
\begin{enumerate}
    \item Oblique wave-like disturbances optimally located upstream of separation excite the inviscid, inflectional K-H instability at $(1\beta,1\omega)$ through a linear mechanism that takes advantage of the shear present in the separated shear layer.
    \item At saturation amplitude, the K-H waves interact non-linearly to produce a source of internal forcing at $(2\beta,0\omega)$ that features mainly streamwise vortices.
    \item Streamwise vorticity manifests in the $(2\beta,0\omega)$ response near the reattachment point, indicating the presence of a centrifugal instability.
    \item The presence of $\omega_{x}'$ due to the centrifugal response translates into streamwise vortices that trigger lift-up and, therefore, streaks in the redeveloping boundary layer.
\end{enumerate}
While there are commonalities with the oblique wave-to-streaks mechanism found by \citet{schmid1992new,rigas2021HBM} for attached boundary layers, the presence of separation makes this scenario more complex due to the centrifugal instability.

Finally, it is important to note that the K-H induced forcing is different to the optimal forcing of either the centrifugal mode in figure \ref{fig:centrifugal GSA vs RA} (bottom left) and lift-up in figure \ref{fig:liftup APG forcing} (bottom). In the former, we have observed a strong streamwise vorticity footprint at the separation point, while in the latter a continuous action of streamwise vorticity in the streamwise direction. Despite these differences, the K-H forcing must have some projection on the optimal forcing mode responsible for the centrifugal instability such that this mechanism can be excited. This will be examined in detail in \S\S\ref{subsec:Resolvent-based reconstruction of the weakly non-linear centrifugal instability mechanism}.

\subsection{Resolvent-based reconstruction of the weakly non-linear centrifugal instability mechanism}\label{subsec:Resolvent-based reconstruction of the weakly non-linear centrifugal instability mechanism}
Broadly following the approach of \citet{dwivedi_sidharth_jovanovic_2022} who derived the non-linear $(\pm2\beta,0\omega)$ forcing from weakly non-linear analysis, we investigate the capability of the resolvent basis to approximate the non-linear forcing/response mechanism at $(\pm2\beta,0\omega)$ for low amplitude forcing and low reversed flow. In doing so, we aim to answer the questions: ``is this mechanism low-rank?'' and ``how similar/different is the K-H forcing to the optimal forcing of the $(\pm2\beta,0\omega)$ mode?''. Our procedure, although sharing many commonalities with \citet{dwivedi_sidharth_jovanovic_2022}, employs the mean-flow (not the laminar base-flow) to extract the resolvent operator. This is to account for any changes in the base-flow topology which are readily calculated by HBNS.

Following the linear resolvent framework from \eqref{eq:linear resolvent} to \eqref{eq:generalised EVP resolvent}, the $(2\beta,0\omega)$ response may be written as,
\begin{equation}
    \hat{\boldsymbol{w}}_{2,0} = \mathsfbi{H} \left ( 2\beta,0\omega \right ) \mathsfbi{M} \mathsfbi{P} \hat{\boldsymbol{f}}_{2,0},
    \label{eq:linear resolvent (2,0)}
\end{equation}
\noindent where $\mathsfbi{H} \left ( 2\beta,0\omega \right ) = \left [ -\mathsfbi{A}_{2} (\hat{\boldsymbol{w}}_{0,0}) \right ]^{-1}$ is the resolvent operator extracted from the time- and spanwise-averaged flow at amplitude $A=1\times10^{-7}$ and $\hat{\boldsymbol{f}}_{2,0} = \hat{\boldsymbol{f}}_{\text{K-H}}$ is the K-H forcing acting on the $(2\beta,0\omega)$ harmonic. Considering the set of orthonormal forcing $\left [ \hat{\boldsymbol{f}}_1,\hat{\boldsymbol{f}}_2,\dots,\hat{\boldsymbol{f}}_n \right ]$ and response $\left [ \hat{\boldsymbol{u}}_1,\hat{\boldsymbol{u}}_2,\dots,\hat{\boldsymbol{u}}_n \right ]$ modes, where $\hat{\boldsymbol{u}}_{i}$ contains only the three components of velocity, and the associated energy gains $\left ( \sigma_1,\sigma_2,\dots,\sigma_n \right )$ (analogous to $\mathcal{G}$), and further letting
\begin{equation}
    \alpha_{i} = \frac{\left \langle \hat{\boldsymbol{f}}_{i},\hat{\boldsymbol{f}}_{\text{K-H}} \right \rangle_{\Omega_{\hat{\boldsymbol{f}}}}}{\sqrt{\left \langle \hat{\boldsymbol{f}}_{\text{K-H}},\hat{\boldsymbol{f}}_{\text{K-H}} \right \rangle_{\Omega_{\hat{\boldsymbol{f}}}}}}
    \label{eq:projection coefficient}
\end{equation}
\noindent be the scalar projection coefficient of the $i$-th forcing vector on the K-H forcing, the non-linear $(2\beta,0\omega)$ response can be written as
\begin{equation}
    \hat{\boldsymbol{w}}_{2,0} = \sum_{i=1}^{n\rightarrow\infty} \hat{\boldsymbol{u}}_{i} \sigma_{i} \alpha_{i} \sqrt{\left \langle \hat{\boldsymbol{f}}_{\text{K-H}},\hat{\boldsymbol{f}}_{\text{K-H}} \right \rangle_{\Omega_{\hat{\boldsymbol{f}}}}},
    \label{eq:response reconstruction}
\end{equation}
\noindent and, similarly, the K-H forcing as
\begin{equation}
    \hat{\boldsymbol{f}}_{\text{K-H}} = \sum_{i=1}^{n\rightarrow\infty} \hat{\boldsymbol{f}}_{i} \alpha_{i} \sqrt{\left \langle \hat{\boldsymbol{f}}_{\text{K-H}},\hat{\boldsymbol{f}}_{\text{K-H}} \right \rangle_{\Omega_{\hat{\boldsymbol{f}}}}},
    \label{eq:forcing reconstruction}
\end{equation}
\noindent Approximations of the non-linear K-H forcing $\tilde{\boldsymbol{f}}_{\text{K-H}}$ and response $\tilde{\boldsymbol{w}}_{2,0}$ can be reconstructed with a finite set of modes ($n=N_{m}<\infty$).

A suitable metric to assess the accuracy of the approximation is the relative error, 
\begin{equation}
    \varepsilon_{\hat{\boldsymbol{f}}} = \frac{\sqrt{\left \langle \hat{\boldsymbol{f}}_{\text{K-H}}-\tilde{\boldsymbol{f}}_{\text{K-H}},\hat{\boldsymbol{f}}_{\text{K-H}}-\tilde{\boldsymbol{f}}_{\text{K-H}} \right \rangle_{\Omega_{\hat{\boldsymbol{f}}}}}}{\sqrt{\left \langle \hat{\boldsymbol{f}}_{\text{K-H}},\hat{\boldsymbol{f}}_{\text{K-H}} \right \rangle_{\Omega_{\hat{\boldsymbol{f}}}}}},
    \label{eq:relative error forcing}
\end{equation}
for the K-H forcing, and 
\begin{equation}
    \varepsilon_{\hat{\boldsymbol{w}}} = \frac{\sqrt{\left \langle \hat{\boldsymbol{w}}_{2,0}-\tilde{\boldsymbol{w}}_{2,0},\hat{\boldsymbol{w}}_{2,0}-\tilde{\boldsymbol{w}}_{2,0} \right \rangle_{\Omega_{\hat{\boldsymbol{w}}}}}}{\sqrt{\left \langle \hat{\boldsymbol{w}}_{2,0},\hat{\boldsymbol{w}}_{2,0} \right \rangle_{\Omega_{\hat{\boldsymbol{w}}}}}},
    \label{eq:relative error response}
\end{equation}
for the non-linear $(2\beta,0\omega)$ response. It can also be shown (see \S\ref{app:relative error} for the derivation) that the relative error of the response approximation is upper bounded according to
\begin{equation}
    \varepsilon_{\hat{\boldsymbol{w}}} \leq \frac{\sigma_{N_{m}+1}}{\overline{\sigma}_{N_{m}}} \frac{\varepsilon_{\hat{\boldsymbol{f}}}}{\sqrt{1-\varepsilon_{\hat{\boldsymbol{f}}}^{2}}} \leq \frac{\varepsilon_{\hat{\boldsymbol{f}}}}{\sqrt{1-\varepsilon_{\hat{\boldsymbol{f}}}^{2}}},
    \label{eq:relative error bounds}
\end{equation}
\noindent where
\begin{equation}
    \overline{\sigma}_{N_{m}}^{2} = \frac{\sum_{i=1}^{N_{m}} \sigma_{i}^2 \left| \alpha_{i} \right|^{2}}{\sum_{i=1}^{N_{m}} \left| \alpha_{i} \right|^{2}}, 
    \label{eq:sigma bary N}
\end{equation}
\noindent is the weighted average of the $\sigma_{i}$ distribution in the range $[1,N_{m}]$ and $\sigma_{N_{m}+1}$ is the gain of the first truncated mode sitting outside of the prescribed range. The ratio $\sigma_{N_{m}+1}/\overline{\sigma}_{N_{m}}\leq1$ indicates the distance of the first truncated mode from the weighted average and its value decreases with $N_{m}$. Effectively, the smaller the ratio, the more energy of the forcing/response mechanism is captured by the resolvent basis expansion.

\begin{figure}
    \centering
    \includegraphics[width=\linewidth]{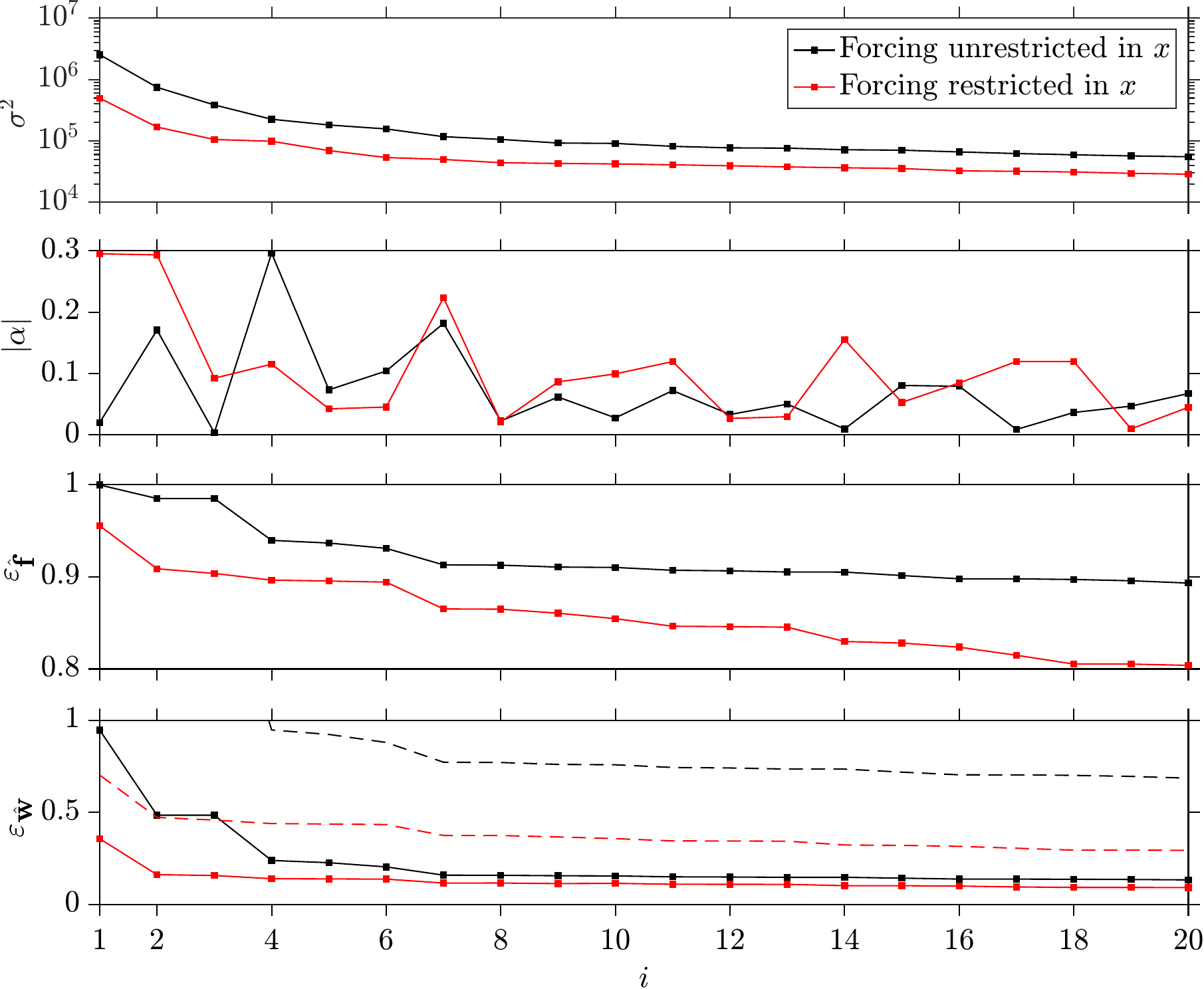}
    \caption{(First row) squared resolvent gain, (second row) modulus of the projection coefficient, (third row) relative error of the K-H forcing approximation, and (fourth row) of the non-linear response approximation with error bound  $\frac{\sigma_{N_{m}+1}}{\overline{\sigma}_{N_{m}}} \frac{\varepsilon_{\hat{\boldsymbol{f}}}}{\sqrt{1-\varepsilon_{\hat{\boldsymbol{f}}}^{2}}}$ plotted with dashed line.}
    \label{fig:reconstruction performance}
\end{figure}

\begin{figure}
    \centering
    \includegraphics[width=\linewidth]{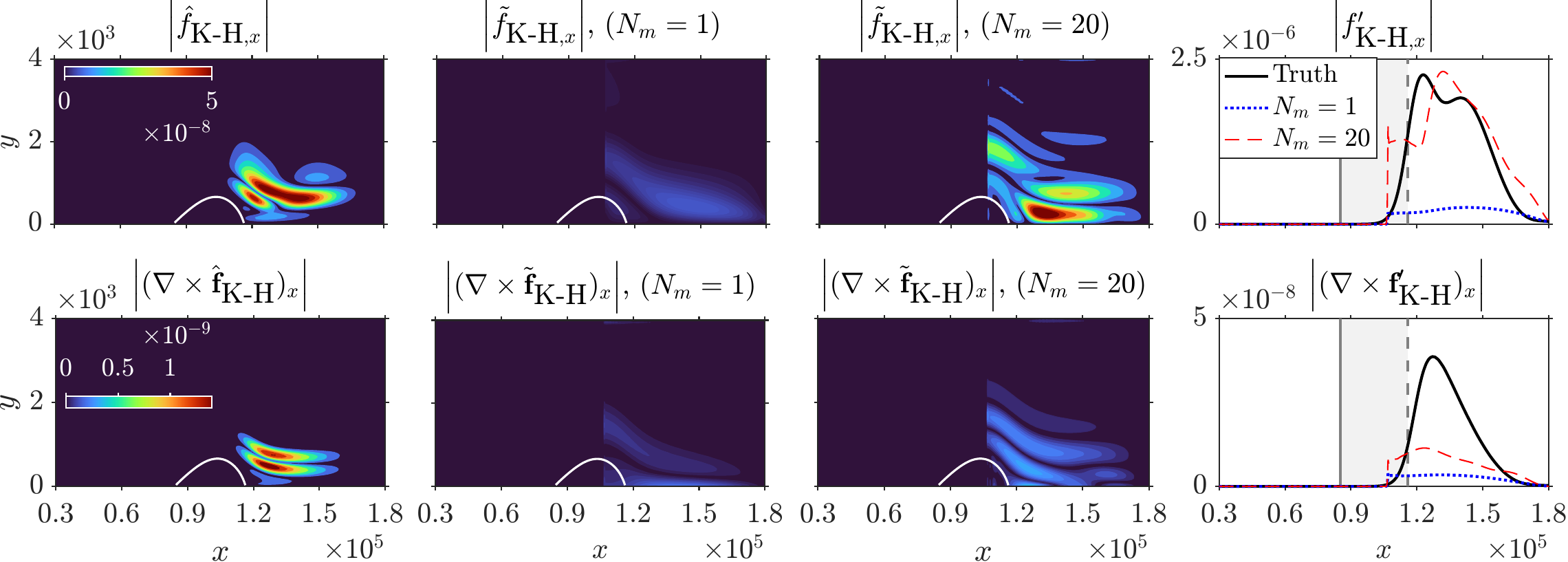}
    \caption{\flavio{Resolvent-based reconstruction of the weakly non-linear ($A=1\times10^{-7}$) K-H forcing $\hat{\boldsymbol{f}}_{\text{K-H}}$ from the restricted forcing case. (First column) true non-linear forcing, reconstruction with (second column) $N_{m}=1$ mode, (third column) $N_{m}=20$ modes and (fourth column) amplitudes. Streamwise components of the forcing (top) and curl of forcing (bottom) shown.}}
    \label{fig:forcing reconstruction}
\end{figure}

The domain of the linear forcing is capped at $y=4\times10^3$ to avoid spurious structures appearing in the far field. We also consider two cases, one where the forcing is not restricted in $x$ and another where it is calculated within $[x_{S}+0.7L_{b},x_{out}]$, i.e. from 70\% of the bubble length to the outlet. This is an indicative region coinciding with incipient shear layer reattachment and the non-linear interaction of K-H waves to mimic the true non-linear mechanism. We outline the reconstruction performance of both test cases in figure \ref{fig:reconstruction performance}. As the behaviour of the squared energy gain in the top row suggests, the linearised operator in \eqref{eq:linear resolvent (2,0)} is not low-rank, implying the underlying mechanism cannot be approximated accurately with a low-order linear model. The restricted forcing case achieves better projections overall, especially within the leading 3 modes -- see second panel in figure \ref{fig:reconstruction performance}. While the behaviour of $\alpha$ is locally erratic, we observe a decay on average, meaning the projection degrades for high-order modes. Overall, the relative error of the forcing approximation $\varepsilon_{\hat{\boldsymbol{f}}}$ is large -- see third panel. The restricted forcing case is clearly better than the unrestricted one, however, even with 20 modes, the error settles around 80\% and 90\%, respectively. These results highlight the limitations of linear RA in modelling the internal stresses produced by the non-linear dynamics of this LSB flow even at low amplitude. In figure \ref{fig:forcing reconstruction} we compare the shape of the true and approximated K-H forcing with streamwise spatial restriction for the streamwise component of both the forcing itself and the curl of the forcing. The ground truth (first column) is characterised by structures emerging in the site of reattachment, which is where saturation of K-H waves occurs -- refer to figure \ref{fig:HBNS21_amplitudes} (middle panels). The approximation with 1 mode (second column) falls short in that both the shape and the amplitude are extremely inaccurate, leading to about 95\% error. The reconstruction improves with 20 modes (third column) but the error of the $x$-component of the curl is still large, which is attributed to the $y$ and $z$ components of the forcing vector -- see figure \ref{fig:reconstruction errors} in \S\ref{app:relative error} for more details. 

\begin{figure}
    \centering
    \includegraphics[width=\linewidth]{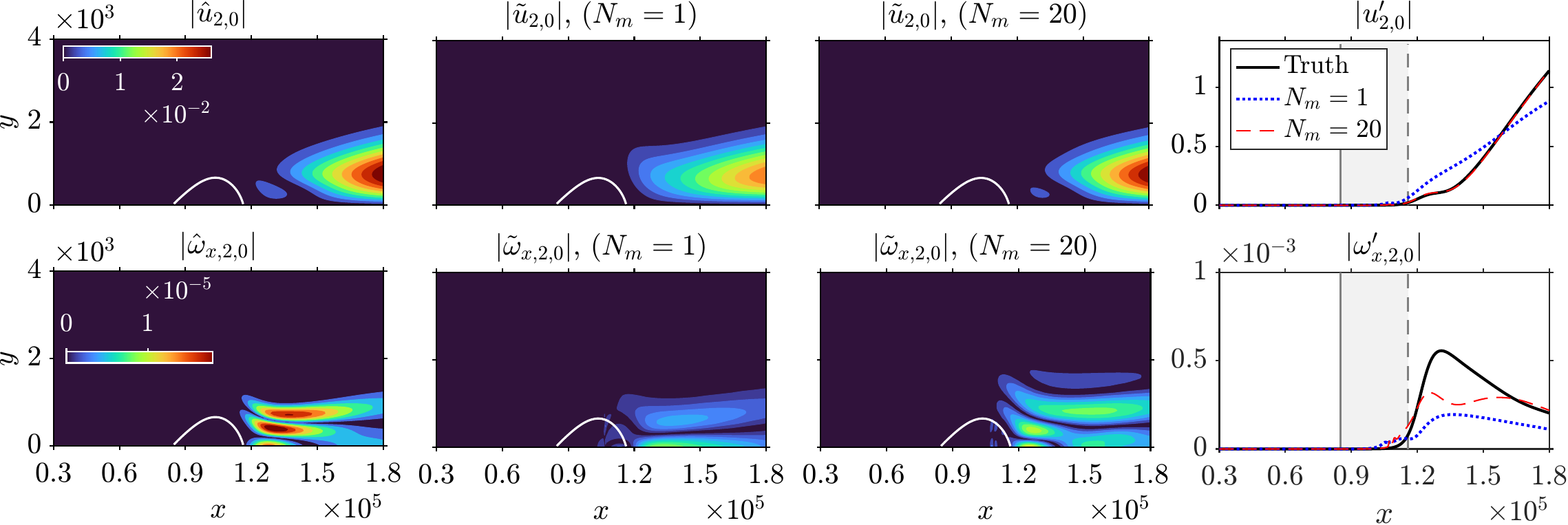}
    \caption{\flavio{Same as figure \ref{fig:forcing reconstruction} but for the non-linear response $\hat{\boldsymbol{w}}_{2,0}$.}}
    \label{fig:response reconstruction}
\end{figure}

Despite the poor reconstruction of the K-H forcing, the approximation of the non-linear $(2\beta,0\omega)$ response is surprisingly good. With only 4 modes in the unrestricted forcing case, we achieve an error of about 25\%, while with only 2 modes in the restricted forcing case, we reach an error of 15\%. With 7 modes the error drops to about 15\% and 10\% in the unrestricted and restricted cases, respectively -- see fourth panel in figure \ref{fig:reconstruction performance}. The $u'$ structure of the response is qualitatively captured by only the first resolvent mode, but with 20 modes the improvement is remarkable since we observe virtually no mismatch in the amplitude (figure \ref{fig:response reconstruction}, top). The streamwise vorticity of the response computed as $\tilde{\omega}_{x,2,0} = \partial_y \tilde{w}_{2,0} - i\beta \tilde{v}_{2,0}$ is less accurate compared to the $u'$ reconstruction (figure \ref{fig:response reconstruction}, bottom). While 1 mode is insufficient to recover the vortical structures even qualitatively, 20 modes improve the accuracy but the error is still above 20\% (refer to figure \ref{fig:reconstruction errors}, bottom). 

The results show that many modes are required to reconstruct the full streamwise vorticity triggered by centrifugal instability. We can therefore conclude that centrifugal instability is not due to a single resolvent mode, but to multiple modes, which all exhibit streamwise vorticity. 

\subsection{Role of centrifugal instability for laminar-turbulent transition}\label{subsec:Role of centrifugal instability for laminar-turbulent transition}
\begin{figure}
    \centering
    \includegraphics[width=0.8\linewidth]{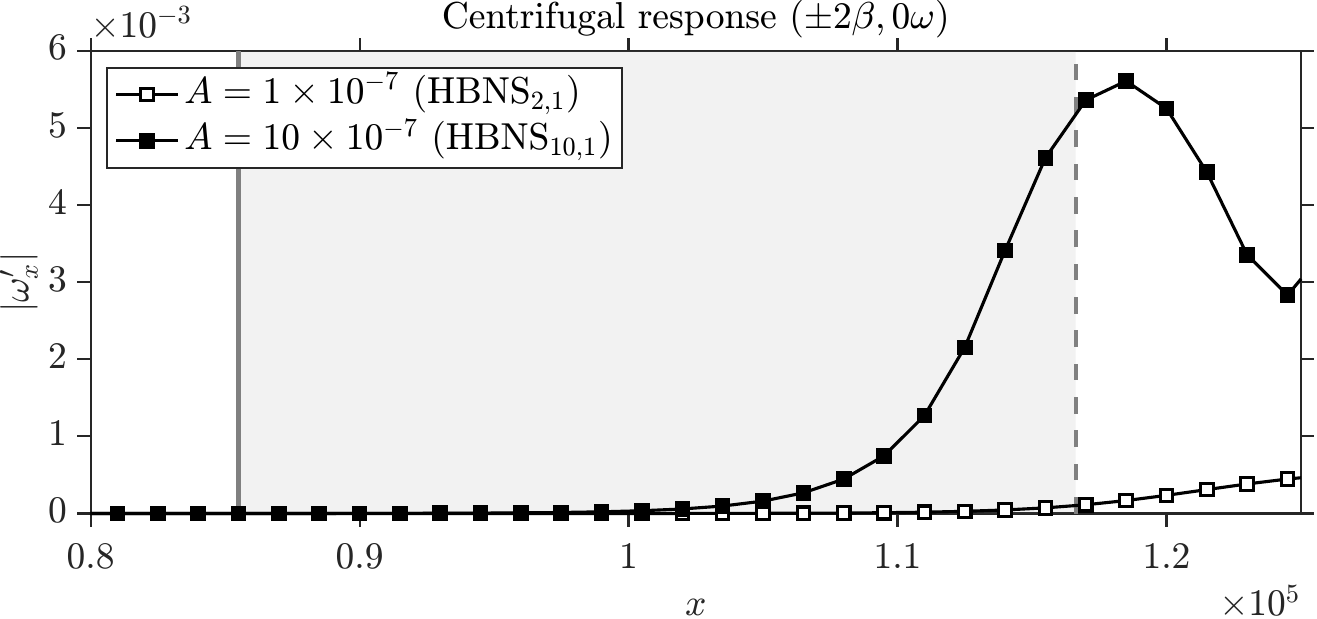}
    \caption{\flavio{Amplitude of $\omega_{x}'$ of the non-linear $(\pm2\beta,0\omega)$ response at low and high forcing amplitudes.}}
    \label{fig:centrifugal instability high amplitude}
\end{figure}

At high forcing amplitude ($A=10\times10^{-7}$), the centrifugal instability arising in region I of figure \ref{fig:HBNS21_amplitudes} becomes more unstable and grows farther upstream inside the separation zone (see figure \ref{fig:centrifugal instability high amplitude}), a feature observed in the global eigenmode representative of modal centrifugal instability of the bubble -- figures \ref{fig:centrifugal GSA vs RA}-\ref{fig:eigenmode amplitudes}.

In the remainder of the manuscript, we describe the role of this instability for breakdown of the main K-H roller shed by the shear layer, which \flavio{paves the way} to turbulence. We highlight that at high amplitude weakly non-linear approximations fail and other methods, such as DNS, become necessary. Our HBNS methodology offers a self-contained tool to also explore high amplitude disturbance mechanisms of the late transition stages. We recall from figure \ref{fig:dCD} that larger systems are required for these regimes, hence, we employ 10 harmonics in space ($M=10$) and 1 in time ($N=1$) for the analysis hereafter.

\subsection{Spanwise vortex distortion and breakdown}\label{subsec:Spanwise vortex distortion and breakdown}
\begin{figure}
    \centering
    \includegraphics[width=\textwidth]{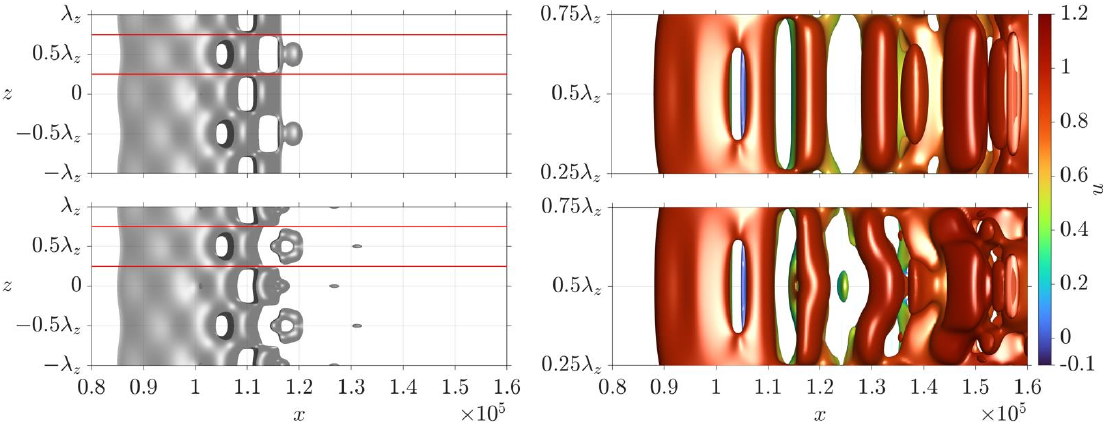}
    \caption{\flavio{Reconstruction of the instantaneous flow at $A=10\times10^{-7}$ from HBNS$_{10,1}$. Isosurfaces of (left) $u=0$ and (right) Q-criterion (iso-value: $4\times10^{-11}$) coloured with $u$ within the spanwise domain marked by the red-solid lines. (Top) time- and spanwise-averaged flow + $(\pm1\beta,\pm1\omega)$ harmonic, (bottom) time- and spanwise-averaged flow + $(\pm1\beta,\pm1\omega)$ + $(\pm2\beta,0\omega)$ + $(\pm3\beta,\pm1\omega)$ harmonics.}}
    \label{fig:instantaneous bubble roll up and vortex distortion}
\end{figure}

In order to describe the sequence of events leading to breakdown of the shear layer, we reconstruct the spatio-temporal flow obtained from HBNS$_{10,1}$ system at $A=10\times10^{-7}$. Spanwise-periodic oscillations of wavelength $\lambda_z=2\pi/\beta$ (approximately equal to the mean separation length, $L_b$) appear in the separation bubble and are ascribed to the K-H mode which is activated in the fore part of the bubble (figure \ref{fig:instantaneous bubble roll up and vortex distortion}, top left). Mutually, a roller of finite span ($=\lambda_z/2$) detaches from the shear layer at $x\approx 1.1\times10^5$ (top right), indicative of the roll-up process \citep{Marxen2013vortex,Michelis2018spanwise,Michelis2018origin,Hosseinverdi2020onset}. Spanwise vortices are shed downstream at fixed temporal frequency equal to the K-H instability frequency, confirming that the main vortex shedding dynamics is driven by the K-H mechanism and hence explaining the success of linear stability analysis in recovering its characteristic frequency \citep{Simoni2013experimentalwakes,Yarusevych2017steady,Ziade2018shear,Michelis2018spanwise,Michelis2018origin}.

\begin{figure}
    \centering
    \includegraphics[width=\textwidth]{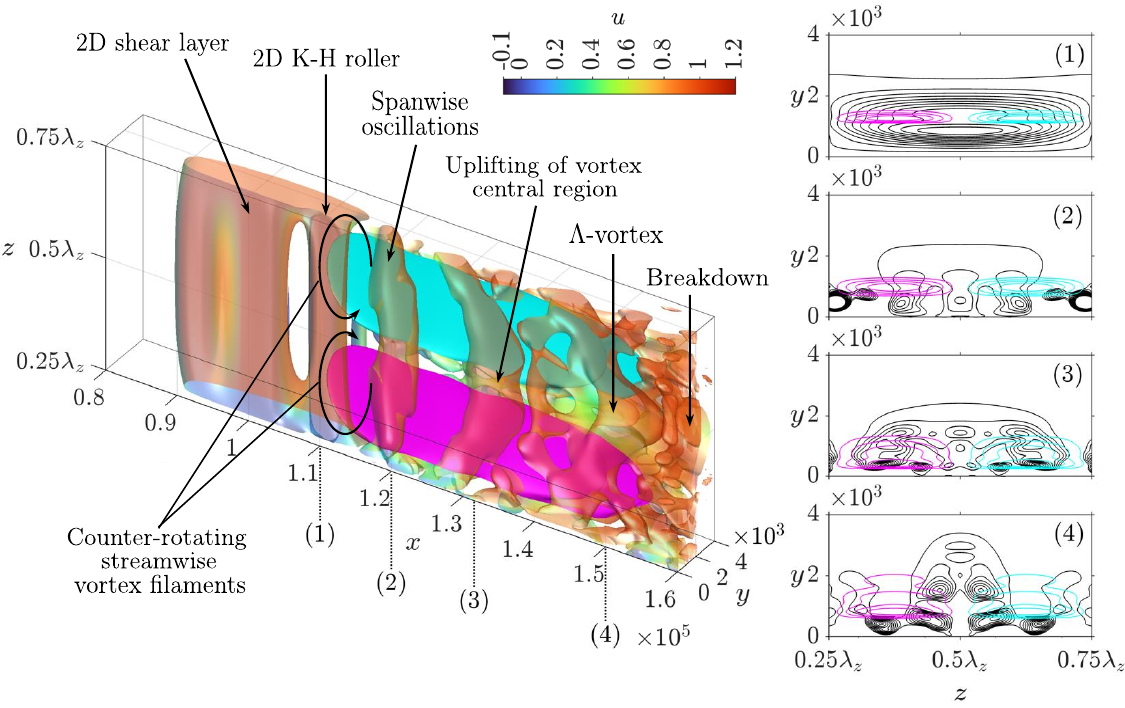}
    \caption{\flavio{(Left) Isosurfaces of Q-criterion (iso-value: $4\times10^{-11}$) coloured with $u$ from the full flow field and of (magenta/cyan) positive/negative $\omega_{x}'$ from the $(\pm2\beta,0\omega)$ mode. (Right) $y$-$z$ planar views extracted at four $x$-stations $[1.1,1.2,1.32,1.5]\times10^5$, labelled with numbers $(1)$ to $(4)$.}}
    \label{fig:physical mechanism of spanwise vortex distortion}
\end{figure}

Spanwise K-H rollers are prone to destabilise due to the action of the centrifugal instability discussed earlier. More specifically, the net effect of the interaction between $(\pm1\beta,\pm1\omega)$ and $(\pm2\beta,0\omega)$ modes, which gives $(\pm3\beta,\pm1\omega)$ through a suitable triad, is the distortion of the spanwise vortex, as seen in the bottom panels of figure \ref{fig:instantaneous bubble roll up and vortex distortion}. 
The K-H rollers developing within the shear layer are progressively distorted into structures that are reminiscent of $\Lambda$-vortices (bottom right of figure \ref{fig:instantaneous bubble roll up and vortex distortion}). The uplift of (negative) spanwise vorticity at $x\approx1.3-1.4\times10^5$, which, in turn, promotes mixing between the boundary layer and the outer flow, is a direct consequence of this mechanism. Therefore, it is observed that the $(\pm1\beta,\pm1\omega) + (\pm2\beta,0\omega) = (\pm3\beta,\pm1\omega)$ triadic interaction is key for the initiation of the main K-H vortex breakdown which paves the way to turbulence. This mechanism is further illustrated in figure \ref{fig:physical mechanism of spanwise vortex distortion}. Here we provide an overview of the complete sequence of events that characterises the breakdown of the K-H vortex and discuss the underlying physical mechanism. On the left, we show the three-dimensional structure of the transitional shear layer and on the right four $y$-$z$ sectional views extracted at $x=[1.1,1.2,1.32,1.5]\times10^5$. At the first station, the K-H roller is essentially 2D. At the second station, the shape of the roller is visibly distorted since spanwise oscillations appear. At the third station, the deformation is enhanced as the roller is no longer aligned in the $z$-direction. The central region is uplifted and stretched away from the sides along the streamwise direction. At the fourth station we observe a $\Lambda$-shaped structure. The mechanism by which the roller deforms is driven by the centrifugal instability, appearing in the figure as a pair of counter-rotating streamwise vortices whose distance from the centres is $\lambda_z/4$, i.e. approximately one quarter of the mean separation length, $L_b$. These vortices act through the braid regions (located between two consecutive rollers) and induce upward flow that interacts with the spanwise roller, causing its distortion and disintegration \citep{Marxen2013vortex,kumar2023}. This process shares some similarities with secondary flow structures, called ``rib'' vortices, which are responsible for the breakdown of spanwise rollers in mixing layers and wakes \citep{lopez_bulbeck1993,sun_schrijer_scarano_van_oudheusden2014,yang2019}. While we do not observe vortex loops revolving around the rollers, the mechanism whereby streamwise vorticity disintegrates the main vortex is present in our results, the origin of streamwise vortices being a centrifugal-type instability of convective nature discussed earlier.

\begin{figure}
    \centering
    \includegraphics[width=0.9\textwidth]{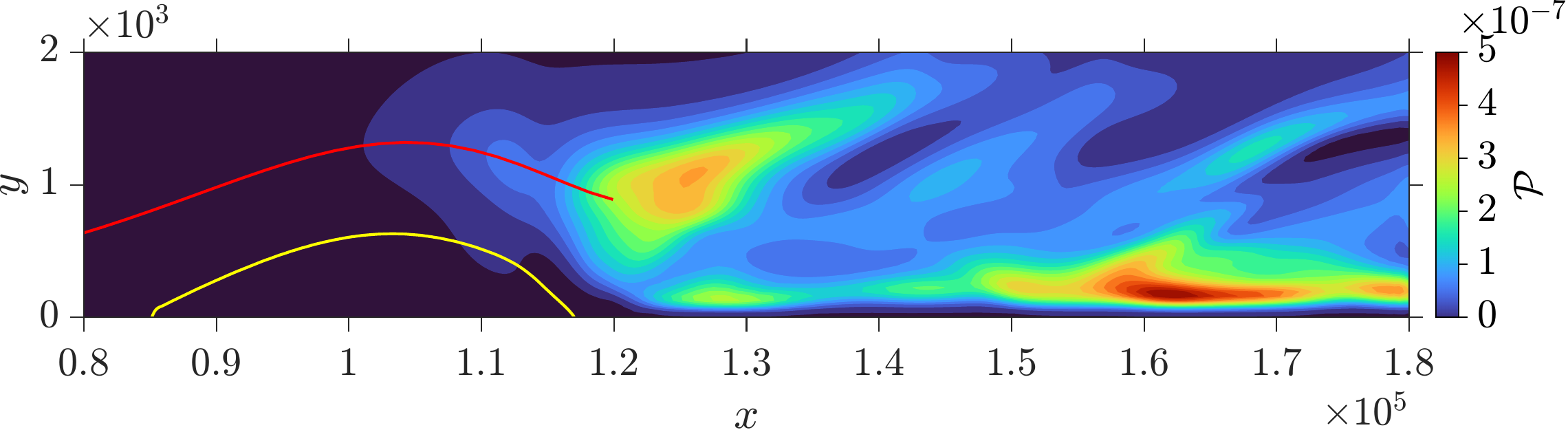}
    \caption{\flavio{Contours of overall production of turbulent kinetic energy $\mathcal{P} = -\overline{u_{i}^{'}u_{j}^{'}}\frac{\partial \left \langle u_i \right \rangle_{z,t}}{\partial x_j}$. The dividing streamline of the time- and spanwise-averaged separation bubble and inflection line are plotted in yellow and red, respectively.}}
    \label{fig:productionTKE}
\end{figure}

The complete breakdown of $\Lambda$-structures establishes transition to turbulence of the shear layer. Notably, when the large scales of the flow breakdown to turbulence energy spreads to infinitely many (smaller) scales all the way down to the Kolmogorov microscales \citep{Pope2000turbulent,George2013}. Capturing these structures is outside the scope of this work since the proposed methodology aims at characterising the transitional stages of the flow given a finite set of harmonics retained in the model. This means we are able to resolve only part of the energy spectrum, the remainder being the truncation error associated with the expansion \eqref{eq:Fourier expansion}. Regardless of this limitation, the HBNS$_{10,1}$ system is able to capture the final stage of the $\Lambda$-vortex breakdown -- see figure \ref{fig:physical mechanism of spanwise vortex distortion}. 
Additional insight into the main sites of turbulence production is offered by plotting the production of turbulent kinetic energy in figure \ref{fig:productionTKE}. Two regions of the flow are identified, one is along the inflection line near incipient shear layer reattachment, which is where shear layer roll-up and K-H roller shedding take place. The correct estimation of production at this station is paramount for the prediction of the bubble topology and, consequently, the generation of turbulence in the redeveloping boundary layer downstream of reattachment, which constitute two major problems in RANS turbulence models applied to separated transitional shear layers \citep{bernardos2019algebraic,bernardos2019prediction}. Along the vortex shedding line ($y\approx2\delta^{*}$), but substantially downstream of reattachment, there are local patches of production, which match the locations of $\Lambda$-vortex formation ($x\approx1.4-1.5\times10^5$) and breakdown ($x\approx1.6-1.7\times10^5$). Much closer to the wall ($y\approx\delta^{*}/5$), the thin layer of turbulence production is related to near-wall turbulence.


\subsection{Integral quantities\label{subsec:Integral quantities}}
\begin{figure}
    \centering
    \includegraphics[width=\linewidth]{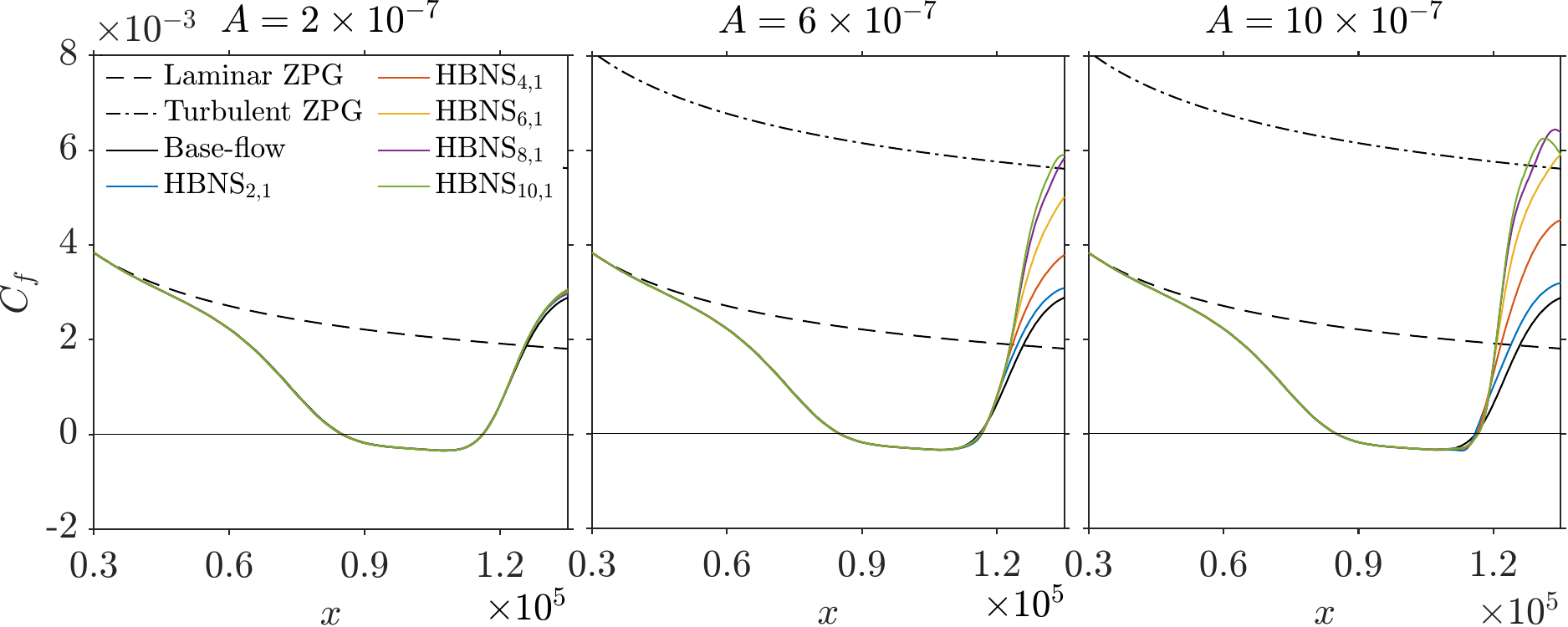}
    \caption{Skin friction coefficient evaluated for different QL-in-$\omega$ systems ($N=1,M>1$) and three forcing amplitudes $A=[2,6,10]\times10^{-7}$. The laminar and turbulent (from \citet{White1991}) ZPG boundary layer profiles are plotted with black dashed and dashed-dotted lines.}
    \label{fig:Cf vs amplitude}
\end{figure}

Although the accurate estimation of the boundary layer transition point is an extremely arduous task, it is often sought to inform turbulence models used in steady simulations for which it is clearly impossible to predict transition otherwise \citep{bernardos2019algebraic,bernardos2019prediction,dellacasagrande_lengani_simoni_ubaldi2024}. Being equipped with spatio-temporal information of the flow, the transition point is yet not fixed in time/space and a single location cannot be identified. Nevertheless, we employ boundary layer integral analysis on the time- and spanwise-averaged flow to possibly draw a connection between the dynamic events of transition described thus far and the statistical characteristics of the boundary layer. In figure \ref{fig:Cf vs amplitude} we plot the skin friction coefficient $C_f$ of the time- and spanwise-averaged flow for five HBNS system architectures: HBNS$_{2,1}$, HBNS$_{4,1}$, HBNS$_{6,1}$, HBNS$_{8,1}$ and HBNS$_{10,1}$. The reference laminar (Blasius) and turbulent \citep{White1991} curves for ZPG boundary layer are also superimposed together with the baseline $C_f$ from the laminar base-flow.

Firstly, the turbulent curve is reached only by systems \flavio{with $M\geq8$}, in agreement with the analysis in \S\ref{subsec:quasi-linear-in-omega and HBNS-in-beta systems}. When the truncation error is large, the turbulent energy cascade cannot take place and the skin friction levels remain close to the baseline laminar curve (systems HBNS$_{2,1}$ and HBNS$_{4,1}$). The HBNS$_{6,1}$ system contains the minimum number of harmonics necessary to reach the turbulent curve at $A=10\times10^5$, but trajectories of $C_f$ at high forcing amplitudes converge only for $M \geq 8$. In this case, the critical amplitude is found to be $A=6\times10^{-7}$. As the amplitude is progressively increased, the coordinate where the skin friction crosses the turbulent curve moves towards the mean reattachment point. A similar trend is also observed for increasing system's order, suggesting that in the limit of zero truncation error, transition and reattachment should collapse to the same point, giving rise to the turbulent reattachment scenario observed in \citep{hosseinverdi_fasel_2019,Hosseinverdi2020onset}.

\begin{figure}
\centering
\begin{subfigure}{\textwidth}
  \centering
  \includegraphics[width=0.9\linewidth,right]{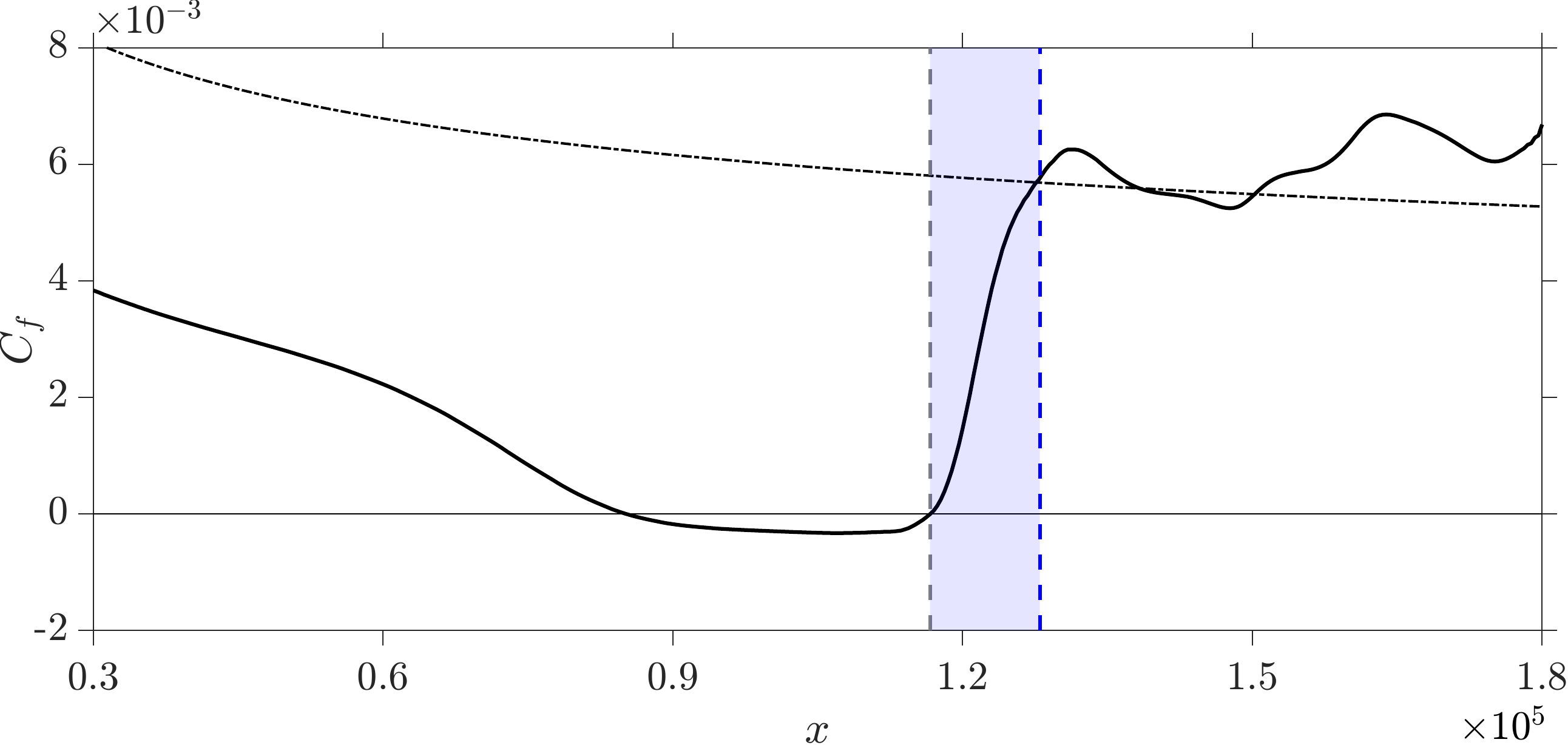}
\end{subfigure}\\
\vspace{2pt}
\begin{subfigure}{\textwidth}
  \centering
  \includegraphics[width=0.9\linewidth,right]{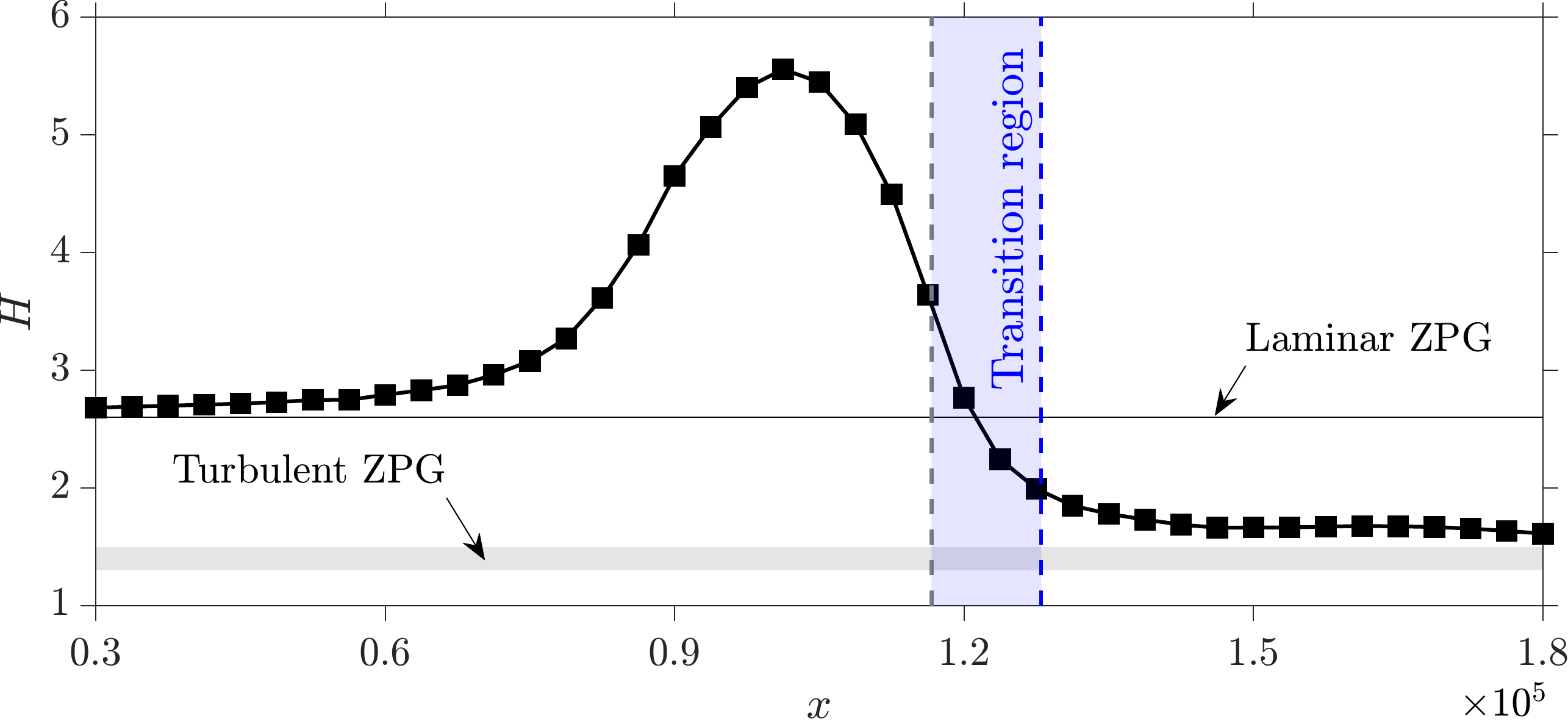}
\end{subfigure}\\
\vspace{1pt}
\begin{subfigure}{\textwidth}
  \centering
  \includegraphics[width=0.86\linewidth,right]{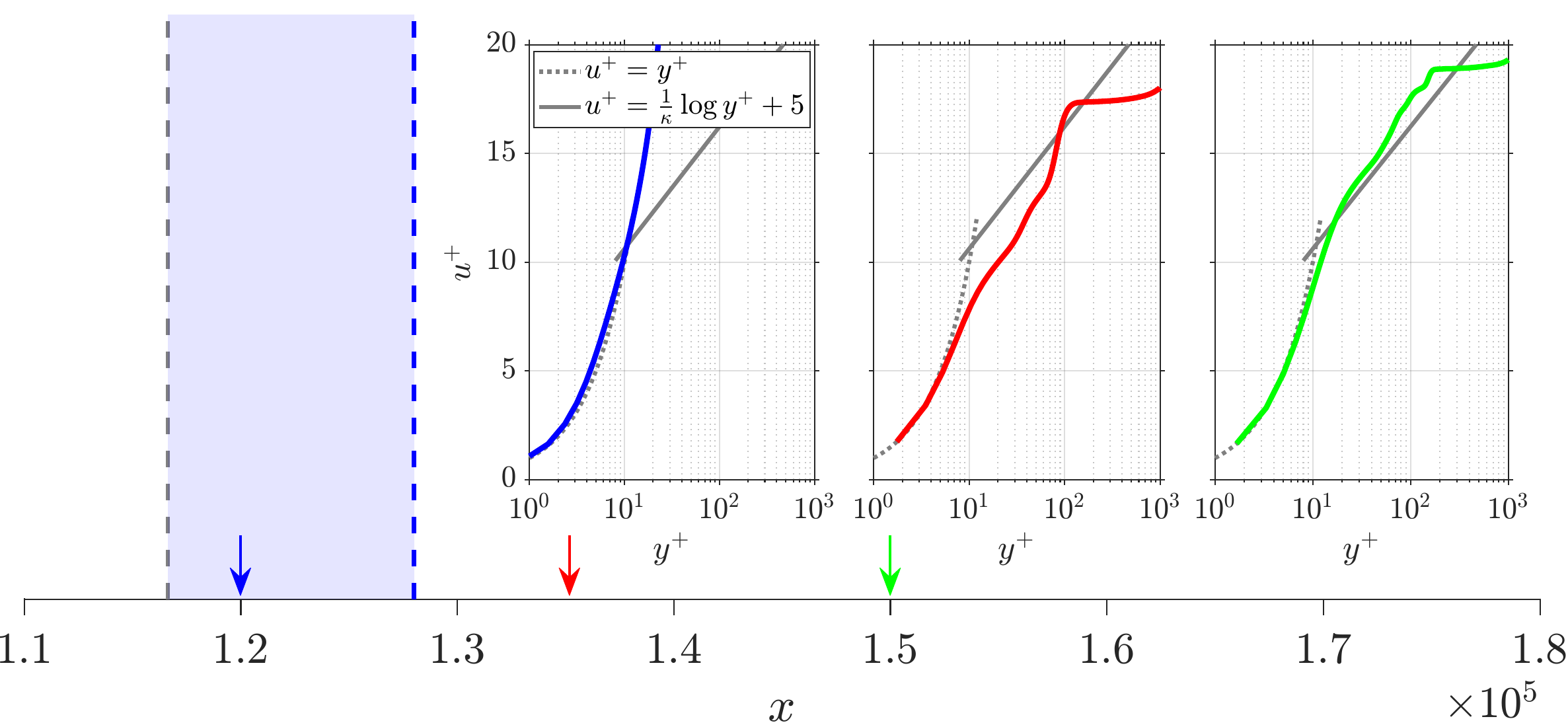}
\end{subfigure}
\caption{\flavio{(Top) Skin friction coefficient, (middle) shape factor and (bottom) boundary layer mean velocity profiles at $A=10\times10^{-7}$. Mean reattachment and the streamwise coordinate at which the skin friction reaches the turbulent curve are plotted with grey-dashed and blue-solid lines, respectively. Profiles extracted at: (blue) $x=1.2\times10^5$, (red) $x=1.35\times10^5$ and (green) $x=1.5\times10^5$.}}
\label{fig:integral}
\end{figure}

From the skin friction results we notice some robustness of the HBNS solutions in the prediction of $C_f$, at least in the transitional regime. Beyond this stage, i.e. where $C_f$ has crossed the turbulent curve, the turbulent reattached boundary layer developing downstream of the bubble requires an intractably large number of harmonics in both $z$ and $t$ to resolve the full spectrum of turbulence down to the dissipative scales. This is where the current state-of-the-art HBNS approach loses robustness and accuracy. An irregular, wavy behaviour of $C_f$ is observed in the top panel of figure \ref{fig:integral}, which also displays the statistical state of the boundary layer downstream of the transitional region. Despite the oscillations, the skin friction remains around the turbulent level. 

Another parameter we consider is the boundary layer shape factor $H=\delta^{*}/\theta$ plotted in the middle panel of figure \ref{fig:integral}. The shape factor gradually departs from the laminar value ($H=2.6$ for ZPG boundary layer, \citep{Pope2000turbulent}) due to the influence of the pressure gradient. High values of $H$ are reached in the separated zone ($H_{max} \approx 5.5$), later followed by a sharp drop due to the coexisting effects of boundary layer reattachment and onset of transition. In the transition region the shape factor crosses the laminar reference level and marches towards the turbulent range ($1.3 \leq H \leq 1.5$ for ZPG boundary layer, \citep{Pope2000turbulent}). Downstream of the transition region the shape factor settles around $H=1.6$, which is, together with the $C_f$ curve, another indication of a turbulent-like boundary layer in the post-reattachment region.

Finally, we show the boundary layer velocity profile in inner units extracted at three $x$-stations marked with blue ($x=1.2\times10^5$), red ($x=1.35\times10^5$) and green ($x=1.5\times10^5$) arrows in the bottom panel of figure \ref{fig:integral}. The profiles are superimposed on the viscous sublayer ($y^{+}<5$) and log-law ($30<y^{+}<300$) curves. A turbulent boundary layer is known to follow the universal log-law profile in the log region \citep{Pope2000turbulent}, hence we look at whether and how accurately the extracted profiles follow this trend. The velocity profile at $x=1.2\times10^5$ is extracted from the early-transitional region, and, in agreement with $C_f$ and $H$, it is not turbulent yet - in fact, it is still very close to the laminar state since the $u^{+} = y^{+}$ law is followed accurately in the viscous sublayer, but the log-layer slope is missing. The boundary layer is definitely not turbulent at this stage, also because, from a spatio-temporal point of view, the process of K-H roller distortion and breakdown is still at an early stage (refer to figures \ref{fig:instantaneous bubble roll up and vortex distortion}-\ref{fig:physical mechanism of spanwise vortex distortion}). At $x=1.35\times10^5$, the profile bends towards the log-law curve, portending the breakdown event, yet, the slope is not perfectly recovered. At this station, the K-H vortex has deformed sufficiently to form a $\Lambda$. Ultimately, the log-layer slope is captured more accurately at $x=1.5\times10^5$, suggesting a more pronounced turbulent character of the boundary layer. Contextually, the $\Lambda$-vortex is on the verge of breaking into small-scale turbulence.

It should be noted that a fully developed turbulent profile may not be achievable within our computational domain since the boundary layer relaxes to the turbulent equilibrium several separation bubble lengths downstream of the reattachment point, estimated to be around six by \citet{alam_sandham_2000}. \citet{hosseinverdi_fasel_2019} hypothesise this slow relaxation may be due to the persistence of coherent vortex structures reminiscent of the large-scale K-H rollers in the boundary layer downstream of reattachment. The final breakup of these large scales into small vortical structures generated near the wall yields profiles in better agreement with the fully developed turbulent boundary layer characteristics.

\section{Conclusions\label{sec:conclusions}}
Non-linear I/O analysis using the framework of the HBNS equations \citep{rigas2021HBM} was applied for the first time to spatially developing, non-parallel flows featuring flow separation. Specifically, we have studied the optimal route to turbulence of a convectively unstable laminar shear layer forming above a short separation bubble with peak reversed flow of 2\%. This baseline was shown to be stable to intrinsic global instabilities, making it suitable for the investigation of external disturbance/noise-driven transition.

Preliminary linear analysis by means of the resolvent approach revealed that boundary layers subject to APG promote the amplification of the K-H instability, even in the absence of separated flow (case C in figures \ref{fig:gain maps}-\ref{fig:T-S and K-H linear resolvent}). It was shown that this inviscid, shear-driven mechanism originating from the unstable inflectional boundary layer velocity profile, dominates over other convective instabilities, such as T-S waves, which are instead dominant in ZPG boundary layers. On the low-frequency end of the spectrum of disturbances, streamwise velocity streaks appear at $\beta=100\times10^{-5}$, as a result of the non-modal, transient, lift-up mechanism, typical of transitional boundary layers \citep{jacobs_durbin_2001,Balamurugan_Mandal_2017,rigas2021HBM}. In the presence of separation, the most unstable spanwise wavenumber for streaks increases to $\beta=150\times10^{-5}$ and the lift-up process, although preserving its fundamental physics, occurs over two distinct regions of the flow, one upstream of separation and the other in the post-reattachment redeveloping boundary layer. Although being very damped in globally stable separation bubbles compared to other convective disturbances, linear analyses also detected the modal centrifugal instability at low $\beta$ reported in literature \citep{theofilis2000,Gallaire2007three-dimensional,Rodriguez_gennaro_juniper_2013,Rodriguez2021self-excited}.

Linear stability analysis being limited only to the initial linear stage of the transition process, we employed the HBNS framework to describe the following stages where non-linearities govern the behaviour of the shear layer towards turbulence. Furthermore, we performed optimisation on the forcing shape in order to identify an efficient route to turbulence at a significantly lower cost compared to DNS or experiments where optimisation is computationally expensive, if not prohibitive.

At low forcing amplitude, we observed that K-H waves self-interact to produce a 3D, steady, Görtler-type mode that displays a strong streamwise vorticity signal near the reattachment point and induces streamwise velocity streaks by lift-up effect in the reattached boundary layer. This scenario shows similarities but also differences with respect to the mechanisms identified by \cite{schmid1992new,rigas2021HBM} for attached boundary layers. The oblique T-S instability is replaced here by oblique K-H, but both mechanisms generate a streamwise rotational forcing, this forcing being stronger in the case of K-H instability due to the fact that K-H instability is a stronger inviscid instability mechanism than the viscous T-S mechanism. Another striking difference is that in attached boundary layers streaks arise from the lift-up effect only, while in the separated case centrifugal instability first generates vortical forcing structures, which then transform into streamwise streaks by the lift-up effect. Hence, there are two inviscid mechanisms, the K-H and centrifugal mechanisms, that overall make the separated case more unstable. Moreover, it appeared that centrifugal instability does not just manifest within a single resolvent mode, but is recovered across many modes, which cumulatively reconstruct the streamwise vorticity of the non-linear response.

\begin{figure}
    \centering
    \includegraphics[width=\textwidth]{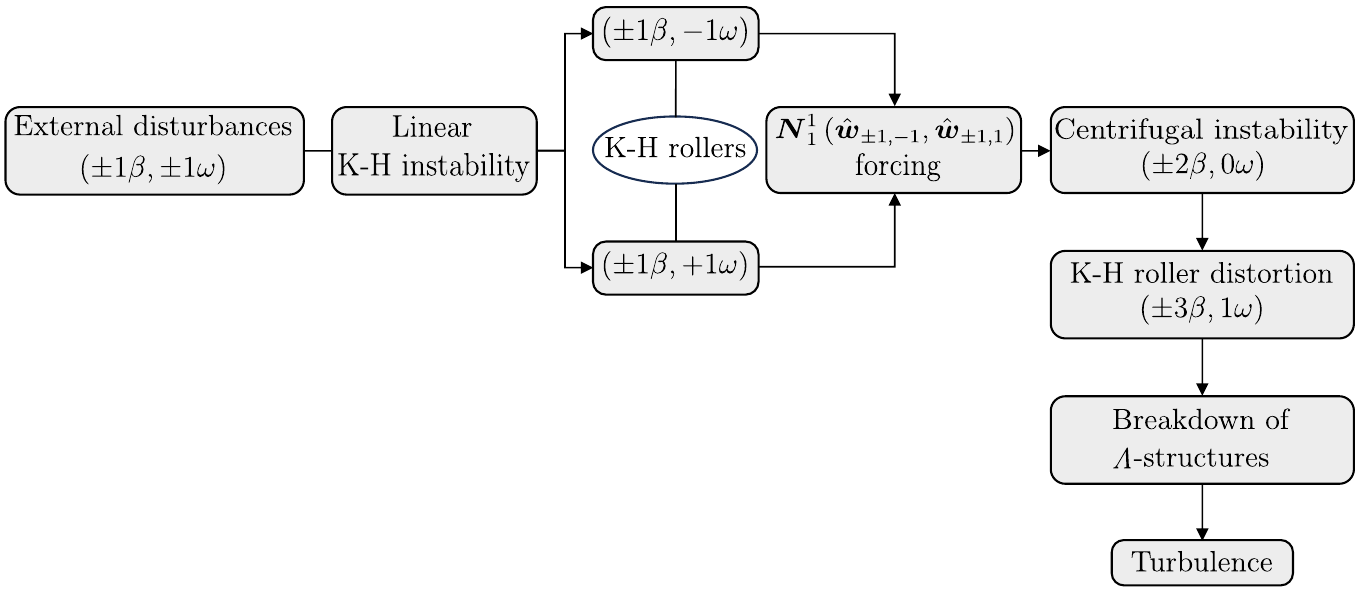}
    \caption{Summary of instability mechanisms leading to optimal transition to turbulence of the separated shear layer.}
    \label{fig:overview}
\end{figure}

At high forcing amplitude, our study revealed that transition of the separated shear layer is achieved by progressive three-dimensional distortion of the K-H rollers through a series of mechanisms that we summarise in the diagram \ref{fig:overview} and briefly recall below:
\begin{enumerate}
    \item Infinitesimal, three-dimensional disturbances force the K-H instability $(1\beta,1\omega)$ through a linear, non-modal mechanism that is driven by shear.
    \item The saturation of the K-H instability around the mean shear layer reattachment point leads to:
    \begin{itemize}
        \item Shear layer roll-up (a K-H roller of finite span detaches from the shear layer),
        \item The non-linear self-interaction of K-H waves, which generates forcing in the $(2\beta,0\omega)$ harmonic. Although not being the optimal source of excitation, it is sufficient to excite an inviscid centrifugal instability.
    \end{itemize}
    \item The centrifugal $(2\beta,0\omega)$ instability consists of counter-rotating, streamwise-oriented, stationary vortices of spanwise wavelength $\lambda_z=2\pi/2\beta$. Through the generation of upwash/downwash of streamwise momentum, they distort the K-H rollers, which develop spanwise oscillations.
    \item The distortion of K-H rollers under the action of streamwise vortices progressively leads to $\Lambda$-structures. They appear concurrently with the emergence of the $(3\beta,1\omega)$ mode, which is the result of the interaction between the $(2\beta,0\omega)$ and $(1\beta,1\omega)$ harmonics.
    \item The disintegration of $\Lambda$s marks the breakdown event and transition to turbulence.
\end{enumerate}
Overall, the fundamental connection between the K-H vortex breakup and the onset of turbulence we describe is substantially in agreement with \citet{Marxen2013vortex,kumar2023}.

We found that shear layer transition can be achieved robustly with quasi-linear-in-$\omega$ and non-linear-in-$\beta$ expansions with $M\geq8$ harmonics, implying that $2\omega$ harmonics do not play any significant role. This is valid prior to the establishment of turbulence, at which stage the energy spectrum becomes broadband and continuous. Integral boundary layer quantities confirm the occurrence of transition over a short distance away from the reattachment point, highlighting the important link between the transition and reattachment processes in separated shear layers \citep{Yarusevych2017steady,Karp2020suppression,Hosseinverdi2020onset,kumar2023}. Although our method does not resolve the turbulent scales, turbulent-like velocity profiles are obtained in the reattached boundary layer, and the log-law slope recovered quite accurately. Mismatches in the slope may originate from the prolonged influence of the pressure gradient even after the shear layer has reattached and the truncation error of the harmonic expansions at the fully turbulent stage. A remedy to the latter that we propose is the coupling of closure models with the existing framework, which is left to future work.

\flavio{Lastly, we note that the present work does not model the leading edge of the geometry, and therefore does not capture instabilities or modifications to existing modes that may arise due to leading-edge receptivity. A natural extension of this study would be to consider a more realistic aerofoil configuration or explicitly model the leading edge, as in \citet{Brandt_Sipp_Pralits_Marquet_2011}, to include the influence of upstream receptivity mechanisms on the transition process.}

\section*{Acknowledgements.}
This work was funded by the Air Force Office of Scientific Research (AFOSR)/European Office of Aerospace Research and Development (EOARD) (Award FA8655-21-1-7009).

\section*{Declaration of Interests.} The authors report no conflict of interest.

\section*{Author ORCIDs.}
\orcidicon{0000-0002-2576-0685} F. Savarino \href{https://orcid.org/0000-0002-2576-0685}{https://orcid.org/0000-0002-2576-0685};

\orcidicon{0000-0002-2808-3886} D. Sipp \href{https://orcid.org/0000-0002-2808-3886}{https://orcid.org/0000-0002-2808-3886};

\orcidicon{0000-0001-6692-6437} G. Rigas \href{https://orcid.org/0000-0001-6692-6437}{https://orcid.org/0000-0001-6692-6437}.

\appendix
\section{Validation of the LSB numerical set-up and comparison with literature}\label{app:validation of the LSB numerical set-up and comparison with literature}
\begin{figure}
\centering
    \includegraphics[width=\textwidth]{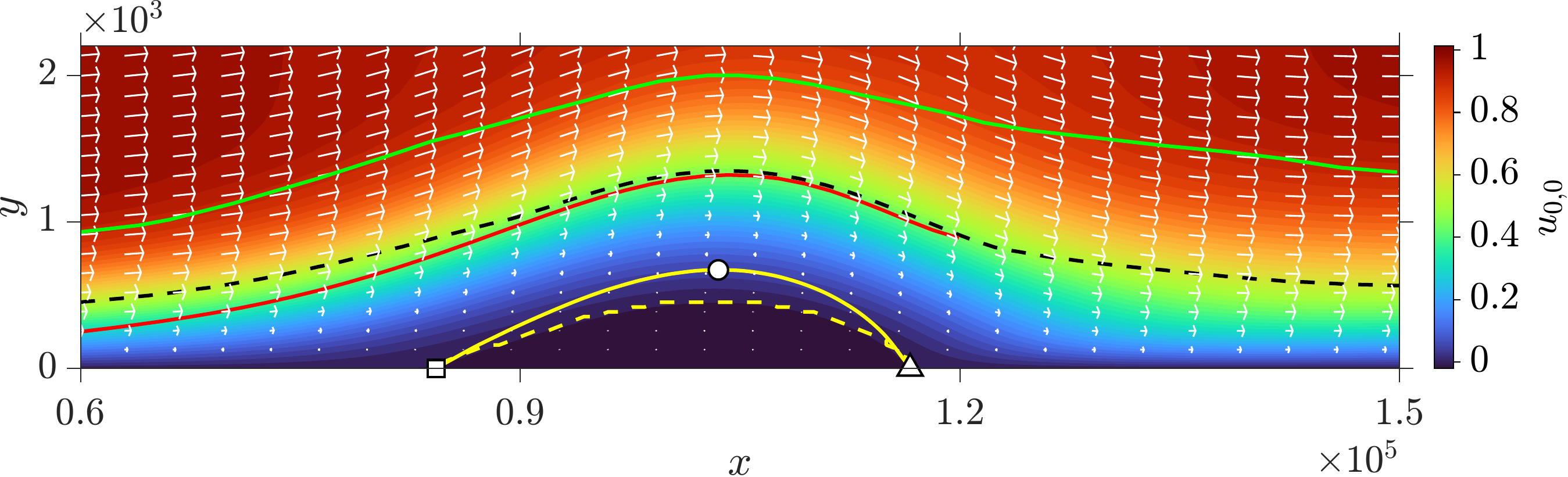}
    \caption{Close-up view of the LSB (case D). Contours of streamwise velocity and velocity vector field $(u_{0,0},v_{0,0})$ displayed. Green-solid: $0.99U_{\infty}$ boundary layer thickness. Black-dashed: displacement thickness. Red-solid: inflection line on the shear layer. Yellow-solid: LSB dividing streamline. Yellow-dashed: $u=0$ isoline. Separation $x_S$, maximum bubble height $h_{b}$ and reattachment $x_R$ are indicated with square, circle and triangle markers.}
    \label{fig:bubble_geometry}
\end{figure}

\begin{figure}
    \centering
    \includegraphics[width=0.85\textwidth]{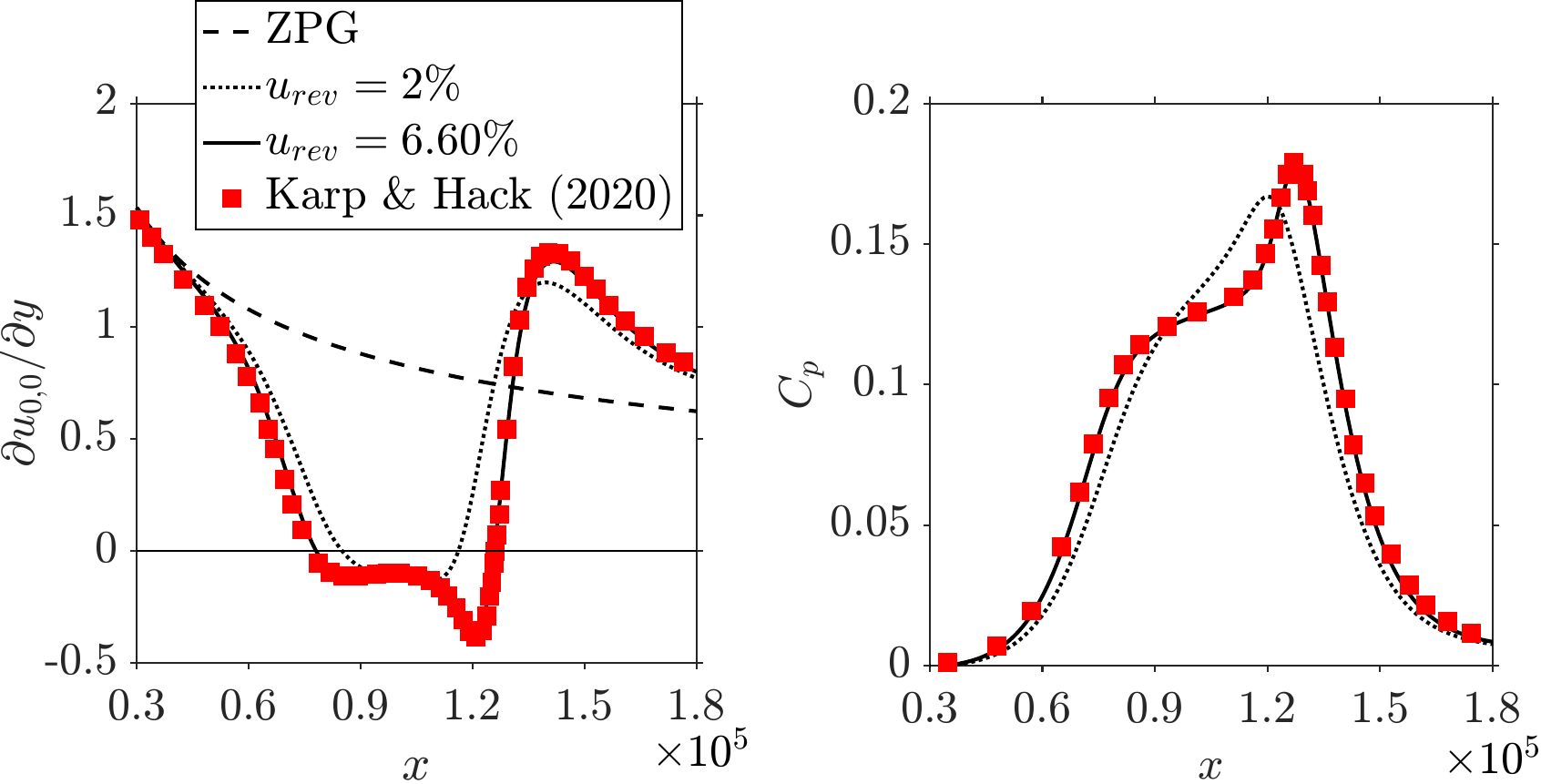}
    \caption{Validation of the LSB numerical set-up. (Left) Streamwise velocity gradient at $y=0$ scaled by $\mathrm{Re}=800$ for direct comparison with \citet{Karp2020suppression}. (Right) Pressure coefficient at $y=0$.}
    \label{fig:validation}
\end{figure}

A detailed view of our model LSB (case D in figure \ref{fig:base-flow}) is displayed in figure \ref{fig:bubble_geometry}. From this figure we can extract the key topological features that allow comparison with other LSBs available in the literature. The dividing streamline from which we determine the size of the bubble is calculated using the definition of zero mass-flux through said streamline, $\int_{0}^{y_d} u_{0,0} \: \mathrm{d}y = 0$ \citep{fitzgerald_mueller1990}. The height of the bubble is $h_{b} = 672$, which is approximately $1.34 \delta_{\mathrm{B}_{S}}$, $\delta_{\mathrm{B}_{S}}$ being the Blasius boundary layer thickness at separation $x=x_S$. We mark this point with the circle marker in figure \ref{fig:bubble_geometry}. The location of the maximum bubble height is $x_{h_{b}} = 1.04\times10^5$. Another meaningful parameter is the bubble length, which is commonly expressed as $\mathrm{Re}_{L_{b}} = U_{\infty}L_{b}/\nu = 0.31\times10^5$, where $L_{b} = x_R - x_S$ is the distance between the reattachment and separation points (marked with triangle and square markers in figure \ref{fig:bubble_geometry}, respectively). These locations are, by definition, where the streamwise velocity gradient computed at the wall (left panel in figure \ref{fig:validation}) vanishes. From the streamwise velocity gradient plot we notice the LSB base-flow departs from the ZPG boundary layer early on due to the influence of the APG. Well before separation ($0.3\times10^5<x<0.8\times10^5$) the flow loses momentum near the wall (resulting in lower shear) and, eventually, the boundary layer becomes inflectional. The inflection line, shown in figure \ref{fig:bubble_geometry}, is the locus of the inflection points emerging in the boundary layer velocity profile at each streamwise location. At separation, $x_S = 0.85\times 10^5$, the flow splits into two parts: the recirculatory flow contained within the LSB dividing streamline and the separated shear layer \citep{gaster1963,horton1968}. Beyond the separation point the wall velocity gradient is negative due to the presence of reversed flow. In the range $x_S<x<1.1\times10^5$ a pressure plateau appears in the $C_p$ curve plotted in the right panel of figure \ref{fig:validation}. This is associated with the dead-air region located in the fore portion of the separation bubble. Shortly after, there is a pressure rise corresponding to the intensification of reversed flow, which forms a vortex core towards the aft portion of the bubble. The transition from the pressure plateau to the pressure rise is less pronounced in smaller bubbles, as observed from comparison of the two LSB cases in figure \ref{fig:validation}. To force flow reattachment at $x_R = 1.16\times 10^5$, wall-normal velocity blowing is applied at the top far field to generate a favourable pressure gradient (FPG); refer to the boundary condition in figure \ref{fig:domain}. At the onset of reattachment, the velocity gradient (hence the skin friction coefficient) increases sharply and overshoots the ZPG solution. Because reattachment is driven by the externally imposed pressure gradient and not by any transition mechanisms in a nominally laminar flow, the reattached boundary layer recovers the ZPG Blasius solution a long distance downstream of the reattachment point, therefore indicating the flow remains laminar throughout.

Good agreement with the DNS data of \citet{Karp2020suppression} is observed in figure \ref{fig:validation} for the $u_{rev}=6.60\%$ LSB. Notably, the location of the bubble is sub-critical to local modal instability of planar T-S waves since $\mathrm{Re}_{x_{S}} < \mathrm{Re}_{x,crit}$, where $\mathrm{Re}_{x,crit} = 1.22\times10^5$ is the Reynolds number at which T-S waves undergo amplification due to the boundary layer receptivity process \citep{schmid_henningson2001stability}. Although the placement of the bubble on the plate may have an impact on boundary layer transition, this aspect is not investigated.

\begin{table}
  \begin{center}
    \def~{\hphantom{0}}
    \begin{tabular}{cccccc}
      $\mathrm{Re}_{x_{S}}$ & $\mathrm{Re}_{\delta^{*}_{S}}$ & $\mathrm{Re}_{\theta_{S}}$ & $\mathrm{Re}_{x_{R}}$ & $\mathrm{Re}_{\delta^{*}_{R}}$ & $\mathrm{Re}_{\theta_{R}}$\\ \vspace{10pt}
      $0.85\times10^5$ & 917.38 & 231.76 & $1.16\times10^5$ & 1,079.5 & 261.79\\ 
      $\mathrm{Re}_{L_{b}}$ & $x_{h_{b}}$ & $h_{b}$ & $L_{b}/h_{b}$ & $P=(\theta_{S}^{2}/\nu) (\Delta U/\Delta X)$ &\\
      $0.31\times10^5$ & $1.04\times10^5$ & 672 & 45.42 & -0.13 &\\
    \end{tabular}
    \caption{Characteristics of the laminar separation bubble of figure \ref{fig:bubble_geometry}. The Reynolds numbers are computed by multiplying $U_{\infty}/\nu$ by the relevant length scale.}
    \label{tab:bubble}
  \end{center}
\end{table}

In table \ref{tab:bubble} we summarise all the measured geometric properties of the LSB. Comparison with existing literature on the classification of LSBs \citep{marxen_henningson_2011,mohamed_aniffa_caesar_dabaria_mandal_2023} helps us conclude that our baseline bubble is ``short'', meaning the effects of separation on the surrounding pressure field are local, as opposed to ``long'' bubbles. Application of the two-parameter bursting criterion by \citet{Gaster1969}, for which we compute the pressure gradient parameter $P=(\theta_{S}^{2}/\nu) (\Delta U/\Delta X) = -0.13$ ($\Delta U$ is the change of streamwise velocity at the edge of the boundary layer over the separation distance $\Delta X$) and the Reynolds number based on the momentum thickness at separation $\mathrm{Re}_{\theta_{S}} = 231.76$, also confirms that our baseline bubble is indeed short since $P$ and $\mathrm{Re}_{\theta_{S}}$ fall below the bursting line (see figure 23 in \citet{Gaster1969} and figure 8 in \citet{Russell1979}).

\begin{figure}[h!]
    \centering
    \includegraphics[width=\linewidth]{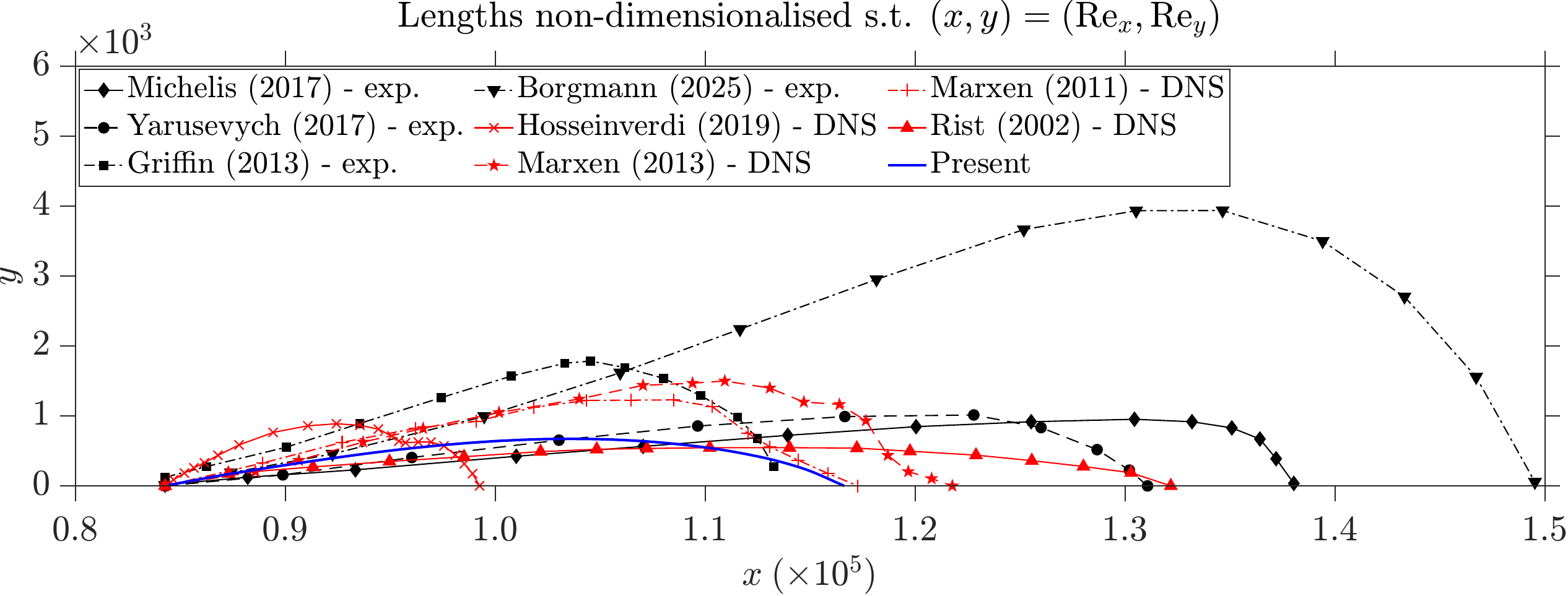}
    \caption{Comparison of our model LSB ($u_{rev}=2\%$) with experimental and numerical LSBs from the literature. For the experimental studies the unforced LSB test case was considered. All lengths extracted from the original source are non-dimensionalised to match our non-dimensionalisation and the location of the separation point is enforced to be $x_S$ to allow direct comparison.}
    \label{fig:bubble comparison}
\end{figure}

An additional comparison targeting the characteristic $\delta_{x_{S}}/L_{b}$ and $\delta_{x_{S}}/h_{b}$ ratios highlights the commonalities of our numerical LSB with experimental LSBs. In \citet{Michelis2017impulsiveforcing}, $\delta_{x_{S}}/L_{b}=0.056$ and $\delta_{x_{S}}/h_{b}=3.174$, in \citet{kumar2023}, these ratios amount to 0.033 and 1.2, \citet{diwan_ramesh_2009} report 0.05 and 1.2, while in the extensive experimental database of \citet{dellacasagrande_lengani_simoni_ubaldi2024} they range from 0.041 to 0.084 and from 0.6 to 3.24, respectively. From our LSB we obtain $\delta_{x_{S}}/L_{b}=0.049$ and $\delta_{x_{S}}/h_{b}=2.317$, which fall into the typical range found in the literature. Nevertheless, obtaining similar bubble shapes across experimental and high-fidelity numerical investigations is very challenging, as observed in figure \ref{fig:bubble comparison}, since LSBs are inherently unstable at transitional $\mathrm{Re}$ numbers $10^3-10^5$ and therefore sensitive to the flow set-up. This means any differences in $\mathrm{Re}$ number, location of the bubble, boundary conditions, disturbance environment, control/no-control, etc., will likely yield dissimilar bubble topologies. Although the comparison attempted in figure \ref{fig:bubble comparison} by extraction of several dividing streamline profiles from the cited studies is by no means exhaustive and should not be considered a one-to-one evaluation, we notice a significant spread across both experimentally- and numerically-derived LSBs.

Overall, these comparisons with literature suggest our LSB can be used as a representative model for the investigation of instabilities and transition. In fact, most of the aforementioned studies report the K-H instability mode as the most unstable primary mechanism developing in the separated shear layer, although the associated growth rates are clearly sensitive to flow-specific conditions.
\section{Global Stability}\label{app:global stability}
Linear global stability analysis is performed on the LSB base-flow for the identification of primary global instability. As figure \ref{fig:base-flow} shows, the LSB can be either convectively unstable, meaning that the generation of turbulence requires continuous external excitation, or globally unstable if a self-excited perturbation mode naturally manifests without need of external forcing. The threshold between these two regimes can be determined by solving a generalised eigenvalue problem to find the unstable eigenmode.

A range of spanwise wavenumbers is investigated to account for three-dimensional eigenmodes. In order to track the trajectory of any unstable mode as the spanwise wavenumber is varied, the shift parameter, initialised at the origin of the complex plane for the first calculation, is updated to the location of the least stable eigenvalue found in the eigenspectrum. In our search for global instability we also vary the peak reversed flow parameter $u_{rev}$ which ultimately corresponds to separation bubbles of varying size. 

\begin{figure}
    \centering
    \includegraphics[width=0.9\textwidth]{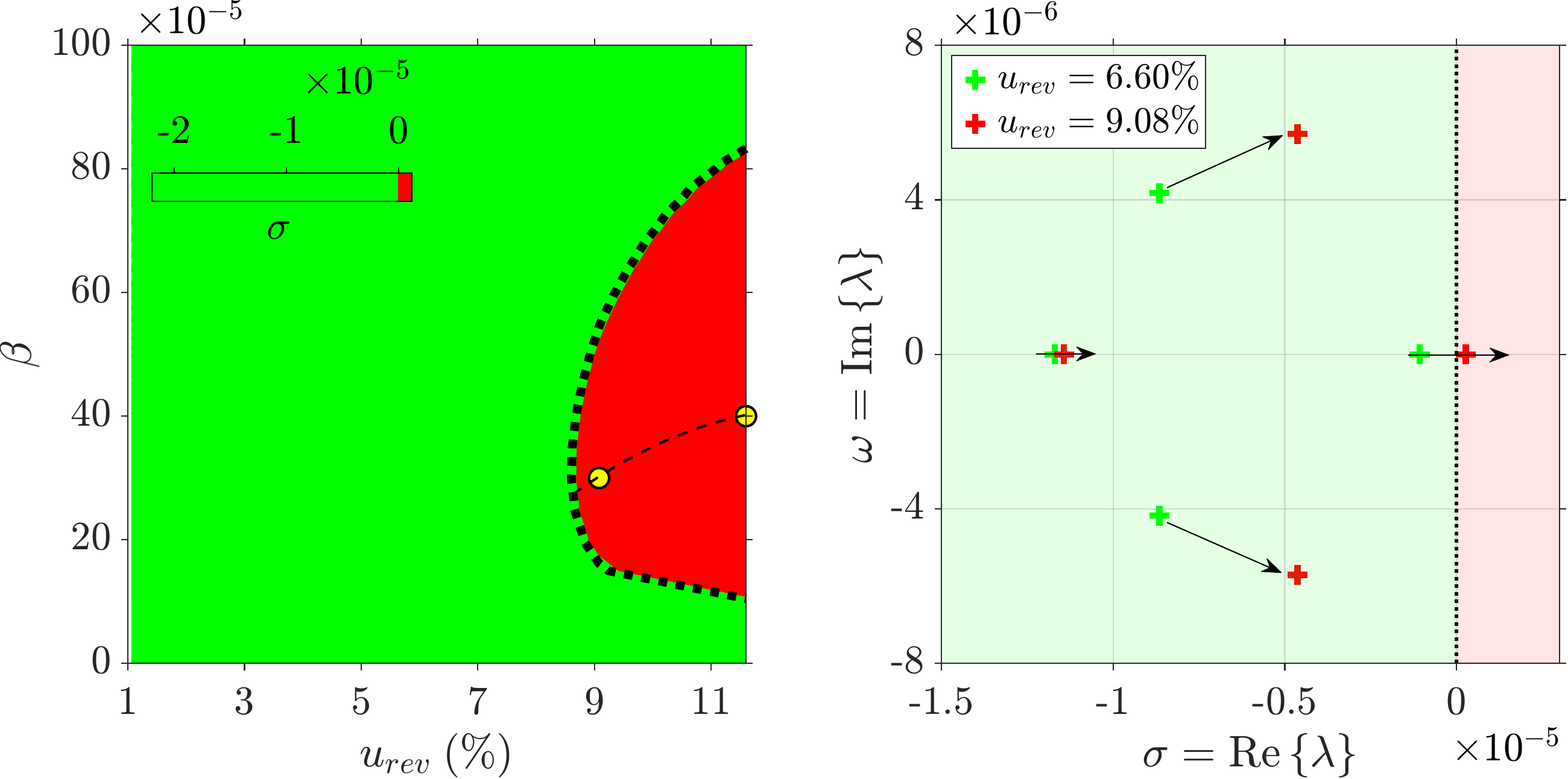}
    \caption{(Left) Contours of temporal amplification ($\sigma=\mathrm{Re} \left \{ \lambda \right \}$) in the parametric space $(u_{rev},\beta)$. Marginal stability axis: black-dotted line. Spanwise wavenumber at which maximal amplification occurs, $\beta_{max}$: yellow circle. Trajectory of $\beta_{max}$: black-dashed line. (Right) Eigenvalue spectrum for $\beta=36\times10^{-5}$ in the neighbourhood of the marginal stability axis for a stable and an unstable LSB.}
    \label{fig:eig spectrum}
\end{figure}

Figure \ref{fig:eig spectrum} (left) shows the temporal amplification of the least stable eigenvalue computed for several base-flows and spanwise wavenumbers. The LSB base-flow is globally unstable to a three-dimensional mode ($\beta\neq0$) and the instability onset is around $u_{rev}\approx9\%$. Conversely, the base-flow is always stable to two-dimensional disturbances. We also notice $\beta_{max}$ is lightly affected by the the bubble size, in agreement with \citet{Rodriguez_gennaro_juniper_2013,Rodriguez2021self-excited}. Even though an exact comparison cannot be made due to the different base-flow topology, \citet{Rodriguez2021self-excited} found $\beta_{max} = 36 \times 10^{-5}$, which is in the range observed in our results. For turbulent shear layers, the optimal spanwise wavenumber decreases, as reported in \citet{Cura_Hanifi_Cavalieri_Weiss_2024}, where $\beta$ is approximately 1/3 to 2/3 of our values. 

In figure \ref{fig:eig spectrum} (right) we plot the 4 least stable eigenvalues of two LSB base-flows. As the recirculation in the bubble becomes sufficiently strong, the least stable eigenvalue crosses the marginal stability axis and the system experiences global resonance at time rate $\sigma>0$. The unstable mode is clearly non-oscillatory since $\omega=0$. 

\begin{figure}
    \centering
    \includegraphics[width=\textwidth]{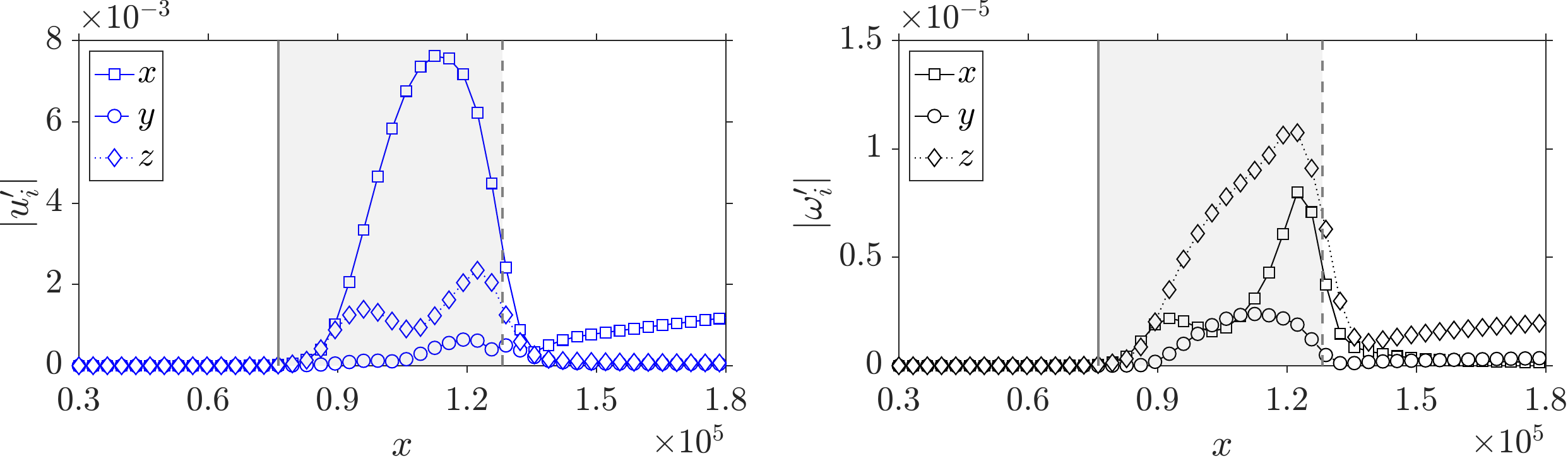}
    \caption{Component-wise amplitudes of (left) velocity and (right) vorticity disturbances of the global eigenmode computed for the unstable LSB with peak reversed flow $u_{rev}=9.08\%$ at $\beta=36\times10^{-5}$.}
    \label{fig:eigenmode amplitudes}
\end{figure}

\begin{figure}
    \centering
    \includegraphics[width=0.9\textwidth]{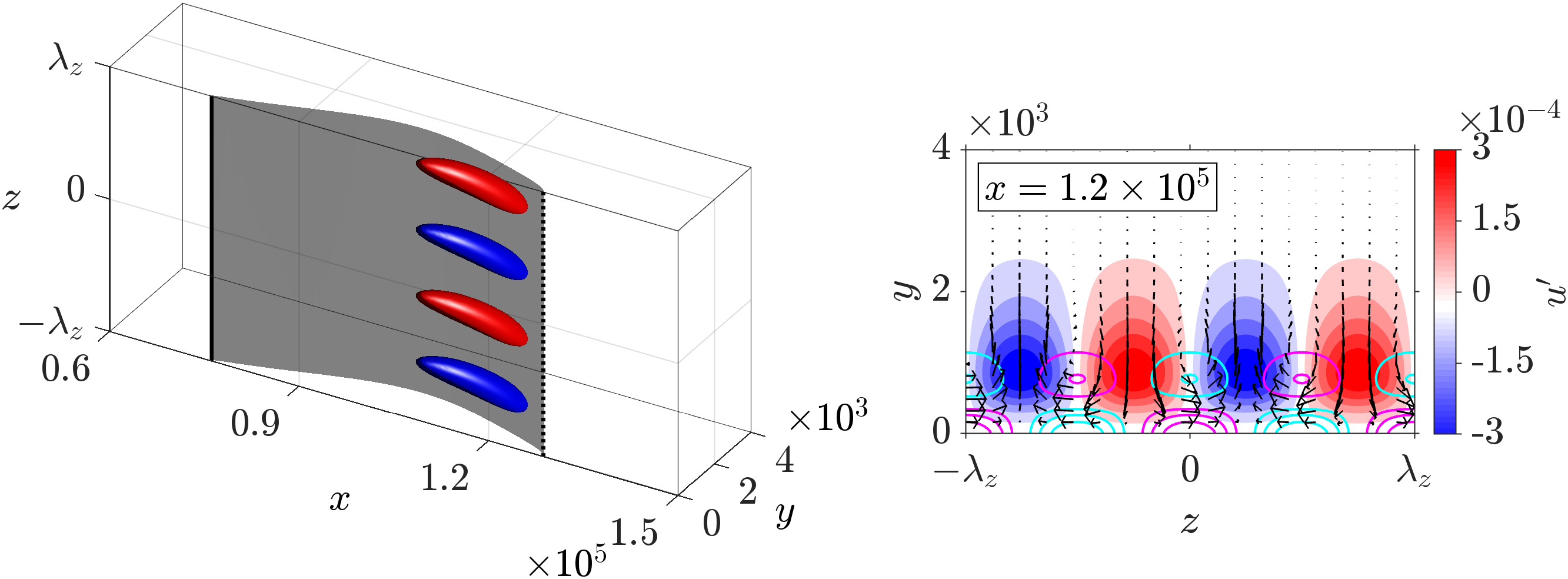}
    \caption{3D structure of the global eigenmode of figure \ref{fig:eigenmode amplitudes}. (Left) Isosurfaces of (red) positive and (blue) negative $u’$ superimposed on the $u=0$ isosurface of the LSB. (Right) $y$-$z$ planar view at $x=1.2\times10^5$ showing contours of $u’$, vectors from the ($v’,w'$) velocity disturbance field and contours of $\omega_{x}'$ (magenta: into the page, cyan: out of the page).}
    \label{fig:eigenmode shape}
\end{figure}

The computed perturbation displays a strong streamwise velocity disturbance component in the aft part of the separation zone (figure \ref{fig:eigenmode amplitudes}), which we visualise with $u'$ isosurfaces in the left panel of figure \ref{fig:eigenmode shape}. These structures sit just above the separation bubble. The amplitudes also show two distinct peaks in $w'$ and $\omega_{x}'$, one in the fore and one in the aft part of separation. The velocity disturbance field in the $y$-$z$ plane (shown in the right panel of figure \ref{fig:eigenmode shape}), illustrates the connection between streamwise vorticity and the alternation of positive/negative $u'$ disturbances associated with downwash/upwash. Moreover, this dynamics leads to transverse velocity disturbance $w'$, as indicated by the two peaks in the amplitudes of $w'$ and $\omega_{x}'$ in figure \ref{fig:eigenmode amplitudes}, also noticed by \citet{Gallaire2007three-dimensional} in separation bubbles behind a bump. These characteristics suggest the presence of a global, self-excited, centrifugal instability originally found by \citet{theofilis2000} for incompressible separation bubbles with $u_{rev}$ in the range 7-8\%. More rigorous analysis such as the Rayleigh discriminant criterion or the geometric optics (WKB) method along streamlines would further support our conclusion but we refer the reader to \citet{sipp_lauga_jacquin1999,sipp_jacquin2000,Gallaire2007three-dimensional} for this. Lastly, it is important to note that while this instability is of centrifugal nature, it should not be confused with the Görtler instability which is convective and therefore requires external forcing to be excited \citep{Gortler1941,Saric1994}.
\section{Effect of adverse pressure gradient on Tollmien-Schlichting waves}\label{app:effect of adverse pressure gradient on Tollmien-Schlichting waves}
\begin{figure}
    \centering
    \includegraphics[width=\linewidth]{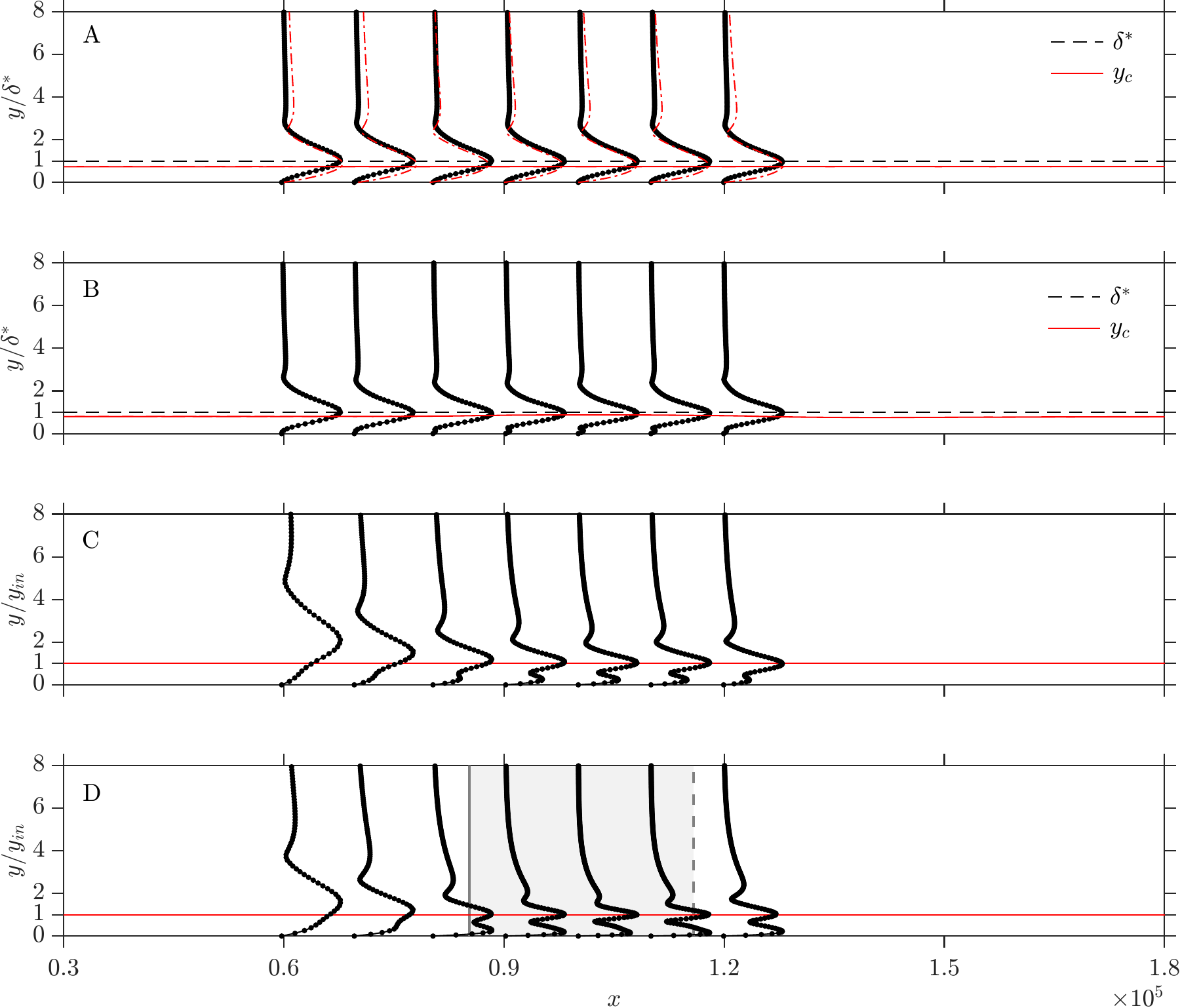}
    \caption{Shape functions (black circle markers on solid lines) of $\left | u'\right | = \sqrt{\textrm{Re}\left\{ \hat{u}\right\}^2+\textrm{Im}\left\{ \hat{u}\right\}^2}$ extracted at seven $x$-stations from the most unstable 3D waves calculated by linear resolvent. The profiles are normalised such that their amplitude is in the range $[0,1]$ and the $y$-coordinate is normalised for each $x$-station by either the displacement thickness (A and B) or the inflection line (C and D). The critical layer $y_c$ is superimposed in A and B. Red dashed-dotted lines in A refer to the 2D mode ($\beta=0$).}
    \label{fig:shape_functions}
\end{figure}

Similarly to figure \ref{fig:T-S and K-H linear resolvent}, we here look into the optimal spatial modes computed by linear RA on the four base-flows A, B, C and D (figure \ref{fig:base-flow}) representative of different APG conditions. We show the amplitude of the streamwise velocity disturbance from the unsteady modes summarised in table \ref{tab:linear resolvent} at several streamwise locations in figure \ref{fig:shape_functions}.

For a ZPG boundary layer (case A) we retrieve the typical dual-lobe profile in both 2D and 3D T-S waves, the latter showing a less pronounced upper lobe in proportion to the amplitude of the lower lobe \cite{Xu_Mughal_Gowree_Atkin_Sherwin_2017}. The main peak is located around the critical layer height where the phase speed of the wave matches the mean velocity \cite{schmid_henningson2001stability,Brandt_Sipp_Pralits_Marquet_2011}.

In mild APG conditions (case B) the profiles are broadly similar to the ZPG case, the main difference being a small local peak located deep inside the viscous sublayer (below the displacement thickness) which arises in the APG region and dampens in the FPG region.

In strongly inflectional boundary layers (cases C and D), the shape of incoming waves changes dramatically as they convect through the APG region. As thoroughly explained in \cite{diwan_ramesh_2009}, the progressive emergence of an inflectional velocity profile due to the APG creates an efficient site of disturbance production around the inflection point, well visible in the disturbance mode shape as a localised strong peak along the inflection line. This feature is ascribed to the inviscid K-H instability. Another lobe appears close to the wall originating from viscous shear \cite{diwan_ramesh_2009}.

Overall, our resolvent analysis correctly captures the complex scenario according to which viscous (T-S) waves in the upstream boundary layer get convected downstream and feed the inviscid inflectional (K-H) instability that governs the linear dynamics of the separated shear layer.
\section{Lift-up mechanism affected by the adverse pressure gradient}\label{app:lift-up mechanism affected by the adverse pressure gradient}
\begin{figure}
    \centering
    \includegraphics[width=\textwidth]{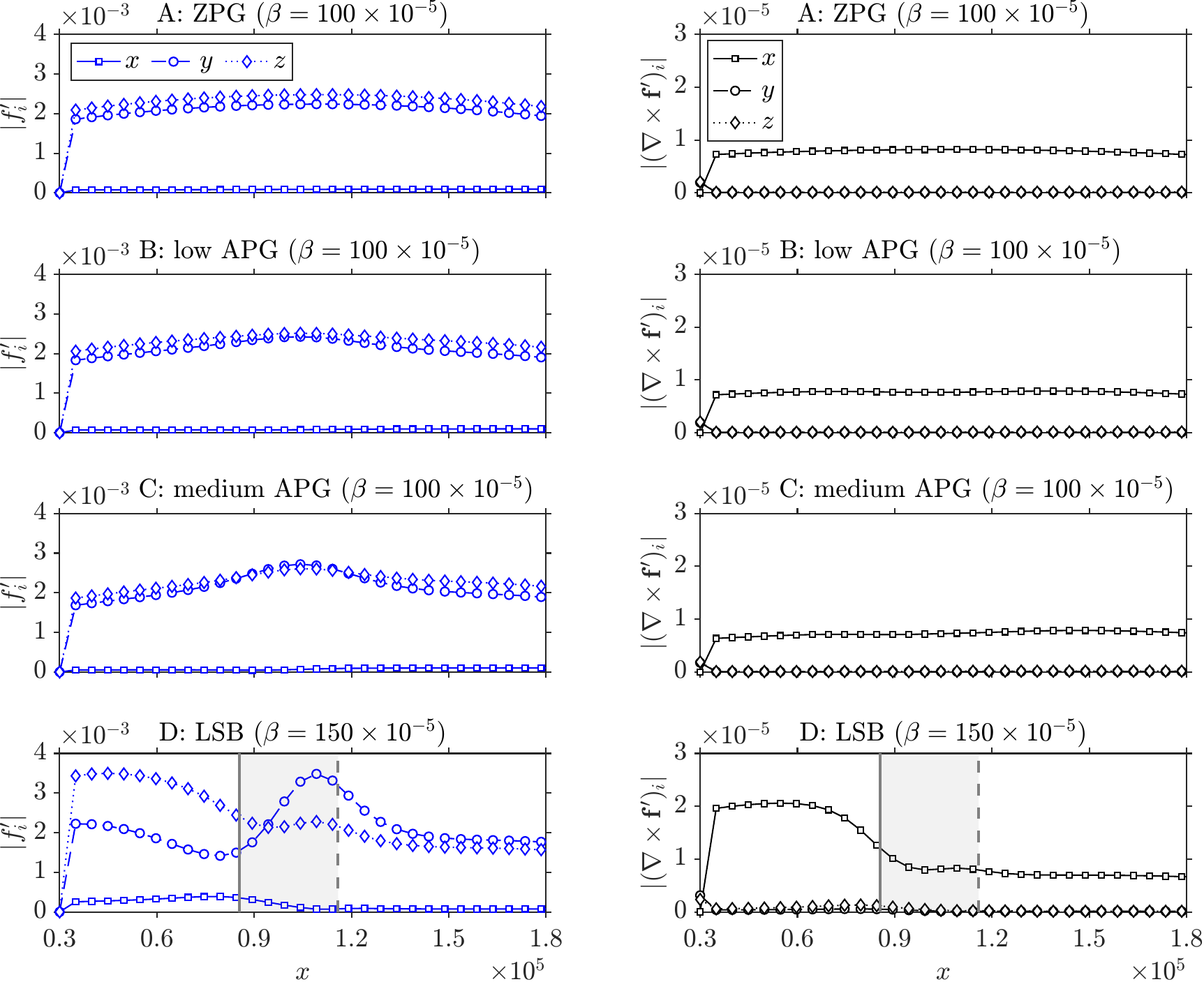}
    \caption{Lift-up mechanism in APG boundary layers. Amplitudes of optimal resolvent forcing (left) and curl of forcing (right) for four boundary layer flows: (first row) ZPG, (second row) low APG, (third row) medium APG and (fourth row) LSB. These modes are steady $(\omega=0)$.}
    \label{fig:liftup APG forcing}
\end{figure}

\begin{figure}
    \centering
    \includegraphics[width=\textwidth]{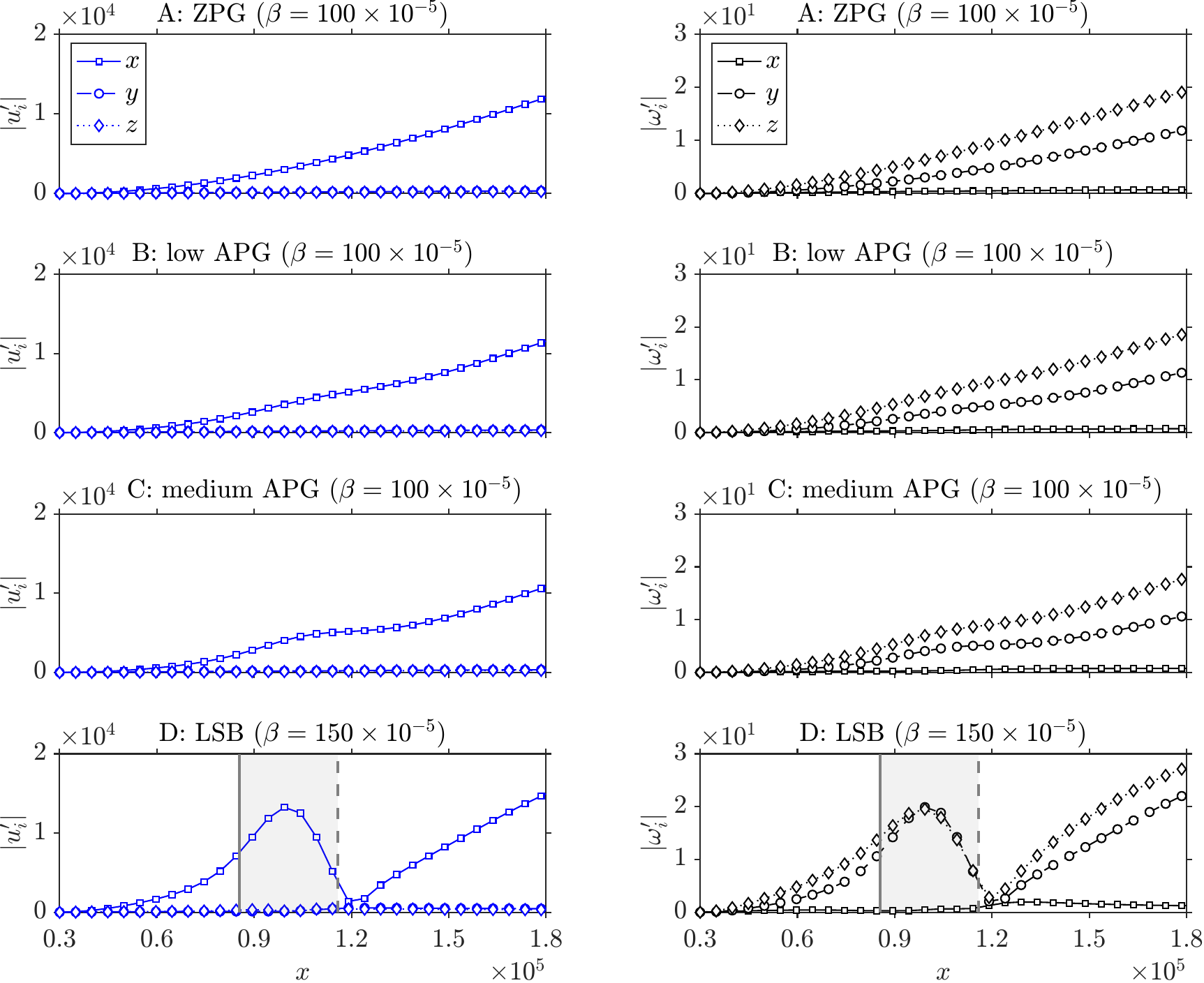}
    \caption{Lift-up mechanism in APG boundary layers. Amplitudes of optimal resolvent response velocities (left) and vorticities (right) for four boundary layer flows: (first row) ZPG, (second row) low APG, (third row) medium APG and (fourth row) LSB. These modes are steady $(\omega=0)$.}
    \label{fig:liftup APG response}
\end{figure}

The component-wise structures of the optimal forcing-response resolvent mode ascribed to lift-up in attached and separated boundary layers are visualised in figures \ref{fig:liftup APG forcing} and \ref{fig:liftup APG response}. The pressure gradient increases from top to bottom. Notably, streamwise vortical forcing excites streamwise velocity response for any pressure gradient considered, demonstrating that the lift-up mechanism retains the same physics for both attached and separated cases. The main difference imparted by the bubble is that the process occurs both upstream of separation and in the post-reattachment region, leading to $u'$ response structures both within the separated zone and in the redeveloping boundary layer -- for this reason referred to as ``double'' lift-up in contrast to classic lift-up in attached boundary layer flow \citep{andersson_brandt_bottaro_henningson_2001,schmid_henningson2001stability,jacobs_durbin_2001,Balamurugan_Mandal_2017,rigas2021HBM}.
\section{Rayleigh discriminant criterion for centrifugal instability}\label{app:rayleigh discriminant criterion for centrifugal instability}
Centrifugal instabilities may originate not only from modal resonance in globally unstable flows (\S\S\ref{subsec:global stability}, \S\ref{app:global stability}) but also from convective mechanisms (Görtler instability) in flows featuring either geometric curvature \citep{Gortler1941,Saric1994,li_malik_1995} or pressure gradient-induced curvature \citep{Hildebrand2018,shinde2019,shinde2020}.

Here we introduce a sufficient criterion for the onset of short-wave centrifugal instability. This method is effectively an extension of the classic Rayleigh discriminant criterion that also accounts for rotating frames of reference \citep{sipp_lauga_jacquin1999,sipp_jacquin2000}.

Defining the absolute shear and curvature vorticity as $\overline{\omega_{z}}+2\Omega$ and $V/\mathcal{R}+\Omega$, where $\overline{\omega_{z}}$ is the vorticity due to pure shear, $\Omega$ is the rotation of the frame of reference, $V=\sqrt{\overline{u}^2+\overline{v}^2+\overline{w}^2}$ is the velocity magnitude and $\mathcal{R}$ is the algebraic radius of curvature associated with the streamline $\Psi$,
\begin{equation}
    \mathcal{R} = \frac{V^3}{\nabla\Psi \cdot \left( \overline{\boldsymbol{u}}\cdot\nabla\overline{\boldsymbol{u}}\right)},
    \label{eq:radius of curvature}
\end{equation}
\noindent we consider the parameter,
\begin{equation}
    \delta\left(x_0,y_0\right) = 2\left(\overline{\omega_{z}}+2\Omega\right)\left(V/\mathcal{R}+\Omega\right),
    \label{eq:delta parameter}
\end{equation}
\noindent defined on the points belonging to the streamline $\Psi_0$. If at any point $\left(x_0,y_0\right)$ the parameter $\delta<0$, the streamline is locally unstable to a centrifugal instability. Clearly, for a fixed frame of reference like in our case, $\Omega=0$ and we recover the classic Rayleigh discriminant $\delta = 2\:\overline{\omega_{z}}\:V/\mathcal{R}$. This criterion essentially states that regions of the streamline where the signs of shear and curvature vorticity are opposite are locally unstable to centrifugal instability. If $\delta<0$ everywhere along the streamline the centrifugal instability is globally unstable. The sign of shear vorticity is determined by the sign of the velocity strain $\partial \overline{v}/\partial x - \partial \overline{u}/\partial y$, while the sign of curvature vorticity depends on the local curvature of the streamline: positive curvature is defined for counter-clockwise spin and negative curvature for clockwise spin.

\begin{figure}
    \centering
    \includegraphics[width=0.55\linewidth]{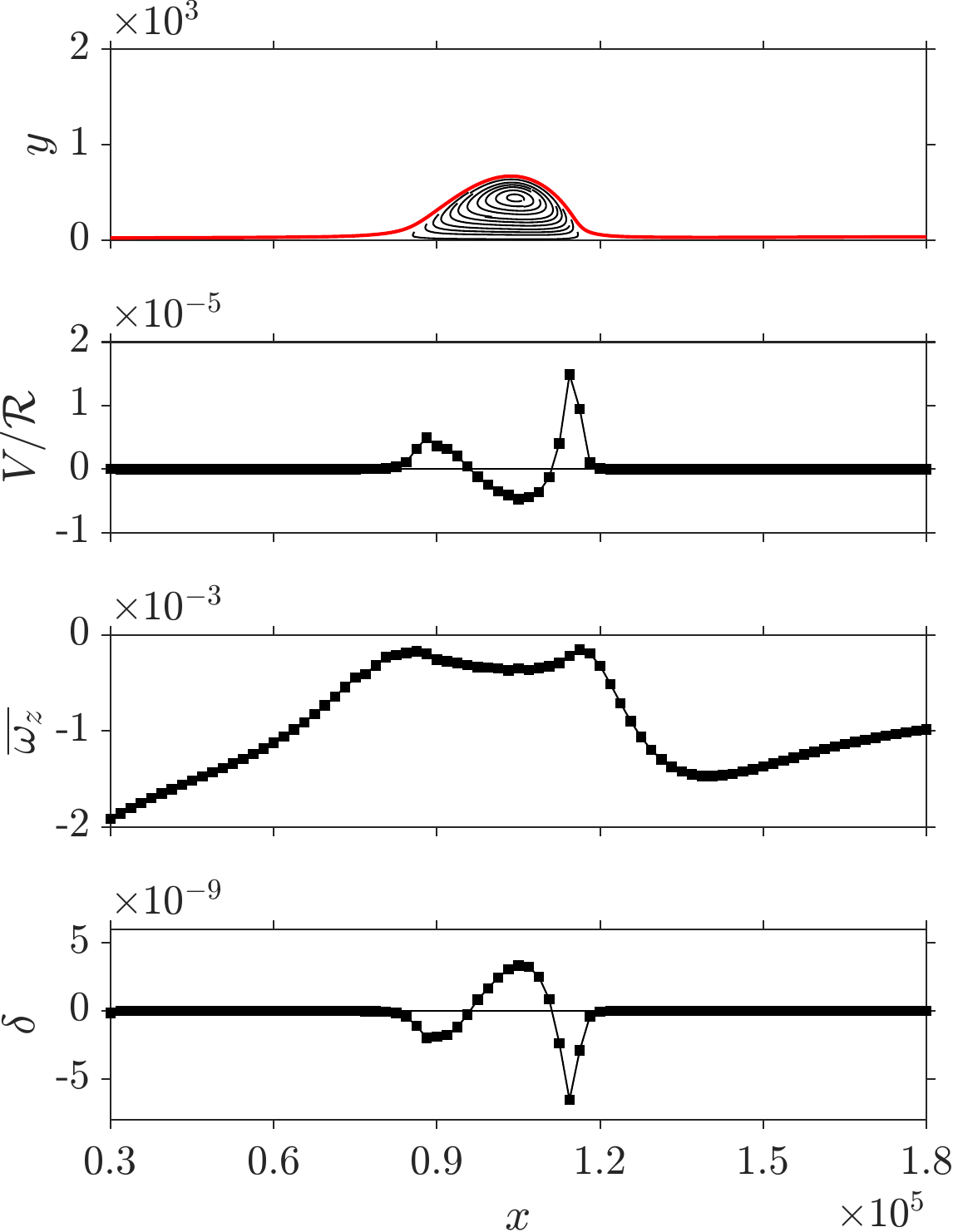}
    \caption{Rayleigh discriminant criterion applied to a streamline of the mean-flow calculated at $A=1\times10^{-7}$ by the HBNS$_{2,1}$ system. First row: streamlines in black and selected streamline for the criterion in red. Second row: curvature vorticity. Third row: shear vorticity. Fourth row: $\delta$ parameter.}
    \label{fig:criterion}
\end{figure}

In figure \ref{fig:criterion} we show the analysis for an open streamline adjacent to the separation bubble, whose location is essentially representative of fluid particles within the shear layer. The streamline topology is obtained from the time- and spanwise-averaged flow solved by the HBNS$_{2,1}$ system at amplitude $A=1\times10^{-7}$. Curvature vorticity (second row) is zero away from the separation bubble, where there is no streamline curvature, and changes sign three times over the separation zone. At the separation and reattachment points the streamline is bent upward (positive curvature) and is otherwise bent downward (negative curvature) around the maximum bubble height. The zeros of $V/\mathcal{R}$ within the separation length are, by definition, the inflection points of the streamline. The shear vorticity (third row) is negative everywhere, which is typical of separated shear layer velocity profiles. This is because the average rotation of the shear layer is clockwise since there is slow moving flow near the wall underneath faster moving flow in the shear layer. As a result, the Rayleigh discriminant $\delta$ (fourth row) is weakly negative around the separation point and strongly negative at reattachment. It is positive around the bubble's apex. These results indicate that the present bubble can support locally unstable centrifugal disturbances at separation and reattachment and also agree with the initial statement that the flow is not globally unstable to a centrifugal mechanism because $\delta$ is not negative everywhere. This was also verified for all either open and closed (within the separation bubble) streamlines, therefore ruling out the existence of a global centrifugal instability for this globally stable LSB.
\section{Forcing produced by the non-linear interaction of a pair of K-H waves}\label{app:forcing produced by the non-linear interaction of a pair of K-H waves}
In the following, we write the explicit terms of the operator $\mathsfbi{N}_{1}^{1} \left ( \hat{\boldsymbol{w}}_{1,-1}, \hat{\boldsymbol{w}}_{1,1} \right )$ which represents the forcing of the harmonic $\hat{\boldsymbol{w}}_{2,0}$ produced by the non-linear interaction of two K-H waves $\hat{\boldsymbol{w}}_{1,-1}$ and $\hat{\boldsymbol{w}}_{1,1}$.

Taking advantage of the spanwise ($z$) symmetry of the solution to the HBNS system, we know that
\begin{equation}
    \hat{\boldsymbol{w}}_{1,-1} = \hat{\boldsymbol{w}}_{-1,1}^{*},
    \label{eq:spanwise symmetry implication}
\end{equation}
\noindent where $\hat{\boldsymbol{w}}_{-1,1}$ is the $z$-symmetric harmonic of $\hat{\boldsymbol{w}}_{1,1}$, implying that $[\hat{u},\hat{v},\hat{w},\hat{p}]_{-1,1}=[\hat{u},\hat{v},-\hat{w},\hat{p}]_{1,1}$. In other words, we can obtain $\hat{\boldsymbol{w}}_{1,-1}$ directly from $\hat{\boldsymbol{w}}_{1,1}$ by applying the symmetry and the complex conjugate conditions.

The operator can therefore be re-written as
\begin{equation}
    \mathsfbi{N}_{1}^{1} \left ( \hat{\boldsymbol{w}}_{-1,1}^{*}, \hat{\boldsymbol{w}}_{1,1} \right ) = \begin{pmatrix} 
    \hat{\boldsymbol{u}}_{-1,1}^{*} \cdot \nabla \hat{\boldsymbol{u}}_{1,1} + \hat{\boldsymbol{u}}_{1,1} \cdot \nabla \hat{\boldsymbol{u}}_{-1,1}^{*} \\ 0 
    \end{pmatrix},
    \label{eq:NL operator}
\end{equation}
\noindent where the two convective terms can be expanded as
\begin{subequations}
    \begin{equation}
        \hat{\boldsymbol{u}}_{-1,1}^{*} \cdot \nabla \hat{\boldsymbol{u}}_{1,1} = 
        \begin{pmatrix}
        \hat{u}_{-1,1}^{*} \frac{\partial \hat{u}_{1,1}}{\partial x} + \hat{v}_{-1,1}^{*} \frac{\partial \hat{u}_{1,1}}{\partial y} + \mathrm{i}\beta \hat{w}_{-1,1}^{*} \hat{u}_{1,1} \\
        \hat{u}_{-1,1}^{*} \frac{\partial \hat{v}_{1,1}}{\partial x} + \hat{v}_{-1,1}^{*} \frac{\partial \hat{v}_{1,1}}{\partial y} + \mathrm{i}\beta \hat{w}_{-1,1}^{*} \hat{v}_{1,1} \\
        \hat{u}_{-1,1}^{*} \frac{\partial \hat{w}_{1,1}}{\partial x} + \hat{v}_{-1,1}^{*} \frac{\partial \hat{w}_{1,1}}{\partial y} + \mathrm{i}\beta \hat{w}_{-1,1}^{*} \hat{w}_{1,1}
        \end{pmatrix},
    \end{equation}
    \begin{equation} 
        \hat{\boldsymbol{u}}_{1,1} \cdot \nabla \hat{\boldsymbol{u}}_{-1,1}^{*} = 
        \begin{pmatrix}
        \hat{u}_{1,1} \frac{\partial \hat{u}_{-1,1}^{*}}{\partial x} + \hat{v}_{1,1} \frac{\partial \hat{u}_{-1,1}^{*}}{\partial y} + \mathrm{i}\beta \hat{w}_{1,1} \hat{u}_{-1,1}^{*} \\
        \hat{u}_{1,1} \frac{\partial \hat{v}_{-1,1}^{*}}{\partial x} + \hat{v}_{1,1} \frac{\partial \hat{v}_{-1,1}^{*}}{\partial y} + \mathrm{i}\beta \hat{w}_{1,1} \hat{v}_{-1,1}^{*} \\
        \hat{u}_{1,1} \frac{\partial \hat{w}_{-1,1}^{*}}{\partial x} + \hat{v}_{1,1} \frac{\partial \hat{w}_{-1,1}^{*}}{\partial y} + \mathrm{i}\beta \hat{w}_{1,1} \hat{w}_{-1,1}^{*}
        \end{pmatrix},
    \end{equation}
    \label{eq:convective terms}
\end{subequations}
\noindent by virtue of the gradient operator $\nabla_{\mathrm{i}m\beta} \equiv \left [ \partial/\partial x, \partial/\partial y, \mathrm{i} m \beta \right ]^{\top}$ valid for spanwise-periodic solutions.

Since
\begin{equation}
    \hat{\boldsymbol{u}}_{1,1} = 
    \begin{pmatrix}
    \hat{u}_{1,1} \\
    \hat{v}_{1,1} \\
    \hat{w}_{1,1}
    \end{pmatrix}
    = 
    \begin{pmatrix}
    \hat{u}_{\mathrm{Re}} + \mathrm{i} \hat{u}_{\mathrm{Im}} \\
    \hat{v}_{\mathrm{Re}} + \mathrm{i} \hat{v}_{\mathrm{Im}} \\
    \hat{w}_{\mathrm{Re}} + \mathrm{i} \hat{w}_{\mathrm{Im}}
    \end{pmatrix}, \hspace{12pt} \hat{\boldsymbol{u}}_{-1,1}^{*} = 
    \begin{pmatrix}
    \hat{u}_{-1,1}^{*} \\
    \hat{v}_{-1,1}^{*} \\
    \hat{w}_{-1,1}^{*}
    \end{pmatrix}
    =
    \begin{pmatrix}
    \hat{u}_{\mathrm{Re}} - \mathrm{i} \hat{u}_{\mathrm{Im}} \\
    \hat{v}_{\mathrm{Re}} - \mathrm{i} \hat{v}_{\mathrm{Im}} \\
    -\hat{w}_{\mathrm{Re}} + \mathrm{i} \hat{w}_{\mathrm{Im}}
    \end{pmatrix},
    \label{eq:direct and c.c. of z-symmetric mode}
\end{equation}
\noindent then
\begin{subequations}
    \begin{equation}
        \begin{split}
        \hat{\boldsymbol{u}}_{-1,1}^{*} \cdot \nabla \hat{\boldsymbol{u}}_{1,1} = 
        \begin{pmatrix}
        \hat{u}_{\mathrm{Re}} \frac{\partial \hat{u}_{\mathrm{Re}}}{\partial x} + \hat{v}_{\mathrm{Re}} \frac{\partial \hat{u}_{\mathrm{Re}}}{\partial y} + \hat{u}_{\mathrm{Im}} \frac{\partial \hat{u}_{\mathrm{Im}}}{\partial x} + \hat{v}_{\mathrm{Im}} \frac{\partial \hat{u}_{\mathrm{Im}}}{\partial y} + \beta (\hat{w}_{\mathrm{Re}} \hat{u}_{\mathrm{Im}} - \hat{w}_{\mathrm{Im}} \hat{u}_{\mathrm{Re}}) \\
        \hat{u}_{\mathrm{Re}} \frac{\partial \hat{v}_{\mathrm{Re}}}{\partial x} + \hat{v}_{\mathrm{Re}} \frac{\partial \hat{v}_{\mathrm{Re}}}{\partial y} + \hat{u}_{\mathrm{Im}} \frac{\partial \hat{v}_{\mathrm{Im}}}{\partial x} + \hat{v}_{\mathrm{Im}} \frac{\partial \hat{v}_{\mathrm{Im}}}{\partial y} + \beta (\hat{w}_{\mathrm{Re}} \hat{v}_{\mathrm{Im}} - \hat{w}_{\mathrm{Im}} \hat{v}_{\mathrm{Re}}) \\
        \hat{u}_{\mathrm{Re}} \frac{\partial \hat{w}_{\mathrm{Re}}}{\partial x} + \hat{v}_{\mathrm{Re}} \frac{\partial \hat{w}_{\mathrm{Re}}}{\partial y} + \hat{u}_{\mathrm{Im}} \frac{\partial \hat{w}_{\mathrm{Im}}}{\partial x} + \hat{v}_{\mathrm{Im}} \frac{\partial \hat{w}_{\mathrm{Im}}}{\partial y} + \underbrace{\beta (\hat{w}_{\mathrm{Re}} \hat{w}_{\mathrm{Im}} - \hat{w}_{\mathrm{Im}} \hat{w}_{\mathrm{Re}})}_{=0}
        \end{pmatrix} \\ 
        + 
        \begin{pmatrix}
        \hat{u}_{\mathrm{Re}} \frac{\partial \hat{u}_{\mathrm{Im}}}{\partial x} + \hat{v}_{\mathrm{Re}} \frac{\partial \hat{u}_{\mathrm{Im}}}{\partial y} - \hat{u}_{\mathrm{Im}} \frac{\partial \hat{u}_{\mathrm{Re}}}{\partial x} - \hat{v}_{\mathrm{Im}} \frac{\partial \hat{u}_{\mathrm{Re}}}{\partial y} + \beta (-\hat{w}_{\mathrm{Re}} \hat{u}_{\mathrm{Re}} - \hat{w}_{\mathrm{Im}} \hat{u}_{\mathrm{Im}}) \\
        \hat{u}_{\mathrm{Re}} \frac{\partial \hat{v}_{\mathrm{Im}}}{\partial x} + \hat{v}_{\mathrm{Re}} \frac{\partial \hat{v}_{\mathrm{Im}}}{\partial y} - \hat{u}_{\mathrm{Im}} \frac{\partial \hat{v}_{\mathrm{Re}}}{\partial x} - \hat{v}_{\mathrm{Im}} \frac{\partial \hat{v}_{\mathrm{Re}}}{\partial y} + \beta (-\hat{w}_{\mathrm{Re}} \hat{v}_{\mathrm{Re}} - \hat{w}_{\mathrm{Im}} \hat{v}_{\mathrm{Im}}) \\
        \hat{u}_{\mathrm{Re}} \frac{\partial \hat{w}_{\mathrm{Im}}}{\partial x} + \hat{v}_{\mathrm{Re}} \frac{\partial \hat{w}_{\mathrm{Im}}}{\partial y} - \hat{u}_{\mathrm{Im}} \frac{\partial \hat{w}_{\mathrm{Re}}}{\partial x} - \hat{v}_{\mathrm{Im}} \frac{\partial \hat{w}_{\mathrm{Re}}}{\partial y} + \beta (-\hat{w}_{\mathrm{Re}}^{2} - \hat{w}_{\mathrm{Im}}^{2})
        \end{pmatrix}\mathrm{i},
        \end{split}
    \end{equation}
    \begin{equation} 
        \begin{split}
        \hat{\boldsymbol{u}}_{1,1} \cdot \nabla \hat{\boldsymbol{u}}_{-1,1}^{*} = 
        \begin{pmatrix}
        \hat{u}_{\mathrm{Re}} \frac{\partial \hat{u}_{\mathrm{Re}}}{\partial x} + \hat{v}_{\mathrm{Re}} \frac{\partial \hat{u}_{\mathrm{Re}}}{\partial y} + \hat{u}_{\mathrm{Im}} \frac{\partial \hat{u}_{\mathrm{Im}}}{\partial x} + \hat{v}_{\mathrm{Im}} \frac{\partial \hat{u}_{\mathrm{Im}}}{\partial y} + \beta (\hat{w}_{\mathrm{Re}} \hat{u}_{\mathrm{Im}} - \hat{w}_{\mathrm{Im}} \hat{u}_{\mathrm{Re}}) \\
        \hat{u}_{\mathrm{Re}} \frac{\partial \hat{v}_{\mathrm{Re}}}{\partial x} + \hat{v}_{\mathrm{Re}} \frac{\partial \hat{v}_{\mathrm{Re}}}{\partial y} + \hat{u}_{\mathrm{Im}} \frac{\partial \hat{v}_{\mathrm{Im}}}{\partial x} + \hat{v}_{\mathrm{Im}} \frac{\partial \hat{v}_{\mathrm{Im}}}{\partial y} + \beta (\hat{w}_{\mathrm{Re}} \hat{v}_{\mathrm{Im}} - \hat{w}_{\mathrm{Im}} \hat{v}_{\mathrm{Re}}) \\
        -\hat{u}_{\mathrm{Re}} \frac{\partial \hat{w}_{\mathrm{Re}}}{\partial x} - \hat{v}_{\mathrm{Re}} \frac{\partial \hat{w}_{\mathrm{Re}}}{\partial y} - \hat{u}_{\mathrm{Im}} \frac{\partial \hat{w}_{\mathrm{Im}}}{\partial x} - \hat{v}_{\mathrm{Im}} \frac{\partial \hat{w}_{\mathrm{Im}}}{\partial y} + \underbrace{\beta (-\hat{w}_{\mathrm{Re}} \hat{w}_{\mathrm{Im}} + \hat{w}_{\mathrm{Im}} \hat{w}_{\mathrm{Re}})}_{=0}
        \end{pmatrix} \\ 
        + 
        \begin{pmatrix}
        -\hat{u}_{\mathrm{Re}} \frac{\partial \hat{u}_{\mathrm{Im}}}{\partial x} - \hat{v}_{\mathrm{Re}} \frac{\partial \hat{u}_{\mathrm{Im}}}{\partial y} + \hat{u}_{\mathrm{Im}} \frac{\partial \hat{u}_{\mathrm{Re}}}{\partial x} + \hat{v}_{\mathrm{Im}} \frac{\partial \hat{u}_{\mathrm{Re}}}{\partial y} + \beta (\hat{w}_{\mathrm{Re}} \hat{u}_{\mathrm{Re}} + \hat{w}_{\mathrm{Im}} \hat{u}_{\mathrm{Im}}) \\
        -\hat{u}_{\mathrm{Re}} \frac{\partial \hat{v}_{\mathrm{Im}}}{\partial x} - \hat{v}_{\mathrm{Re}} \frac{\partial \hat{v}_{\mathrm{Im}}}{\partial y} + \hat{u}_{\mathrm{Im}} \frac{\partial \hat{v}_{\mathrm{Re}}}{\partial x} + \hat{v}_{\mathrm{Im}} \frac{\partial \hat{v}_{\mathrm{Re}}}{\partial y} + \beta (\hat{w}_{\mathrm{Re}} \hat{v}_{\mathrm{Re}} + \hat{w}_{\mathrm{Im}} \hat{v}_{\mathrm{Im}}) \\
        \hat{u}_{\mathrm{Re}} \frac{\partial \hat{w}_{\mathrm{Im}}}{\partial x} + \hat{v}_{\mathrm{Re}} \frac{\partial \hat{w}_{\mathrm{Im}}}{\partial y} - \hat{u}_{\mathrm{Im}} \frac{\partial \hat{w}_{\mathrm{Re}}}{\partial x} - \hat{v}_{\mathrm{Im}} \frac{\partial \hat{w}_{\mathrm{Re}}}{\partial y} + \beta (-\hat{w}_{\mathrm{Re}}^{2} - \hat{w}_{\mathrm{Im}}^{2})
        \end{pmatrix}\mathrm{i},
        \end{split}
    \end{equation}
    \label{eq:convective terms expanded}
\end{subequations}
\noindent and if we add these two terms as in \eqref{eq:NL operator}, we get
\begin{equation}
    \begin{split}
    \mathsfbi{N}_{1}^{1} =
    \begin{pmatrix}
    2 \left [ \hat{u}_{\mathrm{Re}} \frac{\partial \hat{u}_{\mathrm{Re}}}{\partial x} + \hat{v}_{\mathrm{Re}} \frac{\partial \hat{u}_{\mathrm{Re}}}{\partial y} + \hat{u}_{\mathrm{Im}} \frac{\partial \hat{u}_{\mathrm{Im}}}{\partial x} + \hat{v}_{\mathrm{Im}} \frac{\partial \hat{u}_{\mathrm{Im}}}{\partial y} \right ] + 2\beta (\hat{w}_{\mathrm{Re}} \hat{u}_{\mathrm{Im}} - \hat{w}_{\mathrm{Im}} \hat{u}_{\mathrm{Re}}) \\
    2 \left [ \hat{u}_{\mathrm{Re}} \frac{\partial \hat{v}_{\mathrm{Re}}}{\partial x} + \hat{v}_{\mathrm{Re}} \frac{\partial \hat{v}_{\mathrm{Re}}}{\partial y} + \hat{u}_{\mathrm{Im}} \frac{\partial \hat{v}_{\mathrm{Im}}}{\partial x} + \hat{v}_{\mathrm{Im}} \frac{\partial \hat{v}_{\mathrm{Im}}}{\partial y} \right ] + 2\beta (\hat{w}_{\mathrm{Re}} \hat{v}_{\mathrm{Im}} - \hat{w}_{\mathrm{Im}} \hat{v}_{\mathrm{Re}}) \\
    0 \\
    0
    \end{pmatrix} \\ 
    + 
    \begin{pmatrix}
    0 \\
    0 \\
    2 \left [ \hat{u}_{\mathrm{Re}} \frac{\partial \hat{w}_{\mathrm{Im}}}{\partial x} + \hat{v}_{\mathrm{Re}} \frac{\partial \hat{w}_{\mathrm{Im}}}{\partial y} - \hat{u}_{\mathrm{Im}} \frac{\partial \hat{w}_{\mathrm{Re}}}{\partial x} - \hat{v}_{\mathrm{Im}} \frac{\partial \hat{w}_{\mathrm{Re}}}{\partial y} \right ] - 2\beta (\hat{w}_{\mathrm{Re}}^{2} + \hat{w}_{\mathrm{Im}}^{2}) \\
    0
    \end{pmatrix}\mathrm{i},
    \end{split}
    \label{eq:NL operator expanded}
\end{equation}
\noindent which means that the K-H forcing $\hat{\boldsymbol{f}}_{\text{K-H}} = -\mathsfbi{P}^{\top} \mathsfbi{N}_{1}^{1} \left ( \hat{\boldsymbol{w}}_{-1,1}^{*}, \hat{\boldsymbol{w}}_{1,1} \right )$ is a complex 3-component vector with real streamwise ($x$) and wall-normal ($y$) components and imaginary spanwise ($z$) component.
\section{Relative errors $\varepsilon_{\hat{\boldsymbol{w}}}$ and $\varepsilon_{\hat{\boldsymbol{f}}}$}\label{app:relative error}
\subsection{Upper bounds on $\varepsilon_{\hat{\boldsymbol{w}}}$}\label{subsec:upper bounds on epsw}
Here we provide the derivation of the upper bounds of the relative error $\varepsilon_{\hat{\boldsymbol{w}}}$ in \eqref{eq:relative error bounds}. 

Let
\begin{equation}
    \hat{\boldsymbol{w}}_{2,0} - \tilde{\boldsymbol{w}}_{2,0} = \sum_{i=N_{m}+1}^{n\rightarrow\infty} \hat{\boldsymbol{u}}_{i} \sigma_{i} \alpha_{i} \sqrt{\left \langle \hat{\boldsymbol{f}}_{\text{K-H}},\hat{\boldsymbol{f}}_{\text{K-H}} \right \rangle_{\Omega}},
    \label{eq:error vector}
\end{equation}
\noindent be the error between the true non-linear response $\hat{\boldsymbol{w}}_{2,0}$ and the approximation $\tilde{\boldsymbol{w}}_{2,0}$ due to truncation of the linear modes $i=[N_{m}+1,\infty[$. If we substitute this equation and \eqref{eq:response reconstruction} in the definition of the relative error \eqref{eq:relative error response}, we obtain
\begin{equation}
    \varepsilon_{\hat{\boldsymbol{w}}}^2 = \frac{\sum_{i=N_{m}+1}^{n\rightarrow\infty} \sigma_{i}^{2} \left| \alpha_{i} \right|^{2}}{\sum_{i=1}^{n\rightarrow\infty} \sigma_{i}^{2} \left| \alpha_{i} \right|^{2}} \leq \sigma_{N_{m}+1}^{2} \frac{\sum_{i=N_{m}+1}^{n\rightarrow\infty} \left| \alpha_{i} \right|^{2}}{\sum_{i=1}^{n\rightarrow\infty} \sigma_{i}^{2} \left| \alpha_{i} \right|^{2}},
    \label{eq:relative error response squared}
\end{equation}
\noindent and similarly for the relative error associated with the K-H forcing approximation,
\begin{equation}
    \varepsilon_{\hat{\boldsymbol{f}}}^{2} = \frac{\sum_{i=N_{m}+1}^{n\rightarrow\infty} \left| \alpha_{i} \right|^{2}}{\sum_{i=1}^{n\rightarrow\infty} \left| \alpha_{i} \right|^{2}}.
    \label{eq:relative error forcing squared}
\end{equation}
\noindent We can use relations \eqref{eq:relative error response squared}-\eqref{eq:relative error forcing squared} to obtain
\begin{equation}
    \varepsilon_{\hat{\boldsymbol{w}}}^{2} \leq \frac{\sigma_{N_{m}+1}^{2}}{\frac{\sum_{i=1}^{n\rightarrow\infty}\sigma_{i}^{2} \left| \alpha_{i} \right|^{2}}{\sum_{i=1}^{n\rightarrow\infty} \left| \alpha_{i} \right|^{2}}} \varepsilon_{\hat{\boldsymbol{f}}}^{2},
    \label{eq:relative errors related}
\end{equation}
\noindent where
\begin{equation}
    \overline{\sigma}^{2} = \frac{\sum_{i=1}^{n\rightarrow\infty}\sigma_{i}^{2} \left| \alpha_{i} \right|^{2}}{\sum_{i=1}^{n\rightarrow\infty} \left| \alpha_{i} \right|^{2}}.
    \label{eq:sigma bary}
\end{equation}
\noindent Because
\begin{equation*}
    \sum_{i=1}^{n\rightarrow\infty}\sigma_{i}^{2} \left| \alpha_{i} \right|^{2} = \sum_{i=1}^{N_{m}}\sigma_{i}^{2} \left| \alpha_{i} \right|^{2} + \sum_{i=N_{m}+1}^{n\rightarrow\infty}\sigma_{i}^{2} \left| \alpha_{i} \right|^{2},
\end{equation*}
\noindent we obtain
\begin{equation}
    \overline{\sigma}^{2} = \frac{\sum_{i=1}^{N_{m}}\sigma_{i}^{2} \left| \alpha_{i} \right|^{2}}{\sum_{i=1}^{N_{m}} \left| \alpha_{i} \right|^{2}} \frac{\sum_{i=1}^{N_{m}} \left| \alpha_{i} \right|^{2}}{\sum_{i=1}^{n\rightarrow\infty} \left| \alpha_{i} \right|^{2}} + \frac{\sum_{i=N_{m}+1}^{n\rightarrow\infty}\sigma_{i}^{2} \left| \alpha_{i} \right|^{2}}{\sum_{i=N_{m}+1}^{n\rightarrow\infty} \left| \alpha_{i} \right|^{2}} \frac{\sum_{i=N_{m}+1}^{n\rightarrow\infty} \left| \alpha_{i} \right|^{2}}{\sum_{i=1}^{n\rightarrow\infty} \left| \alpha_{i} \right|^{2}},
\end{equation}
\noindent and recalling \eqref{eq:sigma bary N} and letting
\begin{equation}
    \overline{\sigma}_{\infty}^{2} = \frac{\sum_{i=N_{m}+1}^{n\rightarrow\infty}\sigma_{i}^{2} \left| \alpha_{i} \right|^{2}}{\sum_{i=N_{m}+1}^{n\rightarrow\infty} \left| \alpha_{i} \right|^{2}},
\end{equation}
\noindent we get
\begin{equation}
    \overline{\sigma}^{2} = \overline{\sigma}_{N_{m}}^{2} \left ( 1 - \varepsilon_{\hat{\boldsymbol{f}}}^2 \right ) + \overline{\sigma}_{\infty}^{2} \varepsilon_{\hat{\boldsymbol{f}}}^{2} \geq \overline{\sigma}_{N_{m}}^{2} \left ( 1 - \varepsilon_{\hat{\boldsymbol{f}}}^2 \right ).
    \label{eq:sigma bary squared}
\end{equation}
\noindent Finally, the inequality \eqref{eq:relative error response squared} still holds if we inject \eqref{eq:sigma bary squared}, yielding
\begin{equation}
    \varepsilon_{\hat{\boldsymbol{w}}}^{2} \leq \frac{\sigma_{N_{m}+1}^{2}}{\overline{\sigma}_{N_{m}}^{2}} \frac{\varepsilon_{\hat{\boldsymbol{f}}}^{2}}{1 - \varepsilon_{\hat{\boldsymbol{f}}}^{2}} \leq \frac{\varepsilon_{\hat{\boldsymbol{f}}}^{2}}{1 - \varepsilon_{\hat{\boldsymbol{f}}}^{2}},
\end{equation}
\noindent which directly leads to the condition \eqref{eq:relative error bounds}.

\subsection{Component-wise error}\label{subsec:component-wise error}
\begin{figure}
    \centering
    \includegraphics[width=\linewidth]{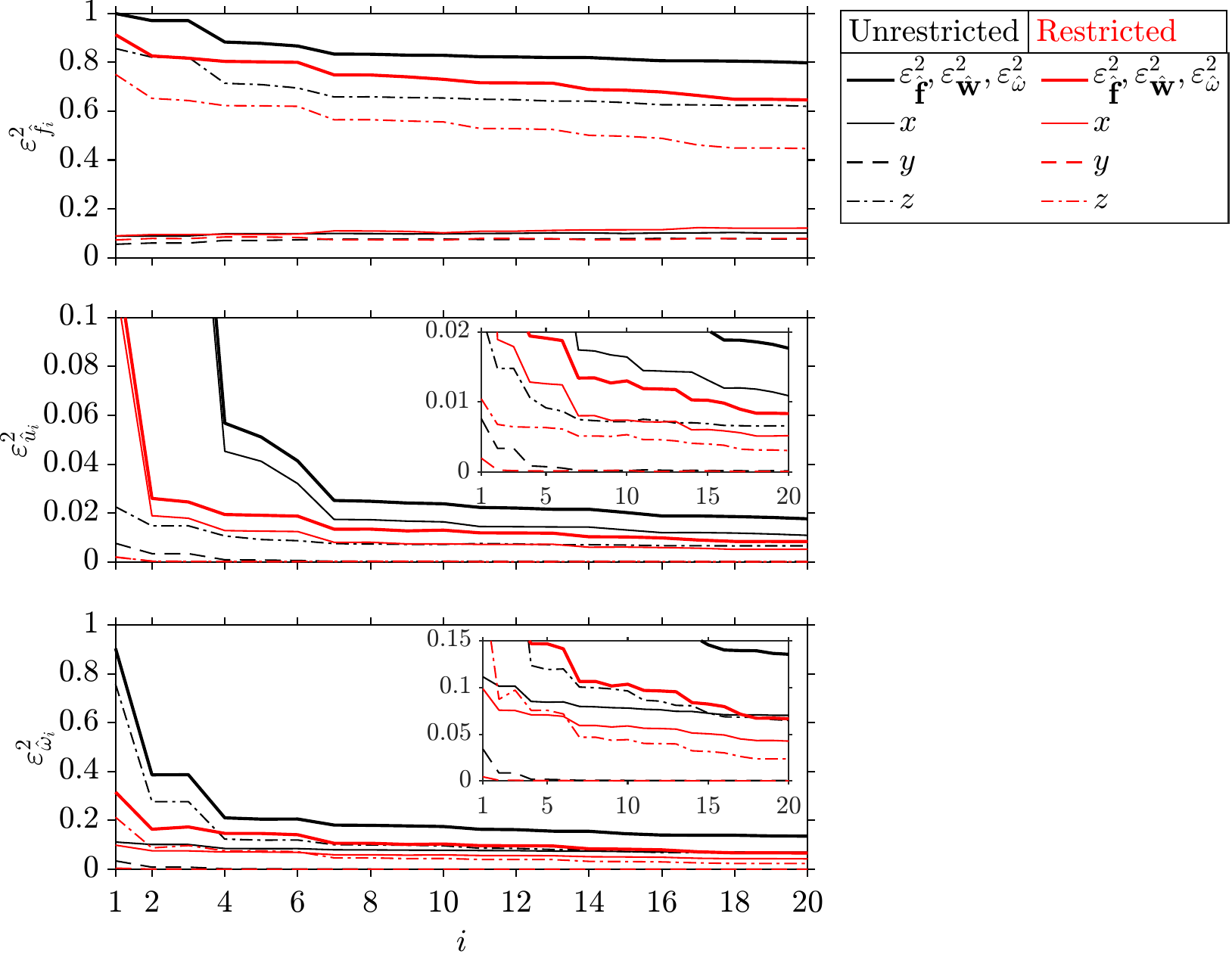}
    \caption{Squared component-wise relative error of the reconstruction of (top) K-H forcing, (middle) non-linear response and (bottom) vorticity computed from the non-linear response.}
    \label{fig:reconstruction errors}
\end{figure}

The contribution of each $x$, $y$ and $z$ component reconstruction to the total (squared) relative error is plotted in figure \ref{fig:reconstruction errors} for the K-H forcing (top), the non-linear response velocities (middle) and vorticities (bottom). The quantities shown are such that the total squared error is the sum of the squares of the component-wise errors. 

For the K-H forcing reconstruction, 70-80\% of the error comes from the $z$-component, while $x$ and $y$ components have equal 10-15\% contribution each. For the non-linear response velocities, the streamwise $\hat{u}$ component contains about 60\% of the error, followed by $\hat{w}$ at 40\% and $\hat{v}$ which has negligible contribution. Finally, for the vorticity computed from the response velocities, $\hat{\omega}_{z}$ and $\hat{\omega}_{x}$ have similar contribution, while $\hat{\omega}_{y}$ carries negligible error.

\bibliographystyle{jfm}
\bibliography{References_FS}

\begin{thebibliography}{77}
\expandafter\ifx\csname natexlab\endcsname\relax\def\natexlab#1{#1}\fi
\def\au#1{#1} \def\ed#1{#1} \def\yr#1{#1}\def\at#1{#1}\def\jt#1{\textit{#1}}
  \def\bt#1{#1}\def\bvol#1{\textbf{#1}} \def\vol#1{#1} \def\pg#1{#1}
  \def\publ#1{#1}\def\arxiv#1{#1}\def\org#1{#1}\def\st#1{\textit{#1}}

\bibitem[Alam(1999)]{alam1999}
{\sc \au{Alam, M.}} \yr{1999} Direct numerical simulation of laminar separation
  bubbles.

\bibitem[Alam \& Sandham(2000)]{alam_sandham_2000}
{\sc \au{Alam, M.} \& \au{Sandham, N.~D.}} \yr{2000}  \at{Direct numerical
  simulation of ‘short’ laminar separation bubbles with turbulent
  reattachment}.  \jt{Journal of Fluid Mechanics}  \bvol{410},  \pg{1–28}.

\bibitem[Andersson {\em et~al.\/}(2001)Andersson, Brandt, Bottaro \&
  Henningson]{andersson_brandt_bottaro_henningson_2001}
{\sc \au{Andersson, P.}, \au{Brandt, L.}, \au{Bottaro, A.} \& \au{Henningson,
  D.~S.}} \yr{2001}  \at{On the breakdown of boundary layer streaks}.
  \jt{Journal of Fluid Mechanics}  \bvol{428},  \pg{29–60}.

\bibitem[Babinsky \& Harvey(2011)]{babinsky_harvey_2011}
{\sc \au{Babinsky, H.} \& \au{Harvey, J.~K.}} \yr{2011} {\em Shock
  Wave-Boundary-Layer Interactions\/}.  \publ{Cambridge University Press}.

\bibitem[Balamurugan \& Mandal(2017)]{Balamurugan_Mandal_2017}
{\sc \au{Balamurugan, G.} \& \au{Mandal, A.~C.}} \yr{2017}  \at{Experiments on
  localized secondary instability in bypass boundary layer transition}.
  \jt{Journal of Fluid Mechanics}  \bvol{817},  \pg{217–263}.

\bibitem[Balay {\em et~al.\/}(1998)Balay, Gropp, McInnes \& Smith]{petsc}
{\sc \au{Balay, S.}, \au{Gropp, W.}, \au{McInnes, L.~C.} \& \au{Smith, B.~F}}
  \yr{1998}  \at{Petsc, the portable, extensible toolkit for scientific
  computation}.  \jt{Argonne National Laboratory}  \bvol{2}~(17).

\bibitem[Bernardos {\em et~al.\/}(2019{\natexlab{{\em a\/}}})Bernardos, Richez,
  Gleize \& Gerolymos]{bernardos2019algebraic}
{\sc \au{Bernardos, L.}, \au{Richez, F.}, \au{Gleize, V.} \& \au{Gerolymos,
  G.A.}} \yr{2019{\natexlab{{\em a\/}}}}  \at{Algebraic nonlocal transition
  modeling of laminar separation bubbles using k- $\omega$ turbulence models}.
  \jt{AIAA Journal}  \bvol{57}~(2),  \pg{553--565}.

\bibitem[Bernardos {\em et~al.\/}(2019{\natexlab{{\em b\/}}})Bernardos, Richez,
  Gleize \& Gerolymos]{bernardos2019prediction}
{\sc \au{Bernardos, L.}, \au{Richez, F.}, \au{Gleize, V.} \& \au{Gerolymos,
  G.A.}} \yr{2019{\natexlab{{\em b\/}}}}  \at{Prediction of separation-induced
  transition on the sd7003 airfoil using algebraic transition triggering}.
  \jt{AIAA Journal}  \bvol{57}~(9),  \pg{3812--3824}.

\bibitem[Borgmann {\em et~al.\/}(2025)Borgmann, Hosseinverdi, Little \&
  Fasel]{Borgmann_Hosseinverdi_Little_Fasel_2025}
{\sc \au{Borgmann, D.}, \au{Hosseinverdi, S.}, \au{Little, J.} \& \au{Fasel,
  H.}} \yr{2025}  \at{Experimental and numerical investigations of transition
  in a pressure-gradient-induced laminar separation bubble}.  \jt{Journal of
  Fluid Mechanics}  \bvol{1007},  \pg{A23}.

\bibitem[Brandt {\em et~al.\/}(2011)Brandt, Sipp, Pralits \&
  Marquet]{Brandt_Sipp_Pralits_Marquet_2011}
{\sc \au{Brandt, L.}, \au{Sipp, D.}, \au{Pralits, J.~O.} \& \au{Marquet, O.}}
  \yr{2011}  \at{Effect of base-flow variation in noise amplifiers: the
  flat-plate boundary layer}.  \jt{Journal of Fluid Mechanics}  \bvol{687},
  \pg{503–528}.

\bibitem[Casacuberta {\em et~al.\/}(2024)Casacuberta, Hickel \&
  Kotsonis]{casacuberta2024}
{\sc \au{Casacuberta, J.}, \au{Hickel, S.} \& \au{Kotsonis, M.}} \yr{2024}
  \at{Passive stabilization of crossflow instabilities by a reverse lift-up
  effect}.  \jt{Physical Review Fluids}  \bvol{9},  \pg{043903}.

\bibitem[Casacuberta {\em et~al.\/}(2022)Casacuberta, Hickel, Westerbeek \&
  Kotsonis]{casacuberta_hickel_westerbeek_kotsonis_2022}
{\sc \au{Casacuberta, J.}, \au{Hickel, S.}, \au{Westerbeek, S.} \&
  \au{Kotsonis, M.}} \yr{2022}  \at{Direct numerical simulation of interaction
  between a stationary crossflow instability and forward-facing steps}.
  \jt{Journal of Fluid Mechanics}  \bvol{943},  \pg{A46}.

\bibitem[Cura {\em et~al.\/}(2024)Cura, Hanifi, Cavalieri \&
  Weiss]{Cura_Hanifi_Cavalieri_Weiss_2024}
{\sc \au{Cura, C.}, \au{Hanifi, A.}, \au{Cavalieri, A.V.G.} \& \au{Weiss, J.}}
  \yr{2024}  \at{On the low-frequency dynamics of turbulent separation
  bubbles}.  \jt{Journal of Fluid Mechanics}  \bvol{991},  \pg{A11}.

\bibitem[Dellacasagrande {\em et~al.\/}(2024)Dellacasagrande, Lengani, Simoni
  \& Ubaldi]{dellacasagrande_lengani_simoni_ubaldi2024}
{\sc \au{Dellacasagrande, M.}, \au{Lengani, D.}, \au{Simoni, D.} \& \au{Ubaldi,
  M.}} \yr{2024}  \at{{An Experimental Database for the Analysis of Bursting of
  a Laminar Separation Bubble}}.  \jt{International Journal of Turbomachinery,
  Propulsion and Power}  \bvol{9}~(1).

\bibitem[Diwan \& Ramesh(2009)]{diwan_ramesh_2009}
{\sc \au{Diwan, S.~S.} \& \au{Ramesh, O.~N.}} \yr{2009}  \at{On the origin of
  the inflectional instability of a laminar separation bubble}.  \jt{Journal of
  Fluid Mechanics}  \bvol{629},  \pg{263–298}.

\bibitem[Duck {\em et~al.\/}(2000)Duck, Ruban, Theofilis, Hein \&
  Dallmann]{theofilis2000}
{\sc \au{Duck, P.~W.}, \au{Ruban, A.~I.}, \au{Theofilis, V.}, \au{Hein, S.} \&
  \au{Dallmann, U.}} \yr{2000}  \at{On the origins of unsteadiness and
  three-dimensionality in a laminar separation bubble}.  \jt{Philosophical
  Transactions of the Royal Society of London. Series A: Mathematical, Physical
  and Engineering Sciences}  \bvol{358}~(1777),  \pg{3229--3246}.

\bibitem[Dwivedi {\em et~al.\/}(2022)Dwivedi, Sidharth \&
  Jovanović]{dwivedi_sidharth_jovanovic_2022}
{\sc \au{Dwivedi, A.}, \au{Sidharth, G.S.} \& \au{Jovanović, M.~R.}} \yr{2022}
   \at{Oblique transition in hypersonic double-wedge flow}.  \jt{Journal of
  Fluid Mechanics}  \bvol{948},  \pg{A37}.

\bibitem[Ehrenstein \& Gallaire(2008)]{ehrenstein_gallaire_2008}
{\sc \au{Ehrenstein, U.} \& \au{Gallaire, F.}} \yr{2008}  \at{Two-dimensional
  global low-frequency oscillations in a separating boundary-layer flow}.
  \jt{Journal of Fluid Mechanics}  \bvol{614},  \pg{315–327}.

\bibitem[Fitzgerald \& Mueller(1990)]{fitzgerald_mueller1990}
{\sc \au{Fitzgerald, E.~J.} \& \au{Mueller, T.~J.}} \yr{1990}  \at{Measurements
  in a separation bubble on an airfoil using laser velocimetry}.  \jt{AIAA
  Journal}  \bvol{28}~(4),  \pg{584--592}.

\bibitem[Gallaire {\em et~al.\/}(2007)Gallaire, Marquillie \&
  Ehrenstein]{Gallaire2007three-dimensional}
{\sc \au{Gallaire, F.}, \au{Marquillie, M.} \& \au{Ehrenstein, U.}} \yr{2007}
  \at{Three-dimensional transverse instabilities in detached boundary layers}.
  \jt{Journal of Fluid Mechanics}  \bvol{571},  \pg{221--33}.

\bibitem[Gaster(1963)]{gaster1963}
{\sc \au{Gaster, M.}} \yr{1963} On the stability of parallel flows and the
  behaviour of separation bubbles.

\bibitem[Gaster(1969)]{Gaster1969}
{\sc \au{Gaster, M.}} \yr{1969}  \bt{The structure and behavior of laminar
  separation bubbles}.  \publ{London: H.M.S.O}.

\bibitem[George(2013)]{George2013}
{\sc \au{George, W.~K.}} \yr{2013} {\em Lectures in Turbulence for the 21st
  Century\/}.  \publ{Chalmers University of Technology}.

\bibitem[G{\"o}rtler(1941)]{Gortler1941}
{\sc \au{G{\"o}rtler, H.}} \yr{1941}  \at{Instabilit{\"a}t laminarer
  grenzschichten an konkaven w{\"a}nden gegen{\"u}ber gewissen
  dreidimensionalen st{\"o}rungen}.  \jt{Zamm-zeitschrift Fur Angewandte
  Mathematik Und Mechanik}  \bvol{21},  \pg{250--252}.

\bibitem[Griffin {\em et~al.\/}(2013)Griffin, Oyarzun, Cattafesta, Tu \&
  Rowley]{griffin_oyarzun_cattafesta_tu_rowley_2013}
{\sc \au{Griffin, J.~C.}, \au{Oyarzun, M.}, \au{Cattafesta, L.~N.}, \au{Tu,
  J.~H.} \& \au{Rowley, C.~W.}} \yr{2013} {\em Control of a canonical separated
  flow\/}.  \publ{American Institute of Aeronautics and Astronautics}.

\bibitem[Hecht(2012)]{freefemHecht}
{\sc \au{Hecht, F.}} \yr{2012}  \at{New development in freefem++}.  \jt{Journal
  of Numerical Mathematics}  \bvol{20}~(3-4),  \pg{251--265}.

\bibitem[Hildebrand {\em et~al.\/}(2018)Hildebrand, Dwivedi, Nichols,
  Jovanović \& Candler]{Hildebrand2018}
{\sc \au{Hildebrand, N.}, \au{Dwivedi, A.}, \au{Nichols, J.~W.},
  \au{Jovanović, M.~R.} \& \au{Candler, G.~V.}} \yr{2018}  \at{Simulation and
  stability analysis of oblique shock-wave/boundary-layer interactions at mach
  5.92}.  \jt{Physical Review Fluids}  \bvol{3}.

\bibitem[Horton(1968)]{horton1968}
{\sc \au{Horton, H.~P.}} \yr{1968} Laminar separation bubbles in two and three
  dimensional incompressible flow.

\bibitem[Hosseinverdi \& Fasel(2019)]{hosseinverdi_fasel_2019}
{\sc \au{Hosseinverdi, S.} \& \au{Fasel, H.~F.}} \yr{2019}  \at{Numerical
  investigation of laminar–turbulent transition in laminar separation
  bubbles: the effect of free-stream turbulence}.  \jt{Journal of Fluid
  Mechanics}  \bvol{858},  \pg{714–759}.

\bibitem[Hosseinverdi \& Fasel(2020)]{Hosseinverdi2020onset}
{\sc \au{Hosseinverdi, S.} \& \au{Fasel, H.~F.}} \yr{2020} Onset of
  three-dimensionality and transition in controlled separation bubbles:
  Secondary instability analysis and direct numerical simulations.  \bt{In {\em
  AIAA Scitech Forum, 2020, January 6, 2020 - January 10, 20\/}}, ,  \vol{vol.
  1 PartF}. Department of Aerospace and Mechanical Engineering, University of
  Arizona, Tucson; AZ; 85721, United States,  \publ{Orlando, FL, United states:
  American Institute of Aeronautics and Astronautics Inc, AIAA}.

\bibitem[Jacobs \& Durbin(2001)]{jacobs_durbin_2001}
{\sc \au{Jacobs, R.~G.} \& \au{Durbin, P.~A.}} \yr{2001}  \at{Simulations of
  bypass transition}.  \jt{Journal of Fluid Mechanics}  \bvol{428},
  \pg{185–212}.

\bibitem[Jahanmiri(2011)]{Jahanmiri2011LaminarSB}
{\sc \au{Jahanmiri, M.}} \yr{2011}  \bt{Laminar separation bubble: Its
  structure, dynamics and control}.  \publ{Chalmers Publication Library}.

\bibitem[Jones {\em et~al.\/}(2010)Jones, Sandberg \&
  Sandham]{Jones2010stabilityaerofoil}
{\sc \au{Jones, L.~E.}, \au{Sandberg, R.~D.} \& \au{Sandham, N.~D.}} \yr{2010}
  \at{Stability and receptivity characteristics of a laminar separation bubble
  on an aerofoil}.  \jt{Journal of Fluid Mechanics}  \bvol{648},  \pg{257--96}.

\bibitem[Karp \& Hack(2020)]{Karp2020suppression}
{\sc \au{Karp, M.} \& \au{Hack, M. J.~P.}} \yr{2020}  \at{Optimal suppression
  of a separation bubble in a laminar boundary layer}.  \jt{Journal of Fluid
  Mechanics}  \bvol{892}.

\bibitem[Klebanoff {\em et~al.\/}(1962)Klebanoff, Tidstrom \&
  Sargent]{klebanoff_tidstrom_sargent_1962}
{\sc \au{Klebanoff, P.~S.}, \au{Tidstrom, K.~D.} \& \au{Sargent, L.~M.}}
  \yr{1962}  \at{The three-dimensional nature of boundary-layer instability}.
  \jt{Journal of Fluid Mechanics}  \bvol{12}~(1),  \pg{1–34}.

\bibitem[Kumar \& Sarkar(2023)]{kumar2023}
{\sc \au{Kumar, R.} \& \au{Sarkar, S.}} \yr{2023}  \at{{Features of laminar
  separation bubble subjected to varying adverse pressure gradients}}.
  \jt{Physics of Fluids}  \bvol{35}~(12),  \pg{124104}.

\bibitem[Li \& Malik(1995)]{li_malik_1995}
{\sc \au{Li, F.} \& \au{Malik, M.~R.}} \yr{1995}  \at{Fundamental and
  subharmonic secondary instabilities of görtler vortices}.  \jt{Journal of
  Fluid Mechanics}  \bvol{297},  \pg{77–100}.

\bibitem[Lopez \& Bulbeck(1993)]{lopez_bulbeck1993}
{\sc \au{Lopez, J.~M.} \& \au{Bulbeck, C.~J.}} \yr{1993}  \at{{Behavior of
  streamwise rib vortices in a three‐dimensional mixing layer}}.  \jt{Physics
  of Fluids A: Fluid Dynamics}  \bvol{5}~(7),  \pg{1694--1702}.

\bibitem[Marxen \& Henningson(2011)]{marxen_henningson_2011}
{\sc \au{Marxen, O.} \& \au{Henningson, D.~S.}} \yr{2011}  \at{The effect of
  small-amplitude convective disturbances on the size and bursting of a laminar
  separation bubble}.  \jt{Journal of Fluid Mechanics}  \bvol{671},
  \pg{1--33}.

\bibitem[Marxen {\em et~al.\/}(2013)Marxen, Lang \& Rist]{Marxen2013vortex}
{\sc \au{Marxen, O.}, \au{Lang, M.} \& \au{Rist, U.}} \yr{2013}  \at{Vortex
  formation and vortex breakup in a laminar separation bubble}.  \jt{Journal of
  Fluid Mechanics}  \bvol{728},  \pg{58--90}.

\bibitem[Mauriello {\em et~al.\/}(2022)Mauriello, Larchev{\^e}que \&
  Dupont]{mauriello2022}
{\sc \au{Mauriello, M.}, \au{Larchev{\^e}que, L.} \& \au{Dupont, P.}} \yr{2022}
  Non-linearities in the low-frequency dynamics of transitional shock wave /
  boundary layer interactions.  \bt{In {\em 56th 3AF International Conference
  on Applied Aerodynamics\/}}. Toulouse, France.

\bibitem[Michelis {\em et~al.\/}(2018{\natexlab{{\em a\/}}})Michelis, Kotsonis
  \& Yarusevych]{Michelis2018spanwise}
{\sc \au{Michelis, T.}, \au{Kotsonis, M.} \& \au{Yarusevych, S.}}
  \yr{2018{\natexlab{{\em a\/}}}}  \at{Spanwise flow development within a
  laminar separation bubble under natural and forced transition}.
  \jt{Experimental Thermal and Fluid Science}  \bvol{96},  \pg{169--79}.

\bibitem[Michelis {\em et~al.\/}(2017)Michelis, Yarusevych \&
  Kotsonis]{Michelis2017impulsiveforcing}
{\sc \au{Michelis, T.}, \au{Yarusevych, S.} \& \au{Kotsonis, M.}} \yr{2017}
  \at{Response of a laminar separation bubble to impulsive forcing}.
  \jt{Journal of Fluid Mechanics}  \bvol{820},  \pg{633--66}.

\bibitem[Michelis {\em et~al.\/}(2018{\natexlab{{\em b\/}}})Michelis,
  Yarusevych \& Kotsonis]{Michelis2018origin}
{\sc \au{Michelis, T.}, \au{Yarusevych, S.} \& \au{Kotsonis, M.}}
  \yr{2018{\natexlab{{\em b\/}}}}  \at{On the origin of spanwise vortex
  deformations in laminar separation bubbles}.  \jt{Journal of Fluid Mechanics}
   \bvol{841},  \pg{81--108}.

\bibitem[Mohamed~Aniffa {\em et~al.\/}(2023)Mohamed~Aniffa, Caesar, Dabaria \&
  Mandal]{mohamed_aniffa_caesar_dabaria_mandal_2023}
{\sc \au{Mohamed~Aniffa, S.}, \au{Caesar, V.S.}, \au{Dabaria, V.} \&
  \au{Mandal, A.C.}} \yr{2023}  \at{Characteristics of geometry-and
  pressure-induced laminar separation bubbles at an enhanced level of
  free-stream turbulence}.  \jt{Journal of Fluid Mechanics}  \bvol{957},
  \pg{A19}.

\bibitem[Na \& Moin(1998)]{Na_Moin_1998}
{\sc \au{Na, Y.} \& \au{Moin, P.}} \yr{1998}  \at{Direct numerical simulation
  of a separated turbulent boundary layer}.  \jt{Journal of Fluid Mechanics}
  \bvol{374},  \pg{379–405}.

\bibitem[O'Meara \& Mueller(1987)]{o_meara1987}
{\sc \au{O'Meara, M.~M.} \& \au{Mueller, T.~J.}} \yr{1987}  \at{Laminar
  separation bubble characteristics on an airfoil at low reynolds numbers}.
  \jt{AIAA Journal}  \bvol{25}~(8),  \pg{1033--1041}.

\bibitem[Pope(2000)]{Pope2000turbulent}
{\sc \au{Pope, S.~B.}} \yr{2000} {\em Turbulent Flows\/}.  \publ{Cambridge
  University Press}.

\bibitem[Rigas {\em et~al.\/}(2021)Rigas, Sipp \& Colonius]{rigas2021HBM}
{\sc \au{Rigas, G.}, \au{Sipp, D.} \& \au{Colonius, T.}} \yr{2021}
  \at{Nonlinear input/output analysis: application to boundary layer
  transition}.  \jt{Journal of Fluid Mechanics}  \bvol{911}.

\bibitem[Rist(2002)]{rist2002}
{\sc \au{Rist, U.}} \yr{2002}  \at{On instabilities and transition in laminar
  separation bubbles}.  \jt{Proceedings CEAS Aerospace Aerodynamics Research
  Conference} .

\bibitem[Rist \& Maucher(2002)]{rist_maucher_2002}
{\sc \au{Rist, U.} \& \au{Maucher, U.}} \yr{2002}  \at{Investigations of
  time-growing instabilities in laminar separation bubbles}.  \jt{European
  Journal of Mechanics - B/Fluids}  \bvol{21}~(5),  \pg{495--509}.

\bibitem[Rodriguez \& Gennaro(2015)]{Rodriguez2015secondary}
{\sc \au{Rodriguez, D.} \& \au{Gennaro, E.~M.}} \yr{2015}  \at{On the secondary
  instability of forced and unforced laminar separation bubbles}.  \jt{Procedia
  IUTAM}  \bvol{14},  \pg{78--87}.

\bibitem[Rodriguez {\em et~al.\/}(2021)Rodriguez, Gennaro \&
  Souza]{Rodriguez2021self-excited}
{\sc \au{Rodriguez, D.}, \au{Gennaro, E.~M.} \& \au{Souza, L.~F.}} \yr{2021}
  \at{Self-excited primary and secondary instability of laminar separation
  bubbles}.  \jt{Journal of Fluid Mechanics}  \bvol{906},  \pg{A13 (34 pp.)}.

\bibitem[Rodríguez {\em et~al.\/}(2013)Rodríguez, Gennaro \&
  Juniper]{Rodriguez_gennaro_juniper_2013}
{\sc \au{Rodríguez, D.}, \au{Gennaro, E.~M.} \& \au{Juniper, M.~P.}} \yr{2013}
   \at{The two classes of primary modal instability in laminar separation
  bubbles}.  \jt{Journal of Fluid Mechanics}  \bvol{734},  \pg{R4}.

\bibitem[Russell(1979)]{Russell1979}
{\sc \au{Russell, J.~M.}} \yr{1979}  \bt{Length and bursting of separation
  bubbles: A physical interpretation}.  \publ{NASA. Langley Res. Center The
  Sci. and Technol. of Low Speed and Motorless Flight}.

\bibitem[Sandham(2008)]{sandham_2008}
{\sc \au{Sandham, N.~D.}} \yr{2008}  \at{Transitional separation bubbles and
  unsteady aspects of aerofoil stall}.  \jt{The Aeronautical Journal (1968)}
  \bvol{112}~(1133),  \pg{395–404}.

\bibitem[Sansica {\em et~al.\/}(2016)Sansica, Sandham \&
  Hu]{Sansica_Sandham_Hu_2016}
{\sc \au{Sansica, A.}, \au{Sandham, N.~D.} \& \au{Hu, Z.}} \yr{2016}
  \at{Instability and low-frequency unsteadiness in a shock-induced laminar
  separation bubble}.  \jt{Journal of Fluid Mechanics}  \bvol{798},
  \pg{5–26}.

\bibitem[Saric(1994)]{Saric1994}
{\sc \au{Saric, W.~S.}} \yr{1994}  \at{Görtler vortices}.  \jt{Annual Review
  of Fluid Mechanics}  \bvol{26}~(1),  \pg{379--409}.

\bibitem[Schmid \& Henningson(1992)]{schmid1992new}
{\sc \au{Schmid, P.J.} \& \au{Henningson, D.S.}} \yr{1992}  \at{A new mechanism
  for rapid transition involving a pair of oblique waves}.  \jt{Physics of
  Fluids A: Fluid Dynamics}  \bvol{4}~(9),  \pg{1986--1989}.

\bibitem[Schmid \& Henningson(2001)]{schmid_henningson2001stability}
{\sc \au{Schmid, P.~J.} \& \au{Henningson, D.~S.}} \yr{2001} {\em Stability and
  Transition in Shear Flows\/}.  \publ{Springer}.

\bibitem[Shinde {\em et~al.\/}(2020)Shinde, McNamara \& Gaitonde]{shinde2020}
{\sc \au{Shinde, V.}, \au{McNamara, J.} \& \au{Gaitonde, D.}} \yr{2020}
  \at{Control of transitional shock wave boundary layer interaction using
  structurally constrained surface morphing}.  \jt{Aerospace Science and
  Technology}  \bvol{96},  \pg{105545}.

\bibitem[Shinde {\em et~al.\/}(2019)Shinde, McNamara, Gaitonde, Barnes \&
  Visbal]{shinde2019}
{\sc \au{Shinde, V.}, \au{McNamara, J.}, \au{Gaitonde, D.}, \au{Barnes, C.} \&
  \au{Visbal, M.}} \yr{2019}  \at{Transitional shock wave boundary layer
  interaction over a flexible panel}.  \jt{Journal of Fluids and Structures}
  \bvol{90},  \pg{263--285}.

\bibitem[Simoni {\em et~al.\/}(2013)Simoni, Ubaldi \&
  Zunino]{Simoni2013experimentalwakes}
{\sc \au{Simoni, D.}, \au{Ubaldi, M.} \& \au{Zunino, P.}} \yr{2013}
  \at{Experimental investigation of the interaction between incoming wakes and
  instability mechanisms in a laminar separation bubble}.  \jt{Experimental
  Thermal and Fluid Science}  \bvol{50},  \pg{54--60}.

\bibitem[Simoni {\em et~al.\/}(2014)Simoni, Ubaldi \&
  Zunino]{Simoni2014experimentalLSB}
{\sc \au{Simoni, D.}, \au{Ubaldi, M.} \& \au{Zunino, P.}} \yr{2014}
  \at{Experimental investigation of flow instabilities in a laminar separation
  bubble}.  \jt{Journal of Thermal Science}  \bvol{23}~(3),  \pg{203--214}.

\bibitem[Sipp \& Jacquin(2000)]{sipp_jacquin2000}
{\sc \au{Sipp, D.} \& \au{Jacquin, L.}} \yr{2000}  \at{{Three-dimensional
  centrifugal-type instabilities of two-dimensional flows in rotating
  systems}}.  \jt{Physics of Fluids}  \bvol{12}~(7),  \pg{1740--1748}.

\bibitem[Sipp {\em et~al.\/}(1999)Sipp, Lauga \&
  Jacquin]{sipp_lauga_jacquin1999}
{\sc \au{Sipp, D.}, \au{Lauga, E.} \& \au{Jacquin, L.}} \yr{1999}
  \at{{Vortices in rotating systems: Centrifugal, elliptic and hyperbolic type
  instabilities}}.  \jt{Physics of Fluids}  \bvol{11}~(12),  \pg{3716--3728}.

\bibitem[Sun {\em et~al.\/}(2014)Sun, Schrijer, Scarano \& van
  Oudheusden]{sun_schrijer_scarano_van_oudheusden2014}
{\sc \au{Sun, Z.}, \au{Schrijer, F. F.~J.}, \au{Scarano, F.} \& \au{van
  Oudheusden, B.~W.}} \yr{2014}  \at{{Decay of the supersonic turbulent wakes
  from micro-ramps}}.  \jt{Physics of Fluids}  \bvol{26}~(2),  \pg{025115}.

\bibitem[Symon {\em et~al.\/}(2018)Symon, Rosenberg, Dawson \&
  McKeon]{Symon2018nonnormality_resolvent}
{\sc \au{Symon, S.}, \au{Rosenberg, K.}, \au{Dawson, S. T.~M.} \& \au{McKeon,
  B.~J.}} \yr{2018}  \at{Non-normality and classification of amplification
  mechanisms in stability and resolvent analysis}.  \jt{Physical Review Fluids}
   \bvol{3}.

\bibitem[Tani(1964)]{tani1964}
{\sc \au{Tani, I.}} \yr{1964}  \at{Low-speed flows involving bubble
  separations}.  \jt{Progress in Aerospace Sciences}  \bvol{5},  \pg{70--103}.

\bibitem[White(1991)]{White1991}
{\sc \au{White, F.M.}} \yr{1991} {\em Viscous fluid flow\/}.  \publ{McGraw-Hill
  New York}.

\bibitem[Xu {\em et~al.\/}(2017)Xu, Mughal, Gowree, Atkin \&
  Sherwin]{Xu_Mughal_Gowree_Atkin_Sherwin_2017}
{\sc \au{Xu, H.}, \au{Mughal, S.~M.}, \au{Gowree, E.~R.}, \au{Atkin, C.~J.} \&
  \au{Sherwin, S.~J.}} \yr{2017}  \at{Destabilisation and modification of
  tollmien–schlichting disturbances by a three-dimensional surface
  indentation}.  \jt{Journal of Fluid Mechanics}  \bvol{819},  \pg{592–620}.

\bibitem[Yang(2019)]{yang2019}
{\sc \au{Yang, Z.}} \yr{2019}  \at{Secondary instability of separated shear
  layers}.  \jt{Chinese Journal of Aeronautics}  \bvol{32}~(1),  \pg{37--44}.

\bibitem[Yarusevych \& Kotsonis(2017)]{Yarusevych2017steady}
{\sc \au{Yarusevych, S.} \& \au{Kotsonis, M.}} \yr{2017}  \at{Steady and
  transient response of a laminar separation bubble to controlled
  disturbances}.  \jt{Journal of Fluid Mechanics}  \bvol{813},  \pg{955--990}.

\bibitem[Yeh {\em et~al.\/}(2020)Yeh, Benton, Taira \&
  Garmann]{yeh_benton_taira_garmann2020}
{\sc \au{Yeh, C.-A.}, \au{Benton, S.~I.}, \au{Taira, K.} \& \au{Garmann,
  D.~J.}} \yr{2020}  \at{Resolvent analysis of an airfoil laminar separation
  bubble at {Re} = 500,000}.  \jt{Physical Review Fluids}  \bvol{5},
  \pg{083906}.

\bibitem[Zhang {\em et~al.\/}(2019)Zhang, Rowley, Wu, Meneveau \&
  Mittal]{Zhang2019input-output}
{\sc \au{Zhang, H.}, \au{Rowley, C.}, \au{Wu, W.}, \au{Meneveau, C.} \&
  \au{Mittal, R.}} \yr{2019} Input-output analysis of a separated flow past a
  flat plate.  \bt{In {\em 2019 AIAA Science and Technology Forum and
  Exposition (SciTech), 7-11 Jan. 20\/}},  \pg{p. 11 pp.} Princeton University,
  Mechanical and Aerospace Engineering, Princeton, NJ 08544, United States,
  \publ{Reston, VA, USA: American Institute of Aeronautics and Astronautics}.

\bibitem[Zhang \& Samtaney(2016)]{Zhang2016biglobal}
{\sc \au{Zhang, W.} \& \au{Samtaney, R.}} \yr{2016}  \at{{BiGlobal linear
  stability analysis on low-Re flow past an airfoil at high angle of attack}}.
  \jt{Physics of Fluids}  \bvol{28}~(4).

\bibitem[Ziade {\em et~al.\/}(2018)Ziade, Feero, Lavoie \&
  Sullivan]{Ziade2018shear}
{\sc \au{Ziade, P.}, \au{Feero, M.~A.}, \au{Lavoie, P.} \& \au{Sullivan,
  P.~E.}} \yr{2018}  \at{Shear layer development, separation, and stability
  over a low-reynolds number airfoil}.  \jt{Journal of Fluids Engineering,
  Transactions of the ASME}  \bvol{140}~(7).

\end{thebibliography}
\end{document}